\documentclass[english,12pt]{article}
\usepackage[T1]{fontenc}
\usepackage[utf8]{inputenc}
\usepackage[english]{babel}
\usepackage{microtype}
\usepackage[top=1.25in, bottom=1.25in, left=1.25in, right=1.25in]{geometry}
\usepackage[pdftex]{graphicx}
\usepackage{amsmath,amssymb,amsopn,dsfont,bm,mathrsfs,marvosym,stmaryrd,array}
\usepackage[width=\textwidth]{caption}
\usepackage{threeparttable,multirow,longtable,booktabs}
\usepackage{lmodern}
\usepackage{float, placeins,flafter}
\usepackage{verbatim}
\usepackage{color}
\usepackage{natbib}
\usepackage{xr}
\usepackage{lscape}
\usepackage{afterpage}
\usepackage{numprint}
\usepackage{subfig}
\usepackage[hidelinks]{hyperref}

\newtheorem{thm}{Theorem}[section]

\newtheorem{rmk}[thm]{Remark}
\newtheorem{lem}{Lemma}[section]

\newtheorem{hyp}{Assumption}

\newcommand{\ind}[1]{\mathds{1}\left\{#1\right\}}
\newcommand{\E}{\mathbb E}
\newcommand{\V}{\mathbb V}
\newcommand{\R}{\mathbb R}
\newcommand{\N}{\mathbb N}

\newcommand{\Supp}{\text{Supp}}
\newcommand{\indep}{\perp \!\!\! \perp}
\newcommand{\eps}{\varepsilon}

\newcolumntype{C}[1]{>{\centering\let\newline\\\arraybackslash\hspace{0pt}}m{#1}}


\title{~\\[-4.5cm]
Instrument-Free Demand Estimation Using Relative Prices Variation, with an Application to Railway Transportation\thanks{We thank people at iDTGV for providing us the data. We also thank Philippe Chon\'e, Pierre Dubois, Carlos Daniel Santos, Laurent Lamy, Lars Nesheim, Matt Shum and seminar participants at the California Institute of Technology, University of Mannheim, Rice University, Sciences-Po, Texas A\&M, Toulouse School of Economics, Yale University and EARIE in Athens, the 3rd CREST-ECODEC Conference,  the 5th French Econometrics Conference, the 2022 European Winter Meeting of the Econometric Society, the Microeconometrics Class of 2020/2021 Conference at Duke University, the IAAE in Oslo, the ESG in Bristol for their comments.}}

\date{}
\begin{document}

\author{Xavier D'Haultf\oe{}uille%
\thanks{CREST, ENSAE, Institut Polytechnique de Paris, xavier.dhaultfoeuille@ensae.fr.} \and Ao Wang\thanks{University of Warwick and CAGE, ao.wang@warwick.ac.uk.} \and Philippe F\'evrier\thanks{CREST, ENSAE, Institut Polytechnique de Paris, fevrier@ensae.fr.} \and Lionel Wilner\thanks{CREST, ENSAE, Institut Polytechnique de Paris, lionel.wilner@ensae.fr.} }
\maketitle ~\vspace{-1.3cm}
\begin{abstract}
We develop a new identification strategy for demand estimation when cost shifters may not be available and there are substantial variations in demand over time. This approaches relies on a kind of nonlinear difference-in-differences, in which price elasticities are identified by relating changes over time in relative purchases between two goods to changes in their relative prices. We apply this strategy to the context of French railway transportation. Our price elasticity estimates are in line with those on airlines, but substantially higher, in absolute terms, than those generally obtained on railway transportation. We then use our demand estimation to compare the current pricing with several counterfactual pricing strategies. Our results suggest similar or better performance of the actual revenue management compared to optimal uniform pricing, but also substantial losses compared to the optimal pricing strategy. Finally, we highlight the key role of revenue management in acquiring information when demand is uncertain.
\end{abstract}


\newpage

\section{Introduction}

Demand estimation in the presence of price endogeneity is a very old but still active research topic in economics. A common strategy is to use instrumental variables that exogenously shift prices but do not directly affect nor depend on demand. In some cases, however, this strategy may fail. For instance, such instruments may be difficult to find or are simply not available because either marginal costs are close to zero or, more generally, they do not substantially vary over the  period under consideration. 
In this article, we develop a novel, instrument-free demand estimation strategy that identifies price sensitivity from within-market relative price variation. 
Specifically, we show how price elasticities can be identified by relating changes in relative purchases to changes in relative price across periods, in the spirit of difference-in-differences. 

\medskip
Our strategy relies on three main conditions. To detail them, consider two goods in a given market. First, we impose a multiplicative separability on consumers' intensity of arrival. In particular, whereas we leave unrestricted the pattern of consumers' arrival over time, we assume that it is the same for the two goods, an analogue of the parallel trend assumption in difference-in-differences. Second, we suppose the two goods do not compete, in the sense that their demands are independent once conditioned on market-specific variables. 
Third, we suppose that total purchases are a sufficient statistic for times at which prices change. 
The latter assumption is credible when prices change at deterministic times or change simultaneously for both goods, as in our application. Even if these restrictions may be strong, we show that they are jointly testable.  Once price elasticities are identified, we can further partially identify the demand for each good at any given price, using in particular consumers' rationality. 



\medskip
This strategy fits naturally transportation goods such as trains serving multiple destinations. This setting features flexible traveler's arrival, non-competing destinations, and simultaneous fare management. More broadly, these conditions are also  plausible for perishable goods, e.g., rental cars with two- and six-seaters, hospitality with single-occupied and family rooms, and connecting flights. In such contexts, a ``good'' would correspond to, e.g., hotel rooms at a given date, or a flight or train ticket at a given date (e.g., Paris-Marseille on March 1, 2008), whereas the corresponding ``market'' would be the set of hotel rooms, flight or train tickets for that same date.

\medskip
We apply our method to high-speed rail transportation, by estimating demand for train tickets issued by iDTGV, a subsidiary of the French railway monopoly (SNCF). Our framework is relevant in this context. Like in other sectors with high fixed costs and low variable costs (e.g., ticketing industry), finding a marginal cost shifter is hard if not impossible. Moreover, the French railway adopts revenue management, a collection of operations research techniques that adjust supply to the random demand for perishable goods and is widely used in a number of industries, notably airline and hospitality. During the booking window of a given train, the operator manages ticket prices of intermediate and final destinations  via a finite number of classes with predetermined fares. This simplifies revenue management implementation, but complicates traditional demand estimation due to substantial price stickiness and demand censoring \citep{Swan_90,Lee_90,Stefanescu_12}. We overcome these issues and identify price elasticities and other demand primitives by exploiting variations across fare classes in the relative prices of tickets to intermediate and final stops served by the same train. 

\medskip
Our baseline price elasticity estimate of $-4.04$ is comparable to those given in some empirical studies on the airline sector that use sales and prices at flight-day level, but larger in absolute value than the estimates obtained in the literature on railway transportation. This gap may stem from the fact that most of these articles use sales and prices aggregated at the train-destination level, and log-log regressions using our data aggregated at this level also lead to lower price elasticity. Intuitively, when aggregating our data, price variations reflect demand fluctuations, and even train fixed effects cannot account for these. By using routes with identical ticket prices to both stops, we also find suggestive evidence in favor of the analogue of the parallel trend assumption, namely that consumers look for tickets for the intermediate and final destinations at similar moments.

\medskip
We next examine the firm's pricing behavior. Descriptive evidence such as price stickiness, increasing fares over the booking period and low loads on some routes suggests that actual pricing may deviate from the revenue-maximizing benchmark. Using our demand estimates, we quantify this suboptimality by comparing current revenue, consumer surplus, and social welfare with their counterfactual counterparts under optimal pricing (from the firm’s perspective).\footnote{Considering revenues is important when, e.g., deciding whether to construct new railway lines.} As another methodological contribution of the article, we show that for a broad class of alternative pricing strategies, these counterfactual quantities depend only on price elasticities and the parameters governing average demand, not on the timing of consumer arrivals. This theoretical property is important: as often, we do not observe the timing of purchases here. Using the point and partially identified demand parameters, we are then able to obtain informative bounds on the counterfactual parameters of interest.

\medskip
We find that the observed revenue management generates a revenue gain of up to 8.1\% compared to the optimal uniform pricing in our preferred, incomplete information setup. Under the same informational setup, we also estimate a significant revenue loss of at least $9.4\%$, and up to $16.8\%$ compared to the optimal dynamic pricing. The optimal dynamic pricing delivers a nearly full load, compared to an observed load of $84.5\%$, and improves the social welfare by around $10\%$ regardless of the information setup. 

\medskip
Finally, we show that relaxing the observed pricing constraints, e.g., monotonic and predetermined fares,  only explains  a small portion of the improvements. An optimal pricing based on the same constraints as those used by revenue managers secures a revenue at least 7.3\% larger than the current one. Hence, our results suggest that managers may fail to appropriately update their information about demand using past purchases. And as it turns out, learning demand from consumer purchases can compensate almost all revenue and welfare loss due to ex ante uncertainty on demand. Relatedly, demand learning occurs quickly: by observing and learning from the sales of the first half of available capacity, for instance, the firm can already secure more than 97\% of the revenue under complete information. 

\medskip
\paragraph*{Related Literature. }
Our article contributes to the research on demand identification and estimation when classic instrumental variables are unavailable or inapplicable. Several recent studies achieve instrument-free identification by exploiting specific features of demand or supply, such as bundle choices \citep{iaria2020identification}, cost data \citep{byrne2022instrument}, nonlinear price structures \citep{reiss2005household}, or surge pricing \citep{cohen2016using,castillo2025benefits}. 
In the context of revenue management, \cite{garcia2022demand} relies on behavioral assumptions about managerial responses, akin to restrictions on the covariance between demand and supply shocks \citep{mackay2025estimating}, to estimate demand elasticities in a difference-in-differences spirit. In contrast, our strategy applies to markets that exhibit the aforementioned independence condition among goods, such as 
trains serving multiple stops, connecting flights, or rental cars with varying capacities. Nonetheless, our strategy shares some similarities with nonlinear difference-in-differences based on an underlying logit model \citep[see, e.g.][]{blundell2004evaluating,wooldridge2023simple}.  

\medskip
Our article also contributes to the literature on revenue management in economics and management. The operations research tradition of revenue management has primarily focused on deriving optimal pricing rules in static \citep{Littlewood_72,Brumelle_McGill_93} and dynamic \citep{Gallego_vanRyzin_94,aviv2002pricing} settings. 
By contrast, we empirically assess the revenue and welfare implications of observed pricing practices. For this purpose, we extend the theoretical results of \cite{mcafee2008dynamic} from a complete- to an incomplete-information environment, where the firm knows only the distribution of demand parameters and updates it as consumers arrive. We also generalize their results by studying constrained pricing strategies that mirror those implemented in practice. This empirical orientation distinguishes our work from recent research on airline revenue management in economics, such as \cite{williams2022welfare} and \cite{aryal2024price}, which infer demand parameters from optimal dynamic pricing conditions and investigate welfare effects of price discrimination. We do not impose full optimality on observed prices and find that the French railway’s pricing practices may deviate substantially from the theoretical benchmark, leading to potential revenue and welfare losses.
\medskip

Finally, our empirical findings highlight the important role of learning about demand from consumer purchases in dynamic pricing. These findings resonate with recent studies showing that firms can improve their pricing performance by learning from consumer characteristics or realized sales, whether through personalized \citep{dube2023personalized} or updated market information \citep{huang2020learning}. The importance of information and demand learning for dynamic pricing has also been emphasized in the theoretical literature, including \cite{lin2006dynamic}, \cite{aviv2002pricing}, and \cite{den2015dynamic} (see also \cite{den2015dynamicsingle} for a comprehensive survey). To our knowledge, we are the first to quantify these effects using real-world data on dynamic pricing behavior.

\medskip
The rest of the article is organized as follows. Section 2 develops our new identification strategy for demand estimation. Section 3 presents the data, institutional context and estimates the demand for train tickets. We use our estimates in Section 4 to study the optimality of pricing at iDTGV. The appendix includes a  microfoundation of our demand model, additional details on estimation and inference, additional results on the application and the proofs of our theoretical results. 

\section{A new strategy for demand identification}\label{sec:mode}

In this section, we first present  our strategy in a simplified setup. Then,  we present and discuss our two main assumptions, before showing successively how several features of this demand (e.g., price elasticities) can be identified. Estimation and inference are discussed in Appendix \ref{sub:estimation_and_inference}.

\subsection{Intuition behind our results}
\label{sub:intuition}

We consider a simple setup in which  there are two goods $a$ and $b$ in each market $m$; the extension to multiple goods is discussed in Subsection \ref{sub:ident_elast}. In our application, $m$ indexes trains, namely routes such as Paris-Toulouse associated with a departure time (e.g., May 2, 2008) while $a$ and $b$ are two destinations served by this train (e.g., Bordeaux and Toulouse). Other possible examples are hotels ($m$) with at least two types of rooms ($a$ and $b$), or supermarkets ($m$), with $a$ and $b$ being two of the multiple goods sold therein. 

\medskip
The prices of $a$ and $b$ may change over time; let $k\in\{1,...,K\}$ denote a period of time during which the prices $p_{amk}$ and $p_{bmk}$ remain constant within market $m$. We suppose to observe $p_{dmk}$ and the corresponding quantities $Q_{dmk}$ sold, for $d\in\{a,b\}$ and $k\in\{1,...,K\}$. We also impose constraints on the demand. Specifically, we suppose that it evolves similarly for $a$ and $b$ within $m$, which may be viewed as a parallel trend assumption. We also assume that (i) price elasticities for both products ($\eps_a$ and $\eps_b$) remain constant over time and (ii) the two goods do not compete, so that the price of $d'$ does not affect sales of $d\ne d'$. Formally, let us momentarily assume that for any $k\in\{1,...,K\}$ and $d\in\{a,b\}$, we have 
\begin{equation}
\ln Q_{dmk} = \delta_{mk} + h_0(W_m) + X'_{dm}\beta_0 + \eps_d \ln p_{dmk} + \nu_{dm} + \eps_{dmk},
    \label{eq:linear_demand}
\end{equation}
with $E[\eps_{dmk}|W_m, X_{dm},p_{dmk}]=0$ and $p_{dmk}$ (but not $X_{dm}$) may be correlated with $\nu_{dm}$. We  also assume, without loss of generality, that $\sum_k \delta_{mk}=E[\nu_{dm}]=0$. Then, $h_0(W_m)$ represents the log total demand in market $m$. We are interested in the identification of $h_0(\cdot)$, $\beta_0$ and $\eps_d$ (we prove below that $\delta_{mk}$ does not matter for identification of several counterfactual parameters). Consider $\eps_d$ first. Equation \eqref{eq:linear_demand} implies that
\begin{equation}
    \ln\left(\frac{Q_{amk}}{Q_{bmk}}\right) = (X_{am}-X_{bm})'\beta_0 + \eps_a \ln p_{amk} - \eps_b\ln p_{bmk} + \nu_{am}-\nu_{bm}+ \eps_{amk} - \eps_{bmk}.
    \label{eq:linear_panel}
\end{equation}
This may be seen as a linear panel data model. As a result, a fixed effect regression identifies $\eps_a$ and $\eps_b$ as long as there are sufficient independent time variation in the prices of both products. Next, we identify $\beta_0$ by a regression of $\ln(Q_{amk}/Q_{bmk}) - \eps_a \ln p_{amk} + \eps_b\ln p_{bmk}$ on $X_{am}-X_{bm}$. Finally, we identify $h_0(\cdot)$ using 
\begin{equation}
  E\left[\sum_{k=1}^K \ln Q_{dmk} -X'_{dm}\beta_0 + \eps_d \ln p_{dmk}\,\big|\,W_m\right] = h_0(W_m).
  \label{eq:ident_f0_simple}
\end{equation}

We have abstracted from two issues so far. First, demand is often discrete and can be equal to zero, which is incompatible with \eqref{eq:linear_demand}. Second, the length of time periods is random and possibly correlated with demand. We account for the first issue by considering a Poisson process (Assumption \ref{hyp:cons_demand} below). Then, we still obtain an equation close to \eqref{eq:linear_panel}, namely 
\begin{equation}
  \ln\left(\frac{\Pr_{amk}}{\Pr_{bmk}}\right) = (X_{am}-X_{bm})'\beta_0 + \eps_a \ln p_{amk} - \eps_b\ln p_{bmk} + \nu_{am}-\nu_{bm},  
  \label{eq:FE_logit}
\end{equation}
in which $\Pr_{amk}$ (resp. $\Pr_{bmk}$) denotes the probability that a given purchase is for product $a$ rather than $b$ (resp. $b$ rather than $a$). Importantly, we show that \eqref{eq:FE_logit} holds even with random period lengths, as long as the timing of price changes only depends on total purchases of $a$ and $b$ (Assumption \ref{hyp:yield_line} below). Equation \eqref{eq:FE_logit} corresponds to a fixed effect logit model, which implies that we can identify $\eps_a$ and $\eps_b$ under the same variations on prices as above; $\beta_0$ can also be identified using the fact that $(X_{am},X_{bm})$ is independent of $(\nu_{am},\nu_{bm})$. 

\medskip
Finally, the issue of random period lengths implies that we cannot point identify $h_0$ through a moment equality akin to \eqref{eq:ident_f0_simple}. However, we can still obtain moment inequalities instead of equality by using consumers' rationality (namely, that demand decreases when prices grow), implying its partial identification.

\subsection{Setup and main assumptions}

We now present our main assumptions, under which we obtain in particular \eqref{eq:FE_logit}. 
Goods $a$ and $b$ are sold between the normalized dates $t=0$ and $t=1$. We denote by $V_{dm}(A, B)$ the number of consumers who seek to buy product $d=a,b$ in market $m$ during time interval $A\subset[0,1]$, with a willingness to pay belonging to the subset $B$ of $[0,\infty)$.  We first assume:
\begin{hyp}	\label{hyp:cons_demand}
	(Consumers' demand) For all $m$, there exists $\eps_a, \eps_b>1$ (independent of $m$), random variables $(\xi_{am},\xi_{bm})$ and a continuous random function $s_m(\cdot)$ defined on $[0,1]$ satisfying $\min_{u\in[0,1]} s_m(u)>0$ almost surely, such that conditional on $\xi_{am}, \xi_{bm}, s_m(\cdot)$:
	\begin{enumerate}
		\item $V_{dm}$ ($d\in\{a,b\}$) is a Poisson process with intensity $I_{dm}(t,p) = \xi_{dm} s_m(t) \eps_d p^{-1-\eps_d}$ for $(t,p)\in [0,1]\times [0,\infty)$. Without loss of generality, we let $\int_0^1 s_m(u)du=1$.
		\item \label{hyp:cons_demand:indep} $V_{am}$ and $V_{bm}$ are independent.
	\end{enumerate}
\end{hyp}
Assumption \ref{hyp:cons_demand}.1 imposes a multiplicative decomposition akin to \eqref{eq:linear_demand}, while accounting for the discrete nature of purchases. The function $s_m(\cdot)$ captures the variations in consumers' arrival over time. It is unrestricted as a function of time, a feature that fits well settings such as transportation and hospitality in which consumer arrival pattern may flexibly change over time and market. On the contrary,  $s_m(\cdot)$ is restricted to be the same for both products; this may not be satisfied if, for instance, travelers to large towns (e.g., Marseille) took their train tickets later than those to smaller towns (e.g., Avignon). The component $\xi_{dm}$ captures a product-market specific, time-invariant term. In our application to trains, one should observe a spike of demand during the Cannes Film Festival, so the Paris-Nice trains departing during these two weeks may face a high, specific demand for Cannes.

\medskip
The parameter $\eps_d$ is the price elasticity, reflecting both consumers' disutility of price and the elasticity of substitution to the outside good (e.g., other modes of transportation in our application). The functional form $p\mapsto p^{-1-\eps_d}$ and the fact that $\eps_d$ does not vary with markets are not essential for identification. On the other hand, that price elasticity does not vary over time is important; see, e.g., \cite{mcafee2008dynamic} for a similar restriction. But as discussed below, this assumption is testable. We test it in the application and consider an augmented model with time-varying elasticities in Section \ref{sec:robustness_checks}. The reason we do not consider this augmented model as our main specification is that it imposes that price elasticity changes only and exactly at the (potentially random) moments when prices changes, which we find somewhat ad hoc.

\medskip
Assumption \ref{hyp:cons_demand}.\ref{hyp:cons_demand:indep} formalizes the idea that the two goods are not substitute. In this sense, our definition of ``markets'' is unusual. Importantly, Assumption \ref{hyp:cons_demand}.\ref{hyp:cons_demand:indep} may still hold when applied to two close goods. For instance, the demand for single and double hotel rooms, or the demand for two destinations served by the same train, may be independent once we control for $(\xi_{am}, \xi_{bm})$. 
On the other hand, the assumption does not hold if some consumers decide on which destination they travel to depending on, e.g., the prices of the corresponding tickets.

\medskip
We now clarify the conditions we require on prices and observed quantities. 
\begin{hyp}\label{hyp:yield_line}
	(Price variations and observed quantities) In each market $m$:
    \begin{enumerate}
        \item \label{hyp:price} The prices for the two goods remain constant over intervals $[\tau_{(k-1)m}, \tau_{km})$, for $k=1,...,K$ ($K\ge 2$), with $\tau_{0m}:=0$, $\tau_{Km}:=1$ and $\tau_{km}$ is a stopping time with respect to the process $t \mapsto N_{am}(t)+N_{bm}(t)$, where $N_{dm}(t)$ is the number of purchases for $d$ made before $t$;
        \item \label{hyp:quantities} We observe the quantities $n_{dkm}:=N_{dm}(\tau_k)-N_{dm}(\tau_{k-1})$ for $(d,k)\in\{a,b\}\times\{1,...,K\}$. 
    \end{enumerate}
\end{hyp}
 
Assumption \ref{hyp:yield_line}.\ref{hyp:price} states that the moments at which prices switch may be related to previous purchases, but only through total purchases $N_{am}(t)+N_{bm}(t)$ rather than purchases specific to $a$ or $b$. When the $(\tau_{km})_{k=1}^{K}$ are predetermined, Assumption \ref{hyp:yield_line} (parts \ref{hyp:price} and \ref{hyp:quantities}) trivially holds. This encompasses settings in which the researcher observes, say, prices and purchases of airline tickets at a daily level \citep{williams2022welfare}, or product prices and purchases in grocery stores weekly (e.g., Nielsen's Retail Scanner Data), provided that prices remain constant during these time intervals. 

\medskip
Assumption \ref{hyp:yield_line}.\ref{hyp:price} also covers more complex situations in which  $(\tau_{km})_{k=1}^{K}$ are random and endogenously set by firms, a typical feature of the quantity-based revenue management. In our empirical application, for instance, prices change simultaneously for all destinations of a given train, so Assumption \ref{hyp:yield_line}.\ref{hyp:quantities} holds. Beyond our application, a rationale behind Assumption \ref{hyp:yield_line} is that pricing is done at an aggregate, market level rather than at the product level.

\medskip
On the other hand, Assumption \ref{hyp:yield_line}.\ref{hyp:price} is more likely to fail if prices change at different moments for $a$ and $b$. Then, the stopping time corresponding to a change in the price of $a$, say, is likely to depend on $t\mapsto N_{am}(t)$ rather than $t\mapsto N_{am}(t)+N_{bm}(t)$.

\medskip
Assumption \ref{hyp:yield_line}.\ref{hyp:quantities} requires that we observe the purchased quantities over the intervals $[\tau_{(k-1)m},\tau_{km})$ ($k=1,...,K$). This assumption holds when the timing and associated prices of all purchases are observed, although we do not require observing this timing, nor the points in time $(\tau_{km})_{k=1}^{K}$ when prices change. 

\subsection{Identification results} 

\subsubsection{Price elasticities}
\label{sub:ident_elast}
We first study identification of price elasticities. Here we consider an asymptotic setup in which the number of markets tends to infinity. In terms of identification, this means that we assume that the distribution of $(n_{akm}, n_{bkm}, p_{akm}, p_{bkm})_{k=1,...,K}$ is known. Assumptions \ref{hyp:cons_demand} and \ref{hyp:yield_line} turn out to be sufficient to identify $(\eps_a,\eps_b)$, as long as there are variations in relative prices (in a sense made precise below) as $k$ varies. To see why, let us first assume, for simplicity, that  $\tau_{km}$ and $\tau_{k+1,m}$ are deterministic. Then, by Assumption \ref{hyp:cons_demand}, $V_{am}\left([\tau_{km},\tau_{(k+1)m}),[p_{akm},\infty)\right)$ and $V_{bm}\left([\tau_{km},\tau_{(k+1)m}),[p_{bkm},\infty)\right)$ are independent conditional on $\xi_{am}, \xi_{bm}$ and $\int_{\tau_{km}}^{\tau_{(k+1)m}} s_m(u)du$. Moreover, they both follow Poisson distributions. Also, by Assumption \ref{hyp:yield_line}.\ref{hyp:price}, $n_{dkm} = V_{dm}\left([\tau_{km},\tau_{(k+1)m}),[p_{dkm},\infty)\right)$.  As a result, given $\xi_{am}$ and $\xi_{bm}$, we have
{\small \begin{equation}\label{eq:binomial}
n_{bkm}|n_{akm}+n_{bkm}=n \sim \text{Binomial}\left(n, \Lambda(\ln(\xi_{bm}/\xi_{am}) -\eps_b  \ln(p_{bkm}) + \eps_a \ln(p_{akm}))\right),	
\end{equation}}
where $\Lambda(x)=1/(1+\exp(-x))$. The term $\ln(\xi_{bm}/\xi_{am})$ may be seen as a market fixed effect. Hence this model boils down to a fixed effect logit model, and $(\eps_a,\eps_b)$ are identified if, basically, we can vary separately $\ln(p_{bkm})$ and $\ln(p_{akm})$ when moving $k$. 

\medskip
In Appendix \ref{sub:proof_of_theorem_ref_thm_ident_eps}, we show that Equation \eqref{eq:binomial} still holds under Assumption~\ref{hyp:yield_line} with stochastic stopping times $(\tau_{km})_{k=1}^K$. We then obtain the following identification result:

\begin{thm}\label{thm:ident_eps}
	If Assumptions \ref{hyp:cons_demand} and \ref{hyp:yield_line} hold, then  Equation \eqref{eq:binomial} holds. Moreover, if, with positive probability, the functions $k\mapsto 1$, $k\mapsto \ln(p_{akm})$ and $k\mapsto \ln(p_{bkm})$ are not collinear, then $(\eps_a,\eps_b)$ are identified.
\end{thm}

Three remarks on this results are in order. First, Equation \eqref{eq:binomial} makes it clear that with enough variations in prices, more flexible price effects than $p\mapsto p^{-1-\eps}$ can be contemplated. Similarly, we can consider elasticities that vary by market characteristics.

\medskip
Second, we can test  Equation \eqref{eq:binomial}, and thus Assumptions \ref{hyp:cons_demand} and \ref{hyp:yield_line}, if purchases are sometimes available at a finer level than prices, in the sense that for some $m$ and $k<K$, $p_{akm}=p_{a(k+1)m}$ and $p_{bkm}=p_{b(k+1)m}$. If so, Equation \eqref{eq:binomial} predicts that the distributions of $n_{akm}|n_{akm}+n_{bkm}$ and $n_{a(k+1)m}|n_{a(k+1)m}+ n_{b(k+1)m}$ are the same, while they may not if, e.g., $s_m(t)$ in Assumption \ref{hyp:cons_demand} depends on $d$.\footnote{Obviously, this prediction does not depend on how prices affect demand: we would obtain the same prediction if $p^{-1-\eps}_d$ were replaced by $\varphi_d(p)$ for any function $\varphi_d(.)$ in Assumption \ref{hyp:cons_demand}.}  The same conclusions hold if $p_{akm}\ne p_{a(k+1)m}$ and $p_{bkm}\ne p_{b(k+1)m}$ but $p_{b(k+1)m}/p_{a(k+1)m}=p_{bkm}/p_{akm}$ and we also assume $\eps_a=\eps_b$. Such tests resemble pre-trends testing in difference-in-differences. 
We conduct a test of this kind in our application. 

\medskip
Third, we have focused for simplicity on two goods but with three goods or more ($d_1,...,d_J$, say), the exact same reasoning applies. In particular, Equation \eqref{eq:binomial} becomes in this case
\begin{equation}\label{eq:multinomial}
(n_{d_1km},...,n_{d_Jkm})\big|\sum_{j=1}^J n_{d_jkm}=n,(\xi_{d_jm})_{j=1\dots J} \sim \text{Multinomial}\left(n, P_{d_1 km},...,P_{d_J km}\right),
\end{equation}
where the probabilities $P_{d_jkm}$ satisfy
$$P_{d_jkm} = \frac{\exp\left(\ln(\xi_{d_jm}) - \eps_{d_j} \ln(p_{d_j k m})\right)}{\sum_{j'=1}^J \exp\left(\ln(\xi_{d_{j'}m}) - \eps_{d_j} \ln(p_{d_{j'} k m})\right)}.$$

\subsubsection{Relative demand parameters}

While the elasticity is of policy relevance, it is not sufficient to identify several counterfactuals of interest. We now discuss the identification of the ``relative'' demand parameters $\ln(\xi_{bm}/\xi_{am})$, which may be seen as a market fixed effect. If markets were large enough, we could let $n$ tend to infinity in Equation \eqref{eq:binomial} and directly identify this fixed effect from that equation. However, $n$ remains small in several settings of interest, and we do not obtain point identification without further restrictions. This motivates the following assumption.
\begin{hyp}\label{hyp:gamma}
	For $d=\{a,b \}$, $\xi_{dm}$ satisfies:
	\begin{itemize}
		\item[(i).] $\xi_{dm}=\exp\{X'_{dm}\beta_0\}g_0(W_m)\eta_{dm}$ where $\eta_{am}$, $\eta_{bm}$ and $(X_{am},X_{bm},W_m)$ are independent and $X_{am}$ (resp. $X_{bm}$, $W_m$) is a vector of observed characteristics for product $a$ (resp. product $b$, market $m$).
		\item[(ii).] $\eta_{dm}\sim\Gamma(\lambda_{d0},1)$, where $\Gamma$ refers to gamma distribution.
	\end{itemize}
\end{hyp}
Assumption \ref{hyp:gamma}(i) specifies $\xi_{dm}$ as the product of a function of $X_{dm}$, $g_0(W_m)$, and a remainder term $\eta_{dm}$. It restricts $\xi_{am}$ and $\xi_{bm}$ to be dependent through the observed variables $(X_{am},X_{bm},W_m)$, rather than $(\eta_{am},\eta_{bm})$. This assumption is plausible provided that sufficient controls are included in $(X_{am},X_{bm},W_m)$. Importantly, we leave the function $g_0(.)$, which determines how market characteristics affect demand, unrestricted. 

\medskip
Assumption \ref{hyp:gamma}(ii) imposes that conditional on $(X_{dm},g_0(W_m))$, $\xi_{dm}$ follows a gamma distribution. Since we include a $d$-specific constant term in $X_{dm}$,  we can normalize the scale parameter of the gamma distribution to $1$. As detailed below, the assumption of a gamma distribution does not matter for identification. It is rather made for computational reasons: that the gamma and Poisson distributions are conjugate significantly simplifies the computation of counterfactuals of interest. But we also consider in our application log-normality as a robustness check for some counterfactuals.

\medskip
The following theorem formalizes the discussion above. In addition to the previous assumptions, identification of $\beta_0$ also relies on a rank condition on the $X_{dm}$, Assumption \ref{hyp:supp_X}, which is presented in Appendix \ref{sub:proof_of_theorem_ref_thm_ident_xi}.

\begin{thm}\label{thm:ident_xi}
	Suppose that Assumptions \ref{hyp:cons_demand}, \ref{hyp:yield_line}, \ref{hyp:gamma}(i) and \ref{hyp:supp_X} hold. Suppose also that with positive probability, the functions  $k\mapsto 1$, $k\mapsto \ln(p_{akm})$ and $k\mapsto \ln(p_{bkm})$ are not collinear. Then: 
    \begin{enumerate}
        \item $\beta_0$ and the distribution of $\eta_{bm}/\eta_{am}$ are identified.
        \item If Assumption \ref{hyp:gamma}(ii) further holds, $(\lambda_{a0},\lambda_{b0})$ are also identified.
    \end{enumerate}
\end{thm}

\subsubsection{Total demand}\label{sec:partial_id}

We now study  the identification of $g_0(W_m)$, the parameters of ``total'' demand  for $d\in\{a,b\}$. Since this term is common to $\xi_{am}$ and $\xi_{bm}$, it drops out of $\xi_{bm}/\xi_{am}$ in Equation \eqref{eq:binomial} and cannot be identified using the strategy behind Theorems \ref{thm:ident_eps} and \ref{thm:ident_xi}. We show how to partially identify $g_0(W_m)$ by building moment inequalities based on the consumer rationality restriction embedded in Assumption \ref{hyp:cons_demand}.1. We discuss uncensored and censored demand. The latter is relevant to perishable goods with limited supply, e.g., flight or train tickets and  hotel rooms.
\paragraph{Uncensored demand.} Under Assumption \ref{hyp:cons_demand}.1, all consumers who bought good $d\in\{a,b\}$ at price $p_{djm}$  would have also bought it at price $p_{dkm}$ if $p_{dkm}\le p_{djm}$. Therefore, for all $k=1,...,K$ and $d\in \{a,b\}$,
\begin{equation}
D_{dm}(p_{dkm};g_0(W_m), X_{dm}) \geq \sum_{j:p_{djm}\ge p_{dkm}} n_{djm},
    \label{eq:moment_ineg}
\end{equation}
where $D_{dm}(p_{dkm};g_0(W_m), X_{dm}):= V_{dm}\left([0,1),[p_{dkm},\infty)\right)$, but
we now index demand by $g_0(W_m)$ and $X_{dm}$ to emphasize its dependence on these variables. Taking conditional expectations on both sides of \eqref{eq:moment_ineg} and using Assumption \ref{hyp:cons_demand}.1, we obtain
\begin{equation}
g_0(W_m)\exp(X_{dm}'\beta_0)\lambda_{d0}\eps_d p_{dkm}^{-1-\eps_d} \ge \E\left[\sum_{j:p_{djm}\ge p_{dkm}} n_{djm}\bigg|W_m, X_{dm}, p_{dkm}\right].
    \label{eq:moment_ineg1}
\end{equation}
By using the law of iterated expectations and optimizing over $k$ and $d$, we finally get
\begin{equation}\label{eq:lower_bound}
g_0(W_m)\geq g_0^L(W_m):=\max_{\substack{d=a,b \\k=1,...,K}}\E\left[\frac{1}{\lambda_{d0}\eps_d} \exp(-X_{dm}'\beta_0)p_{dkm}^{1+\eps_d} \sum_{j:p_{djm}\ge p_{dkm}} n_{djm}\big|W_m\right].
\end{equation}
Though this bound is not sharp, it is convenient as it avoids estimating a conditional expectation depending on several variables as in \eqref{eq:moment_ineg1}.  

\medskip
We can obtain an upper bound on $g_0(W_m)$ by a similar reasoning. Letting $p_{d\overline{K}_mm}$ denote the maximal price set in market $m$ for product $d$, we have, by Assumption \ref{hyp:cons_demand} again,
$$D_{dm}(p_{d\overline{K}_mm};g_0(W_m), X_{dm}) \le \sum_{k=1}^K n_{dkm}.$$ 
Hence, as above
\begin{equation}\label{eq:upper_bound}
g_0(W_m)\le g_0^U(W_m):=\min_{d=a,b} \mathbb{E}\left[\frac{1}{\lambda_{d0}\eps_d}\exp(-X_{dm}'\beta_0)p_{d\overline{K}_mm}^{1+\eps_d} \sum_{k=1}^K n_{dkm}\big| W_m\right].
\end{equation}
We summarize the partial identification of $g_0(W_m)$ in the next result.
\begin{thm}[Partial identification by the law of demand, uncensored demand]\label{thm:partial_id_g}
    Suppose that Assumptions \ref{hyp:cons_demand}.1 and \ref{hyp:gamma} hold. Then, $g_0^L(W_m)$ in \eqref{eq:lower_bound} and $g_0^U(W_m)$ in \eqref{eq:upper_bound} are a lower and upper bounds for $g_0(W_m)$, respectively.
\end{thm}
\paragraph{Censored demand.}  Theorem \ref{thm:partial_id_g} holds if the total number of purchases is unlimited. Yet, demand may be censored in practice. In our application for instance, the number of seats in a train is fixed. In such cases, we can improve the lower bound $g_0^L(W_m)$ as follows. Let $C_m$ denote the maximal number of purchases in market $m$. Then, instead of \eqref{eq:moment_ineg}, we have
\begin{equation}\label{eq:lower_demand}
    D_{dm}(p_{dkm};g_0(W_m), X_{dm})\wedge C_m \ge \sum_{j:p_{djm}\ge p_{dkm}} n_{djm}.
\end{equation}
Even if $Q(g;p,x,c):=\E[D_{dm}(p;g, x)\wedge c]$ does not have a simple form, we can compute it by simulations. We can also show that $Q(\cdot;p,x,c)$ is strictly increasing. Then, letting $Q^{-1}(\cdot;p,x,c)$ denote its inverse, we obtain, instead of \eqref{eq:lower_bound}, 
\begin{equation}\label{eq:lower_bound_censored}
\resizebox{0.92\textwidth}{!}{$
    g_0(W_m)\geq \max_{\substack{d=a,b \\k=1,...,K}}\E\left[Q^{-1}\left(\mathbb{E}\left[\sum_{j:p_{djm}\ge p_{dkm}} n_{djm}\bigg|W_m, X_{dm}, p_{dkm}\right];p_{dkm},X_{dm},C_m\right) \bigg|W_m\right].
    $}
\end{equation} 
On the other hand, it is unclear to us whether an upper bound solely based on demand holds under censoring. Intuitively, the observed demand does not correspond to the unobserved uncensored one that would otherwise serve as a upper bound. To circumvent this issue, in our application we use instead a weak optimality condition on the supply  and obtain a upper bound for $g_0(W_m)$, see Theorem \ref{thm:partial_id_g_censored} below.

\subsubsection{Timing of arrival} 

Identification of the distribution of $s_m(\cdot)$ obviously requires more data than simply the number purchases $(n_{dkm})_{d=a,b, k=1,...,K}$. If the dates of all purchases are known, for instance, features of the distribution of $s_m(\cdot)$ can be identified by comparing the number of purchases between two time intervals on which prices remain constant, adjusting for difference in prices between the two time intervals. In any case, it turns out that many counterfactuals of interest do not depend on $s_m(\cdot)$, a point that we formalize in Theorem \ref{thm:counterf_rev} below.

\section{Application to demand for trains in France}

We now turn to our application. Our final aim is to compare the revenues and consumer surplus under the current yield management and under alternative pricing strategies. To this end, we first estimate in Section \ref{sub:demand_est} the price elasticities as well as relative and total demand parameters, using our strategy above. Before that, we present the institutional background, the data and descriptive statistics in Sections \ref{sub:revenue management} and \ref{sub:data}. We also discuss in these sections why we believe our previous setup is well-suited to this application.

\subsection{Institutional background}\label{sub:revenue management}

We exploit data on iDTGV, a low-cost subsidiary of the French railway monopoly, SNCF, which was created in 2004. 
Its creation was motivated  by low-cost airlines' entry in the French domestic routes (e.g., EasyJet launched the line Paris-Marseille at the beginning of 2004) and the EU rail market liberalization.\footnote{See the quotation from iDTGV's CEO, Paul Sessego, in   \href{https://www.lenouveleconomiste.fr/idtgv-le-laboratoire-marketing-de-la-sncf-1668}{\textit{iDTGV, le laboratoire marketing de la SNCF}}.} 
Before its disappearance in December 2017, iDTGV was highly autonomous. It owned its trains and had a pricing strategy independent from SNCF.\footnote{Due to internal strategic considerations at SNCF, iDTGV was replaced by Ouigo, the new low-cost service at SNCF, in 2013.} Prices were generally lower than the full-rate prices of SNCF, but were also associated with a slightly lower quality of services. Namely, tickets had to be purchased on-line, they were nominative and could not be cancelled. On top of that, they could be exchanged only under some conditions and at some cost.

\medskip
The routes of iDTGV were all between Paris and French towns. For each of those towns and every day, one train was leaving Paris and another coming to Paris. Table \ref{DES1} presents the routes in our data from May 2007 till March 2009. These routes have several stops, but to simplify the analysis, we gather them so as to form a single intermediate stop and a single final stop. We aggregate the cities according to the price schedule. For instance, we group Aix-en-Provence and Avignon together in the Paris-Marseille route since the corresponding prices are always the same. This gathering is consistent with Assumption \ref{hyp:cons_demand}, as our demand model remains valid after aggregation of cities.\footnote{See Remark \ref{rmk:aggregation_dest} for details. We also estimate demand without the grouping. The results are quantitatively similar. See Columns II and IV of Table \ref{tab:binomial}.}

\bigskip
\begin{center}
\small{
\begin{threeparttable}
	\vspace{-1cm}
    \caption{\small Routes with intermediate and final destinations}\label{DES1}		
    \begin{tabular}{lllc}
    \toprule
    Route name     	&Final stop(s)                &Intermediate stop(s)     &Nb. of trains \\
    \midrule
    C\^ote d'Azur  &Cannes,Saint-Rapha\"el,Nice & Avignon			  & 452             \\
    Marseille          &Marseille              &Aix-en-Provence/Avignon          & 453             \\
    Perpignan & Perpignan & N\^imes, Montpellier  & 689    \\
\multirow{2}{*}{C\^ote basque} &St Jean de Luz,Bayonne,& \multirow{2}{*}{Bordeaux} & \multirow{2}{*}{405} \\
 	  			   & Biarritz,Hendaye  	    &  			 &      \\
    Toulouse       & Toulouse               &   Bordeaux         &411  \\
    Mulhouse       & Mulhouse               &Strasbourg           &499  \\
    \midrule
    Total          &                      &                     & 2,909  \\
    \bottomrule
    \end{tabular}
    \begin{tablenotes}
    \footnotesize
    \item \emph{Notes:} The number of observations differs from one route to another because the period we cover accordingly varies from one route to another.
\end{tablenotes}
\end{threeparttable}
}
\end{center}

\medskip
Different routes may share the same intermediate destination. For instance, Bordeaux is the intermediate destination of Paris-C\^ote basque and Paris-Toulouse. Importantly, no tickets were sold between the intermediate and the final destination, e.g. no Bordeaux-Toulouse tickets are sold on the Paris-Toulouse route. The reason for this surprising practice was that tickets were only controlled on the departure platform by an external firm, rather than by iDTGV employees in the trains (as is usually done by SNCF). This tentative measure (among many at iDTGV) aims at reducing costs.

\medskip
The trains are split into economy class and business class cars of fixed sizes. Revenue management was implemented almost independently between the two classes, i.e. under the sole constraint that prices in economy class are always lower than in business class. This constraint was very seldom binding in practice, so we ignore it hereafter. We focus on the economy class, which represents most of the sales (roughly $70\%$ in terms of seats and $67\%$ in terms of revenue). In this category, there are 12 fare classes corresponding to 12 prices sorted in ascending order. The price of a given fare class, at a peak or off peak time and for some origin-destination trip (e.g. Paris-Bordeaux)  remained constant for several months (e.g. from 03/01/2007 to 10/31/2007) before being adjusted marginally, mostly to account for inflation. Contrary to SNCF, iDTGV did not make any third-degree price discrimination, so there was no discount for young people, old people or families.

\medskip
The quantity-based revenue management at iDTGV consists in deciding in real time to maintain the current fare class or to close it and move to a higher one, resulting in a price increase. Coming back to a previous fare is impossible; as a result, there are no 'last minute' sales, i.e., drops in ticket prices for trains that are still non full before departure.\footnote{According to \cite{Mariton2008}, a congressional report on the pricing policy of the SNCF, reopening lower fare classes when approaching the departure may cause strong dissatisfaction among travelers who have  purchased seats with higher prices before, and could thus harm firm's reputation.} Also, revenue managers could decide to never open the first fare classes and begin to sell tickets in a higher fare class. Similarly, the last fare class may never be reached. In practice, revenue management was operated through a Computerized Reservation System (CRS). Before the beginning of sales, it fixes a seat allocation planning for all fare classes, using the history of purchases on past trains. During sales, the CRS uses the number of tickets sold up to now to make recommendations on the size of subsequent fare classes. Similarly to other industries such as hospitality \citep{cho2018optimal}, revenue management managers can nevertheless always intervene, both on the initial and on subsequent seat allocations, according to their experience on past trains.

\medskip
Finally, the revenue management did not use separate fare classes  for a given train with several destinations. For instance, in a Paris-Toulouse train, the closure of the first fare class occurred exactly at the same moment for both Bordeaux and Toulouse. Hence, price changes of Paris-Bordeaux and Paris-Toulouse tickets happened exactly at the same time for any given train. According to discussions with people in the revenue management department, this was to limit the number of decisions to be taken at each moment. This train-fare class-level management is a special case of Assumption \ref{hyp:yield_line}. 

\subsection{Data}\label{sub:data}

For each iDTGV train between May 2007 and March 2009 in economy class and for journeys from Paris to the rest of France, we observe the following characteristics: all stops, departure and arrival times, day of departure (e.g. May 2, 2008) and whether they correspond to a rush hour or not. We also observe the price grid used for each fare class of the train. For each route and type of period (peak time or off peak), there are a limited number of such grids, as they change these grids only a few times during the period we observe (e.g. 3 times for Paris-Toulouse). For every train, we observe the sales in each fare class. On the other hand, we do not observe the purchasing dates, nor the opening moments of each fare class. For a given route, capacity is defined as the maximal number $C$ such that for at least three trains,~$C$ seats were sold.\footnote{We use this definition (rather than the maximal number of seats sold across all trains of a given route) to account for rare cases of overbooked trains. With this definition, we observe 5 cases of overbooking, over the $2,909$ trains of our dataset. Note that capacity can be assumed to be fixed for a given route because the number of coaches in economy class is fixed.}

\medskip
Table \ref{DES2} presents some descriptive statistics on our data. First, we observe a substantial amount of price dispersion within trains. For instance, on the C\^ote d'Azur route, the minimal price paid by consumers, on average, over the different trains (\EUR{19.3}) was three times and a half lower than the average maximal price (\EUR{68.4}). There is also rich variation in the price ratio between final and intermediate stops, at the exclusion of Paris-Marseille and Paris-Mulhouse. As discussed above, this variation is key to identify  price elasticities (Theorem \ref{thm:ident_eps}). The fact that the final and intermediate stops of Paris-Marseille and Paris-Mulhouse have the same price in each fare class also allows to test the parallel condition underlying our identification strategy.

\medskip
The coefficients of variations in load vary between 0.075 and 0.304. Under Assumption \ref{hyp:cons_demand}, if prices were constant and there were neither capacity constraints nor aggregate shocks, the number of purchases would follow a Poisson distribution, with a variance equal to its mean. Then, the coefficient of variation in load for the six routes would be between 0.057 and 0.073. Revenue management and  capacity constraints would likely further reduce these numbers. Hence, the observed coefficients of variations point towards substantial aggregate demand shocks at the train level.

\begin{center}
\small{
	\begin{threeparttable}
		\vspace{-0.8cm}
		\caption{\small Descriptive statistics, economy class, from Paris} \label{DES2}
		\begin{tabular}{lccccccc}
			\toprule
			& & & Coeff. of var. &Avg& Avg min. \& & 
            Final/interm.  \\
            Route &Capacity  &Load  & on load &price&  max. prices & 
             price ratio \\
			\midrule
			C\^ote d'Azur  &324 & 85.4\% &0.175& 49.4 & 19.3, 68.4& 
            [0.943, 1.236]\\
			Marseille      &324 & 96.7\% &0.075&49.5 & 19.0, 70.5 & 
            1\\
			Perpignan      &324 & 89.1\% &0.137& 49.9 & 20.2, 72.6 &
            [1, 1.103]\\
			C\^ote basque   &350  & 65.4\% &0.304& 35.7 & 19.7, 53.3 &
            [1,1.189]\\
			Toulouse        &350  &87.5\% &0.152& 43.6  & 19.4, 67.2&
            [1,1.174]\\
			Mulhouse       &238  &79.4\%&0.184& 34.5 & 19.4, 50.0 &
            1\\
			\bottomrule
		\end{tabular}
		\begin{tablenotes}
			\scriptsize
			\item Notes: Avg min. and max. are the average of the minimal and maximal prices charged for each train, for the final destination. The brackets in the fifth column display the minima and maxima of the ratio of prices between final and intermediate destination. Coeff. of var. is the coefficient of variation. 
		\end{tablenotes}
	\end{threeparttable}
    }
\end{center}

\begin{figure}[H]
\caption{Distribution of load across fare classes}\label{fig:distr_load}
\includegraphics[width=\textwidth]{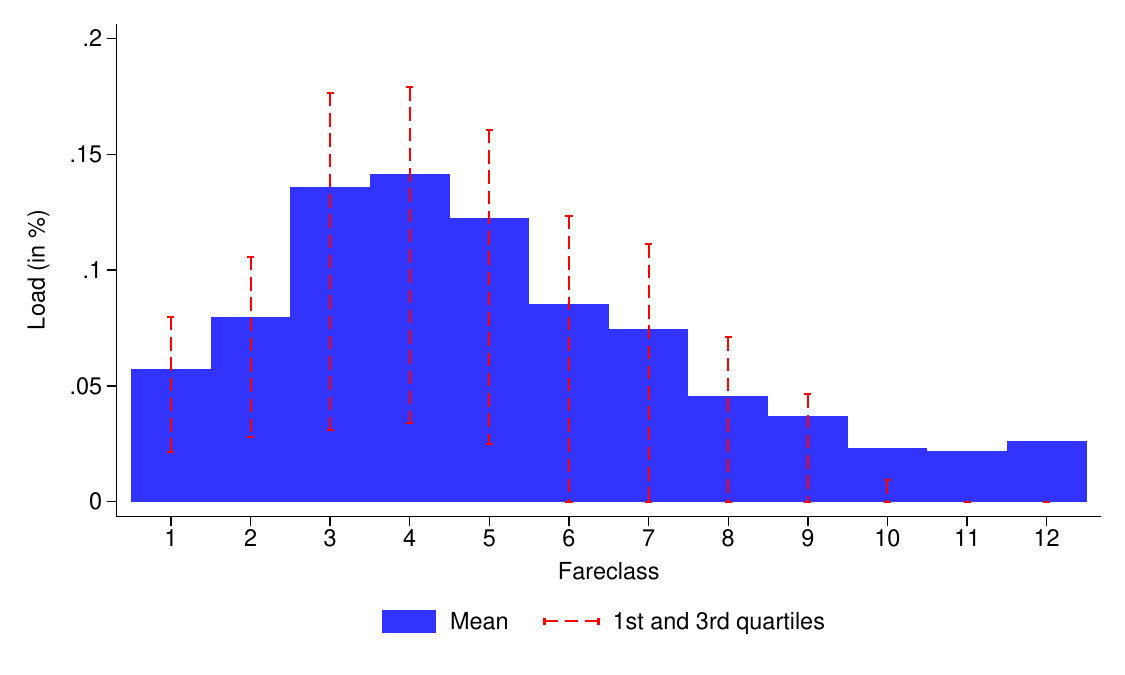}
\end{figure}

\medskip
In a similar vein, Figure \ref{fig:distr_load} shows that there is significant variation in the number of sales for each fare class. In particular,  the first quartiles of the load distribution across trains (lower points of the dashed segment in Figure \ref{fig:distr_load}) are zero for fare classes six to twelve; for fare classes eleven and twelve, the third quartiles (higher points of the dashed segment) are still zero. In other words, for more than $75\%$ trains, either fare class eleven (or twelve) was not open, or it was open but no tickets were sold.

\medskip
Finally, the overall load across all routes is $84.5\%$, with substantial variations on the average load across routes, from 65.4\% only for the C\^ote Basque route to more than 95\% on Paris-Marseille. Note that the firm would rather sell any unsold unit even with a small positive margin right before the booking period ends, hereby increasing its profit (and possibly welfare). The loads in iDTGV's observed practices do not seem consistent with this feature. The constraint of non-decreasing fares in the current system may be reasonable if consumers’ price elasticity rises steeply as departure approaches, as often observed in airline markets. In such environments, monotone pricing can be effective for  screening consumers and managing capacity. However, this constraint also limits iDTGV's ability to sell residual capacity toward the end of the booking period. Moreover, iDTGV’s reliance on predetermined fare classes, and their manual adjustment in response to demand fluctuations, may further reduce pricing efficiency. We quantify the extent of this suboptimality in Section~\ref{sec:opt_pric} in terms of its implications for revenue, load, and welfare.

\subsection{Demand estimation results}\label{sub:demand_est}
We consider the estimation of the price elasticity ($\eps_d$), the coefficients of destination-train specific effects ($\beta_0$), and the parameters of gamma distribution $(\lambda_{a0},\lambda_{b0})$. The variables we include in $W_m$ are route dummies, time dummies for the year and month of the train, whether it occurs during the weekend, on public holidays, on school holidays and whether departure occurs during a peak time. Regarding the variables $X_{dm}$,  we include travel time to $d$ by train $m$, its square, city-specific effects $X_d$ (namely, the population of the urban area of $d$ and whether $d$ is a regional capital) and all interactions $X_{dj}\times W_{mk}$ for all components $X_{dj}$ and $W_{mk}$ of the vectors $X_d$ and $W_m$, respectively. In Appendix \ref{sub:estimation_and_inference}, we provide details on the estimation and inference procedure.

\medskip
The estimates of price elasticities are displayed in the top panel of Table \ref{tab:binomial}. In Column I (our baseline specification), we assume a constant price elasticity, i.e., traveler's sensitivity to iDTGV seat price change is the same across routes and trains. We obtain a price elasticity of $-4.04$. This is comparable to the price elasticity estimates in some empirical studies on the airline sector, which use sales and prices at the flight-day level. For instance, \cite{aryal2024price} estimated an elasticity of $-3.9$ for leisure passengers (p.668); \cite{williams2022welfare} obtained an average price elasticity around $-3.31$ (p.846). In contrast, our estimate is larger in absolute value than those reported in the literature on  rail transportation. We refer for instance to the meta-analysis by \cite{Jevons_05} and the studies of \cite{Wardman_97}, \cite{Wardman_06} and \cite{Wardman_07}, which point to price elasticities in the range $[-1.3;-2.2]$. Unlike ours, most of these studies rely on aggregated data, which are likely to bias price-elasticity estimators upward. We illustrate this point below by running regressions based on our data aggregated at different levels. 

\medskip
In Column II, we allow price elasticity to vary across routes and trains. We find that travelers of routes from Paris to the southwest of France (namely, the routes to C\^ote basque, Toulouse and Perpignan) are less price-sensitive to price than those of other routes. Travelers on weekend or national holidays have a smaller price elasticity (in absolute value) than those on other days.  On the other hand, once controlling for weekend and national holidays, individuals traveling during peak hours appear to have a similar elasticity to the others.

\begin{center}
				\small{
		\begin{threeparttable}
			\caption{{Estimates of $(\eps_d,\beta_0,\lambda_0)$}} \label{tab:binomial}
			\begin{tabular}{lcccc}
				\toprule
				& \multicolumn{2}{c}{Binomial model} & \multicolumn{2}{c}{Multinomial model} \\
				&I&II & III & IV \\
				\midrule
				Minus price elasticity ($-\eps_d$) &  & & &\\
				\quad
				Constant&$\underset{(0.22)}{4.04}$&$\underset{(0.57)}{6.96}$ & $\underset{(0.22)}{4.04}$&$\underset{(0.38
					)}{6.96}$ \\
				\quad Southwest &&$\underset{(0.54)}{-2.09}$ & & $\underset{(    0.18 )}{-2.09}$ \\
				\quad Weekend/national holidays &&$\underset{(0.51)}{-2.46}$ & & $\underset{(0.17  )}{-2.46}$ \\
				\quad Peak hour &&$\underset{(0.48)}{0.37}$ & & $\underset{(0.15 )}{0.37}$ \\
				\midrule
				Destination effects & & & & \\
				\quad     Population (in M. inhabitants)&$\underset{(0.20)}{2.23}$&$\underset{(0.20)}{2.24}$  & $\underset{(0.19)}{2.17}$ & $\underset{(0.19)}{2.18}$ \\
				\quad     Regional capital&$\underset{(0.19)}{0.20}$ &$\underset{(0.19 )}{0.20}$ & $\underset{(0.19)}{0.17}$ & $\underset{(0.20)}{0.17}$ \\
				\quad     Travel time by train (in hours) & $\underset{(0.11)}{ -2.07}$ & $\underset{(0.11 )}{-2.11}$ & $\underset{(0.11)}{-1.60}$ & $\underset{(0.11)}{ -1.64}$ \\
				\quad     Travel time by train, squared&$\underset{(0.66)}{0.34}$&$\underset{(0.66)}{0.35}$ & $\underset{(0.64)}{0.28}$ & $\underset{(0.65)}{0.29}$ \\
				\hline
				Gamma distributions & & & & \\
				$\lambda_{a0}$ (intermediate) &$\underset{( 1.01)}{3.63}$& $\underset{(1.00)}{3.63}$ & $\underset{(1.07)}{3.93}$ & $\underset{(1.06)}{3.93}$ \\	
				$\lambda_{b0}$ (final) & $\underset{(0.40)}{2.62}$ & $\underset{(0.40)}{2.62}$ & $\underset{(0.41)}{2.79}$ & $\underset{( 0.41)}{2.78}$	\\			
				\hline
				Control for $X_d\times W_m$& Yes & Yes & Yes & Yes \\
				$R^2$ of the reg. of $\ln(\xi_{bm}/\xi_{am})$ & $0.502$ & $0.506$ & $0.521$ & $0.525$ \\
				\bottomrule
			\end{tabular}
			\begin{tablenotes}
				\scriptsize
				\item \emph{Notes:} The total number of trains is 2,909. In Columns I and II, with all fare classes, the total number of observations (fare classes $\times$ trains) is 21,988. In Columns III and IV, the total number of observations (fare classes $\times$ trains) is 34,908. Southwest correspond to the lines to C\^ote Basque, Toulouse and Perpignan. Standard errors (under parentheses) are calculated using the bootstrap with $500$ re-sampled datasets.
			\end{tablenotes}
		\end{threeparttable}
	}
\end{center} 

\medskip
The middle panel of Table \ref{tab:binomial} reports the estimates of the components of $\beta_0$ corresponding to travel time and city-specific effects. The effect of the population size and travel time by train are as expected. Larger cities lead to higher demand and a longer travel time decreases demand for train tickets. The effect of travel time may nonetheless be attenuated for long journeys, though the coefficient of the square of travel time is not significant. The bottom panel of Table \ref{tab:binomial}  reports the estimates of the parameters $(\lambda_{a0},\lambda_{b0})$ of the gamma distribution. Intermediate destinations are estimated to have larger uncertainty on demand ($V(\eta_{dm})=\lambda_{d0}$ under the gamma specification), though the difference between the two is not statistically significant.

\medskip
Several routes actually have multiple intermediate or final destinations. As discussed above, Theorem \ref{thm:ident_eps} then implies that the joint distribution of the purchases for these multiple destinations, conditional on the total number of purchases on the train, is multinomial rather than binomial (as is the case when the intermediate and final stops are aggregated). We re-estimate the demand models corresponding to Columns I and II using the multinomial model in Equation \eqref{eq:multinomial}. The results are displayed in Columns III and IV, respectively. The resulting price elasticities are almost identical to those obtained before. The destination effects and the estimates of $(\lambda_{a0},\lambda_{b0})$ are also very similar across specifications.

\paragraph{Test of Assumptions \ref{hyp:cons_demand} and \ref{hyp:yield_line} and $\eps_a=\eps_b$.} Together, these three conditions imply that the proportions $n_{bkm}/(n_{akm}+n_{bkm})$ remain constant through fare classes $k$ satisfying $p_{bkm}=p_{akm}$. A convenient way to check this is to restrict ourselves to two routes, Paris-Marseille and Paris-Mulhouse, for which $p_{bkm}=p_{akm}$ for all $k\in\{1,...,K\}$.  By taking the first fare class as a reference, we simply regress $n_{bkm}/(n_{akm}+n_{bkm})$ on the other 11 fare class dummies and train fixed effects. We then test  whether the coefficients of the fare class dummies are equal to zero.

\medskip
The results are presented in Figure \ref{fig:separability}. As shown by the vertical bars, most coefficients are not significant, despite the large number of observations ($453$ and $499$ for the two routes). For Paris-Marseille (blue bars), the p-value of the joint test is larger than 0.05. For Paris-Mulhouse (red bars), the p-value is lower, but it appears that this result is mostly driven by the last fare classes (the joint test for nullity of the first 10 classes has a p-value of 15\%). The coefficients of the last two fare classes are indeed positive and relatively large for this route, indicating that there  would be more ``late purchasers'' for Mulhouse than for Strasbourg.

\medskip
To see whether this pattern could influence our results beyond this specific route, we re-estimate $\eps$ using only the first 10 fare classes. We obtain a price elasticity of $-4.86$, which is thus somewhat higher in absolute value than the baseline estimate of $-4.04$ obtained with the 12 fare classes in Table \ref{tab:binomial}. However, such a difference does not materially affect our main counterfactuals
results.\footnote{We recompute the counterfactual revenues corresponding to u2, u4, s3, s6, f1 and f2 in Table \ref{tab:supply_main}. The optimal revenues are marginally higher but with differences never exceeding 3.3\% on the lower bounds and 1.2\% on the upper bounds.}

\begin{center}
	\begin{figure}[H]
		\caption{Test of the separability in Assumption \ref{hyp:cons_demand}}\label{fig:separability}
		\begin{center}
				\includegraphics[width=\textwidth]{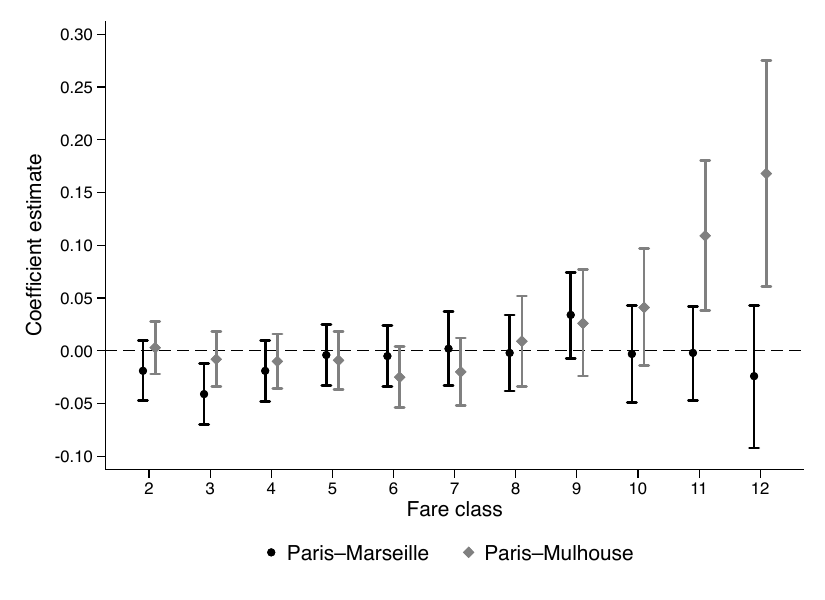}
		\end{center}
	\end{figure}
    \vspace{-1.5cm}
    
    \flushleft\footnotesize{
    \emph{Notes}: Fare class 1 is the reference category. Markers indicate the coefficient estimates, and vertical bars report the corresponding 95\% confidence intervals. The $p$-values for the joint null hypothesis that all fare-class coefficients are zero are $0.053$ for Paris--Marseille and $0.0004$ for Paris--Mulhouse. 
    }
\end{center}

\paragraph{Comparison with log-log regressions.} To better understand our identifying strategy, we compare our results with others based on log-log regressions. Specifically, we consider regressions of $\log(1+q_{dkm})$ on $\log(p_{dkm})$, including various fixed effects. The results are displayed in Table~\ref{tab:td_level}. The last column, where we include train-destination and train-fare class fixed effects, yields a result close to those in Table \ref{tab:binomial}. The identifying strategies are in fact similar. In both cases, we rely on the within-train comparison of the evolutions, over fare classes, of log-quantities and log-prices, between different destinations. An important advantage of our approach over this regression is that it handles the case of zero sales without the need of any ad hoc adjustments such as the addition of one sale, as we do in Table \ref{tab:td_level}, or the removal of observations with zero sales.

\begin{table}[H]
	\caption{Log-log regressions with train-destination or train-destination-fare class     data}\label{tab:td_level}
	\begin{center}
		\footnotesize{
			\begin{threeparttable}
				\begin{tabular}{rcccccccc}
					\toprule
                    Dependent var.&\multicolumn{3}{c}{$\log \left(1+\sum_{k=1}^{12}q_{dkm}\right)$}&\multicolumn{5}{c}{$\log(1+q_{dkm})$}\\ \midrule
                    $\log \bar{p}_{dm}$ &$\underset{(0.028)}{0.290}$& 
                     $\underset{(0.022)}{0.101}$&$\underset{(0.146)}{-1.163}$&&&\\
                    $\log {p}_{dkm}$ &&&&$\underset{(0.011)}{-1.214}$&$\underset{(0.011)}{-1.243}$&$\underset{(0.083)}{-0.472}$&$\underset{(0.011)}{-1.243}$&$\underset{(0.106)}{-4.338}$\\
					\hline
                    Destination FE&&Yes&Yes&Yes&Yes&Yes&&\\
                    Train FE&&&Yes&&Yes&Yes&&\\
                    Fare class FE &&&&&&Yes&&\\
                    Train-dest. FE&&&&&&&Yes&Yes\\
                    Train-fare class FE&&&&&&&&Yes\\
                	\hline
					No. of obs. &$5,803$ 
                    &$5,803$&$5,788$& $69,816$&$69,816$&$69,816$&$69,816$&$69,816$\\	
	               $R^2$&$0.018$ &$0.575$&$0.820$&$0.181$&$0.277$ &$0.378$&$0.297$&$0.938$\\
					\bottomrule
				\end{tabular}
				\begin{tablenotes}\footnotesize
					\item \emph{Notes}: 
                    We define $\bar{p}_{dm}=\sum_{k=1}^{12}q_{dkm}p_{dkm} /\sum_{k=1}^{12}q_{dkm}$. Train fixed effect is route-year-month-day specific. Destination fixed effect is defined as city dummies. 
				\end{tablenotes}
			\end{threeparttable}
		}
	\end{center}
\end{table}

\medskip
Table \ref{tab:td_level} also shows that when failing to include train-destination and train-fare class fixed effects, we obtain very different price-elasticity estimates. In the last but one column, for instance, we consider within-train comparisons in levels, rather than in evolution, of log-prices and log-quantities. The price-elasticity estimate is much larger (-1.24). We obtain similar estimates with coarser fixed effects.

\medskip
We also investigate the effect of considering more aggregate data, at the train-destination level. In empirical research on demand for airline tickets, this is typically the level at which products are defined (e.g., the airline-itinerary level in \cite{bontemps2023price} and \cite{yuan2024network}). We then define prices as the sales-weighted average prices $\sum_{k=1}^{12}q_{dkm}p_{dkm} /\sum_{k=1}^{12}q_{dkm}$. An issue with such prices is that they are endogenous by construction, as they directly depend on purchased quantities. This could explain why we obtain positive estimates (0.29 and 0.10) in the absence of train fixed effects. The situation improves when including such fixed effects, but the price-elasticity estimate remains large (-1.16) and very different from our preferred results.

\section{Implications for train pricing}
\label{sec:opt_pric}

In this section, we use our demand estimates to investigate the optimality of the observed iDTGV pricing strategy, in terms of revenue, load, and social welfare, compared to other possible strategies, from the most basic to the most sophisticated ones. We first study the identification of the corresponding counterfactuals, before presenting our results. We focus our discussion below on counterfactual revenues, but under the microfoundation of the demand for train tickets detailed in Appendix \ref{app:microfoundation}, counterfactuals for social welfare follow directly, see in particular Equation \eqref{eq:social_welfare} therein.

\subsection{Identification of counterfactual pricing strategies}\label{sec:counterfactual_rev}

\paragraph{Alternative pricing strategies and information structure.} The first counterfactual scenario we consider corresponds to the simplest pricing strategy, namely uniform pricing. In that case, the price of each route in a given train is fixed once and for all. In the opposite case,  under ``full'' dynamic pricing, prices can be changed at any time. We also study intermediate pricing strategies, referred to as stopping-time strategies hereafter, where prices can be changed only after a ticket is sold. Finally, we consider constrained stopping-time strategies close to what was implemented in practice, by assuming that only a finite (resp. increasing fares) number of fares are allowed.\footnote{Other forms of constraints may exist beyond prices. For instance, the initial planning of seat allocation among destinations and fare classes  may sometimes be constrained; revenue managers cannot adjust enough this initial allocation to achieve the optimal allocation under complete information. Our analysis abstracts away from such constraints that are unobserved in the data.}

\medskip
As regards information available to the revenue managers, we consider two scenarios:
\begin{enumerate}
	\item (Complete information) Revenue managers fully know the expected demand for each train. Put differently, they observe $\eps$, $s_m(\cdot)$, $\xi_{am}$ and $\xi_{bm}$ for each train $m$;
	\item (Incomplete information) Revenue managers observe $\eps$, $s_m(\cdot)$, $(X_{am},X_{bm},W_m)$ but only $f_{\xi_{am},\xi_{bm}|X_{am},X_{bm},W_m}$. As time goes by, revenue managers update their information on $(\xi_{am},\xi_{bm})$ according to Bayes' rule.
\end{enumerate}
The complete information case should be seen as a benchmark. It is especially useful to quantify the value of information and contrasts the gains of revenue management in both complete and incomplete information setups. The case of incomplete information is probably more realistic. In this scenario, revenue managers know price elasticities and the pattern of $s_m(\cdot)$ over time for each train, but do not know exactly the aggregate demand for each destination ($\xi_{am}$ and $\xi_{bm}$). Assuming  $s_m(\cdot)$ to be known makes particular sense if $s_m(\cdot)$ does not depend on $m$, in which case revenue managers may have learned how consumers arrive over time from previous observations. If the scenario of incomplete information holds in practice, the differences between counterfactual and observed revenues shall be interpreted as potential revenue gains (or losses) of the optimal revenue management under different constraints compared to the actual ones. 

\medskip
In fact, information may be even more limited than we assume here: for instance, managers may not know $s_m(\cdot)$. Our incomplete information setup may also be too pessimistic, in that  unobserved demand shock $\eta_{dm}$ in $\xi_{dm}$ could be serially correlated across departure dates, rather than independent.\footnote{To investigate this, we regressed revenues of given trains on revenues of past trains, controlling for train characteristics (e.g., rush hour, dummies of day, day-in-month, month, year, and route). We find significantly positive coefficients for one-day as well as one-week lagged revenue, while the coefficient decays to near zero and becomes insignificant when using lags of one month or more, suggesting indeed serial correlation in the $(\eta_{dm})_m$.} Such correlation would imply that one can predict future demand shocks based on current observations, creating information spillover in demand learning. This would lead to a more precise prior on the distribution of the aggregate shocks of tomorrow, favoring demand learning and narrowing the revenue gap between incomplete and complete setups. We abstract from such information spillover in this article; it is unclear to us whether one can still compute counterfactuals in such cases. 
Despite these limitations, our exercise can be seen as a first step that relaxes the complete information setup used in the existing empirical literature \cite[see, e.g.,][]{lazarev2013welfare, williams2022welfare, aryal2024price}.  

\paragraph{Counterfactual invariance.} Before stating our identification result on the counterfactuals, it is worth noting that we maintain Assumption \ref{hyp:cons_demand} throughout. This requirement of counterfactual invariance rules out that the distribution of consumers' valuations varies with their arrival time.\footnote{\label{foot:change_s} On the other hand, Theorem \ref{thm:counterf_rev} makes it clear that $s_m(.)$ may depend on the pricing strategy, as long as the firm knows the $s_m(.)$ function corresponding to each pricing strategy.} Yet, with a more sophisticated pricing strategy, consumers with lower valuation may arrive earlier, for instance. Similarly, Assumption \ref{hyp:cons_demand} is not compatible with consumers who wait before purchasing their ticket, as they progressively acquire information on the exact value of their travel. In that scenario indeed, the composition of potential buyers at a given point in time would depend on all the prices that were set before that moment, but also on the anticipated future path of prices. When consumers are not forward-looking, counterfactual invariance is nonetheless a natural and standard choice in the literature \citep[see, e.g.][]{aryal2024price}. Moreover, assuming that consumers are not forward-looking may be reasonable here: in the related setup of airlines, \cite{li2014consumers} (section 6.3, p. 2127) finds that fewer than 13\% of consumers are forward-looking.


\medskip
Another implication of the counterfactual invariance is that the pricing strategies of competitors remain fixed. Thus, our counterfactuals provide partial equilibrium outcomes that isolate the effect of pricing strategy changes on revenue, load, and welfare. Nevertheless, the microfoundation of the demand model (Appendix \ref{app:microfoundation}) allows us to quantify the compensating adjustments in competitors’ prices that would exactly offset these gains or losses (see also footnote \ref{footnote:empirical_CV}).

\medskip
Lastly, the counterfactual invariance overlooks potential  substitutions over time: the demand for other iDTGV trains departing just before or after the train under consideration may change following price changes for that specific train. This concern is nonetheless mitigated by the fact that there is only one iDTGV train per day for each of the six routes we consider.

\paragraph{A general identification result on counterfactual revenues.} The following result clarifies the ingredients needed to identify the revenues of the counterfactual pricing strategies. We condition these revenues on destination $d$ and train $m$, and therefore their characteristics $(X_{dm},W_m)$ and the allocated capacity $C_{dm}$. The proof of Theorem \ref{thm:counterf_rev} can be found in Appendix \ref{sub:theorem_ref_thm_counterf_rev}. Though not stated here, the same result holds for the corresponding counterfactual loads and welfare (see the paragraph ``Welfare'' in Appendix \ref{app:microfoundation}).
\begin{thm}\label{thm:counterf_rev}
	Suppose that Assumptions \ref{hyp:cons_demand}, \ref{hyp:yield_line}, and \ref{hyp:gamma}(i)  hold. Then, for complete or incomplete information setups, the optimal revenues under uniform pricing, full dynamic pricing and stopping-time pricing (with a finite number of (increasing) fares) strategies depend on $\eps_d$ and the distribution of $(\xi_{am},\xi_{bm})$, but not on $s_m(\cdot)$.
\end{thm}

This result is crucial in our context where no information on purchasing dates, and thus on $s_m(\cdot)$, is available, because it shows that this knowledge is in fact not necessary to identify counterfactual revenues.\footnote{On the other hand, the price paths corresponding to counterfactual revenues do depend on $s_m(\cdot)$ in general (see Section 3 
of the Online Appendix for details). Consequently, we cannot identify them given our data in which no information on purchasing dates is available.} 

\medskip
We obtain Theorem \ref{thm:counterf_rev} by constructing a different setup from the true one, with a new time variable and new Poisson demand processes. This new setup is very close to the true one, except that the Poisson processes are now homogeneous: $s_m(.)$ is replaced by the constant function $s_m(.) = 1$. We prove that the optimal revenues are the same in this setup as in the true one. This shows that the optimal revenues do not depend on $s_m(.)$: it does not matter whether consumers arrive early or late, as long as on average, the same number of consumers eventually arrive. The result holds because basically, all the constraints on pricing we consider are independent of time. In this sense, Theorem \ref{thm:counterf_rev} holds beyond the specific scenarios we consider here. But it would fail if time constraints were imposed on the pricing strategies, for instance if a limit on the number of price changes occurring before a given date $t^*<1$ were set.\footnote{Theorem \ref{thm:counterf_rev} would also fail with time-varying elasticity. We discuss in Section 2 
of the Online Appendix the identification of counterfactual revenues in this case.} In Section 1 
of the Online Appendix, we express the counterfactual revenues, loads, and social welfare as a function of the demand parameters when $(\eta_{am},\eta_{bm})$ follows a gamma distribution (Assumption \ref{hyp:gamma}(ii)), allowing us to then bound these revenues.

\paragraph{Upper bound on total demand when demand is censored: A weak optimality condition on supply.} Because demand is censored by the total number of seats available in a train, it is unclear whether an upper bound on $g_0(W)$ can be obtained by solely relying on Assumption \ref{hyp:cons_demand} and inequality \eqref{eq:upper_bound}.  Instead, we rely on an additional condition on supply. This approach may apply beyond our application, provided that the marginal cost of the good is known. 

\medskip
Let $R_m(p_a,p_b)$ denote the maximal revenue for train $m$ under a uniform pricing of $(p_a, p_b)$ for destinations $a$ and $b$ respectively. This maximal revenue is obtained by considering the optimal quotas $C_{am}$ and $C_{bm}$ of tickets sold for destinations $a$ and $b$ respectively, with $C_{am}+C_{bm}=C_m$, the total capacity of train $m$. Then, under Assumptions \ref{hyp:cons_demand}-\ref{hyp:gamma}, we have (see Appendix \ref{proof_thm_limited_rev}, section ``uniform pricing'', for details)
\begin{align}
\mathbb{E}[R_m(p_a,p_b)|W_m] =&\max_{\substack{(C_{am},C_{bm}):\\ C_{am}+C_{bm}=C_m}}\bigg\{\sum_{d\in\{a,b\}}p_d\int_0^\infty \mathbb{E}\bigg[D(\exp\{X_{dm}'\beta_0 \}p_d^{-\eps}g_0(W_m)z) \notag \\
&\hspace{5.2cm} \wedge C_{dm}|W_m\bigg] \times  \gamma_{\lambda_{d0},1}(z)dz\bigg\},
\label{eq:uniform_rev}
\end{align}
where $D(u)\sim \mathcal{P}(u)$ and $\gamma_{\lambda_{d0},1}$ is the density of Gamma distribution $\Gamma(\lambda_{d0},1)$. The supply-side condition we consider is the following:

\begin{hyp}\label{hyp:weak_opt}
	(Weak optimality of actual revenue management) We have
	\begin{equation}\label{eq:moment_ineg3}
	\max_{k=1,...,K}\E\left[{R_m(p_{akm},p_{bkm})}|W_m\right]\leq \E\left[R^{\text{obs}}_m|W_m\right].
	\end{equation}
\end{hyp}

\medskip
Assumption \ref{hyp:weak_opt} is weaker than the usual supply-side assumption in the literature of airline revenue management. The latter often imposes that the firm maximizes its revenue (or profit) in the observed scenario and chooses the optimal dynamic pricing strategy, e.g., \cite{lazarev2013welfare,williams2022welfare,aryal2024price}. Instead, we simply assume that observed revenues are on average higher than those one would have obtained by sticking to a single fare class from the 12 predetermined ones over the whole booking period. 

\medskip
We refrain from imposing strong optimality for several reasons. First, this would result in gains against most simpler pricing strategies, conflicting with our very objective of quantifying the gains or losses of the actual revenue management compared to alternative scenarios. Second and related to the first point, it seems restrictive in our setting to assume that the optimal dynamic strategy was adopted. As discussed in Section \ref{sub:revenue management}, the revenue management applied simplified rules (increasing fares from 12 predetermined fare classes), which can at best approach the optimal solution. Third, given that iDTGV is a subsidiary of publicly-owned SNCF, it may have constraints on, e.g., load or social welfare, which would be violated under revenue maximization but remain compatible with Assumption \ref{hyp:weak_opt}. Finally, there may also exist managerial reasons that lead to the lack of strong optimality in the observed scenario.\footnote{For instance, seat allocation decisions were subject to the manager's manual intervention, which could be a source of suboptimality. See \cite{cho2018optimal,cho2019semi} and \cite{phillips2021pricing} for evidence of suboptimality due to human management.}

\medskip
Let $\theta_0:=(\eps,\beta_0,\lambda_{a0},\lambda_{b0})$. Theorems \ref{thm:ident_eps}, \ref{thm:ident_xi}, and Equation \eqref{eq:uniform_rev} imply that the function $R$ defined by $$R(g_0(W_m);X_{am},X_{bm},\theta_0):=\max_{k=1,...,K}\E\left[{R_m(p_{akm},p_{bkm})}|W_m\right]$$
is identified (here, our notation reflects that $(X_{am},X_{bm})$ is a deterministic function of $W_m$). The weak optimality condition \eqref{eq:moment_ineg3} is equivalent to:
\begin{equation}
R(g_0(W_m);X_{am},X_{bm},\theta_0)\leq \E\left[R^{\text{obs}}_m|W_m\right].
\label{eq:moment_ineg3bis}
\end{equation}
The function $R(.;X_{am},X_{bm},\theta_0)$ is strictly increasing. Denoting by  $R^{-1}(.;$ $ X_{am},X_{bm},\theta_0)$ its inverse, we obtain the following (identified) upper bound for $g_0(W_m)$:
\begin{equation}\label{eq:upper_bound_rev}
g_{0}(W_m)\leq R^{-1}\left(\E\left[R^{\text{obs}}_m| W_m\right];X_{am},X_{bm},\theta_0\right).
\end{equation}
The next result parallels Theorem \ref{thm:partial_id_g} and summarizes the partial identification of $g_0(W_m)$ in the case of censored demand.
\begin{thm}[Partial identification, censored demand]\label{thm:partial_id_g_censored}
     Suppose that Assumptions \ref{hyp:cons_demand}-\ref{hyp:weak_opt} hold. Then,  \eqref{eq:lower_bound_censored} and \eqref{eq:upper_bound_rev} provide lower and upper bounds for $g_0(W_m)$, respectively.
\end{thm}
The width of the bounds in Theorem 
\ref{thm:partial_id_g_censored} depends on the true pricing strategy in the observed scenario, and on the amount of censoring. To see this, suppose that the true pricing is uniform and that most trains are not full (i.e., low demand). Then, the realized purchases on the right-hand side of  \eqref{eq:lower_demand} are close to the demand generated by uniform pricing, i.e., the quantity on the left-hand side. The lower bound  \eqref{eq:lower_bound_censored} is then close to an equality. Furthermore, if the observed pricing is close to the optimal uniform pricing described in Assumption \ref{hyp:weak_opt}, the upper bound \eqref{eq:upper_bound_rev} is also close to an equality. Consequently, the bounds in Theorem  \ref{thm:partial_id_g_censored} are narrower and more informative about $g_0(W_m)$. In contrast, if the observed pricing is close to a dynamic one, or many trains are full, either the lower or the upper bound becomes looser, leading to wider bounds on  $g_0(W_m)$. 

By Theorems \ref{thm:ident_eps}, \ref{thm:ident_xi},  and \ref{thm:partial_id_g_censored}, we can then partially identify average counterfactual revenues in Theorem \ref{thm:counterf_rev} and, under the model in Appendix \ref{app:microfoundation}, other quantities of interest such as consumer surplus and social welfare. 

\subsection{Results}\label{sec:counterfactuals}

We now report counterfactual simulation results on revenue, load, and social welfare under the different pricing strategies and information structures discussed above. We use hereafter the demand specification in Column I of Table \ref{tab:binomial} for counterfactual simulations; robustness checks are considered in the next subsection. 

\paragraph{How does the actual strategy compare to counterfactual pricing strategies?}
Table \ref{tab:supply_main} summarizes our main results on revenues. By Assumption \ref{hyp:weak_opt}, the actual strategy is supposed to achieve a higher expected revenue than any uniform pricing strategy under incomplete information and with prices constrained to belong to the price grid. The gains are however moderate: they range between 0\% and 9.5\% (Scenario u.1). Besides, we cannot exclude that the observed strategy delivers lower revenue than the optimal uniform pricing strategy with unconstrained prices (Scenario u.2). In any case, the gains would be at most 8.1\%.\footnote{We also find that the effect of the grid is higher in the complete information setup. This may arise because for trains with unusually high or low demand, optimal prices are beyond the range of the grid. Such instances are anticipated under complete information only, and then the  constraint induced by the grid binds.} 

\medskip
Observed revenues range between 9.2\% and 16.7\% below the optimal stopping-time pricing strategy (Scenario s.3), and between 9.4\% and 17.1\% below the optimal dynamic pricing strategy (Scenario f.1). Both losses are statistically significant at the 5\% level, with the upper bound of p-values being $0.048$ and $0.047$, respectively.\footnote{These p-values are obtained using a subsampling procedure described in Section \ref{sub:estimation_and_inference} and implemented by drawing $1,000$ subsamples of size $300$, with $50$ trains from each of the $6$ routes.}  

\setlength{\tabcolsep}{0.1em}
\begin{table}[H]
			\caption{Revenues under counterfactual pricing strategies, average over lines}\label{tab:supply_main}
			\vspace{-0.5cm}
	\begin{center}	
	\setlength{\tabcolsep}{5pt}
	\begin{threeparttable}
		\begin{tabular}{lc}
			\toprule
			Scenario& \multicolumn{1}{c}{Point or set estimate (in K\EUR)} \\ 
		    \textbf{Observed pricing strategy}&$
		    {12.21}$\\
			\hline
			\textbf{Uniform pricing strategy} &\\
			u.1\ Incomplete information, constrained prices &$
			{[11.15,12.21]}$ \\
			u.2\ Incomplete information, unconstrained prices &$
			{[    11.29,   12.31]}$\\
			u.3\ Complete information, constrained prices &$
			{[12.56,13.86]}$\\
			u.4\ Complete information, unconstrained prices &
			${[13.23,14.42]}$\\
			\midrule 
			\textbf{Stopping-time pricing strategy} &\\
			s.1\ Incomplete information, 12 increasing fares&${[13.10,14.39]}$\\
			s.2\ Incomplete information, 12 fares&${[13.27,14.57]}$\\
			s.3\ Incomplete information  &$
			{[13.44, 14.66]}$\\
			s.4\ Complete information, 12 increasing fares&${[13.39,14.60]}$\\
			s.5\ Complete information, 12 fares &${[13.47,14.69]}$\\
			s.6\ Complete information &$
			{[13.48,14.70]}$\\
			\midrule
			\textbf{``Full'' dynamic pricing strategy} & \\
			f.1\ Incomplete information  & $
			{[13.47,    14.68]}$\\
			f.2\ Complete information &$
			{[13.50, 14.72]}$\\
			\bottomrule
	\end{tabular}
		\begin{tablenotes}
			\footnotesize
			\item \emph{Notes: } 
			Estimated bounds on counterfactual revenues are the averages across all routes. With ``constrained prices'' (resp. ``unconstrained prices''), optimization is conducted over the actual price grid (resp. over all positive real numbers). 
			\end{tablenotes}
	\end{threeparttable}
\end{center}
\end{table}

We also consider almost the same constraints as in the actual pricing strategy, namely 12 fare classes and increasing prices (Scenario s.1). The only difference between s.1 and actual pricing is that we do not impose a fixed price grid in s.1; nevertheless, as the difference in revenues between u.1 and u.2 shows, the effect of the grid is unlikely to alter much our results. In any case, the observed revenues appear to be at least 6.8\% and up to 15.1\% below the optimal revenue under s.1.\footnote{\label{footnote:empirical_CV}As discussed in Section \ref{sec:counterfactual_rev},
the gains are obtained by keeping competitors' prices the same as in the observed scenario, i.e., a partial equilibrium. We can compute the decrease in competitors' overall prices that offsets the gains. For instance, in the case with 12 increasing prices, we show that the decrease should be at least 6.8\%. See Appendix \ref{app:microfoundation} for more details.}$^{,}$\footnote{Under the constraint of increasing fares, it may be without loss to restrict to constrained stopping-time strategies. Intuitively, in the absence of sales, the manager has no incentives to further decrease demand by increasing the price. The price then remains unchanged. It is only when a sale occurs that the manager has incentives to increase the price (we thank an anonymous referee for pointing this out). Even though we do not prove this conjecture, it is supported by simulations.} Under very limited forms of dynamic pricing strategies, namely two fare classes and increasing prices, we still observe a loss in revenue ranging between 3.6\% and 12.2\%. 

\medskip
In Table \ref{tab:counter_sw_cs}, we report counterfactual consumer surplus and social welfare. Overall, revenue-maximizing stopping-time and full dynamic pricing strategies deliver higher social welfare than the observed one, with an improvement of around $10\%$, regardless of the underlying information structure. Moreover, the welfare loss of the observed strategy is likely to be mainly driven by revenue rather than consumer surplus. For instance, compared to the full dynamic pricing strategy under incomplete information (f.1), the revenue loss explains around $75\%$ of  that in social welfare, while consumer welfare explains around $25\%$. Besides, the actual strategy achieves a median load of $89.2\%$ across all routes, slightly lower than, but not statistically different from, the load under the optimal uniform pricing in the incomplete information setup ($91.7\%$). The median loads of the optimal stopping-time under incomplete and complete information are both greater than $99\%$, significantly higher than the observed load. 

\medskip
Overall, the results above highlight the importance of not imposing strong supply-side optimality conditions in our setting. 
The small difference in revenues between s.1 and s.3 also suggest that the actual pricing constraints may not be the main source of suboptimality. To shed light on the origins of the estimated suboptimalities, we thus discuss below the role of demand learning. We focus hereafter on revenue, as it appears to be the main driver of suboptimality. 

\medskip
\begin{table}[H]
	\begin{center}
		\caption{Estimates of revenues, social welfare, and consumer surplus}\label{tab:counter_sw_cs}
			\begin{threeparttable}
 				\begin{tabular}{lccccc}
					\toprule
					&   \multicolumn{5}{c}{Estimates (in K\EUR)} \\
					Scenario &  Revenue && SW && CS \\
					\hline
					\textbf{Observed} &$12.21$&&$16.23$&&$4.02$\\
					\textbf{Uniform pricing strategy} & & & & &\\
					u.2 Incomplete information &$[11.29,12.31]$&&$[15.00,16.36]$&&$[3.71,4.05]$\\
					u.4 Complete information &$[13.23,14.42]$&&$[17.58,19.16]$&&$[4.35,4.74]$\\
					\textbf{Stopping-time pricing strategy} & & & & &\\
					s.3 Incomplete information &$[13.44,14.66]$&&$[17.86,19.48]$&&$[4.42,4.82]$\\
					s.6 Complete information &$[13.48,14.70]$&&$[17.91,19.54]$&&$[4.43,4.84]$\\
					\textbf{``Full'' dynamic pricing strategy} & & & & &\\
					f.1 Incomplete information &$[13.47,14.68]$&&$[17.90,19.51]$&&$[4.43,4.83]$\\
					f.2 Complete information &$[13.50,14.72]$&&$[17.94,19.56]$&&$[4.44,4.84]$\\
					\hline
				\end{tabular}
				\begin{tablenotes}\footnotesize
					\item \emph{Notes}: Estimated bounds  are the averages across all routes. 
                    See Appendix \ref{app:microfoundation} for the derivations of social welfare and consumer surplus under revenue-maximizing pricing strategies. 
				\end{tablenotes}
			\end{threeparttable}
	\end{center}
\end{table}

\paragraph{The role of information and demand learning.} Manager's information on demand plays a key role in dynamic pricing. In the complete information case, dynamic pricing still improves revenue relative to uniform pricing because of the uncertainty on the demand process. But demand learning under incomplete information, namely the possibility to adjust the pricing strategy as one learns about $(\eta_{am},\eta_{bm})$ from past purchases, plays a much more important role. To shed light on this point, we decompose the variance of the demand under the optimal uniform pricing in incomplete information into two parts:
\begin{align*}
\E\left[\V(D_{dm}(0,1,p^u_{dm})|W_m) \right]= & \mathbb{E}[\V(D_{dm}(0,1,p^u_{dm})|\xi_{dm})] +\E\left[\V(\mathbb{E}[D_{dm}(0,1,p^u_{dm})|\xi_{dm}]|W_m)\right],	
\end{align*}
where $p^u_{dm}$ is the optimal price under uniform pricing for destination $d$ and train $m$. Even though they both involve $g_0(W_m)$, one can show that the two terms in this decomposition are point identified. For intermediate and final destinations respectively, the variation of the demand process (the first term) only explains on average $1.3\%$ and $0.9\%$ of the total variance.

\medskip
In terms of revenues, by moving from uniform pricing to ``full'' dynamic pricing strategies, we find a modest revenue gain of around 2.0\% in the complete information setup.\footnote{This empirical finding is close to the existing simulation results in operations research. For example, \cite{zhao2000optimal} shows a similar improvement by between 2.4\% and 7.3\%.} This figure sharply contrasts with the 19.3\%  gain we estimate under incomplete information by comparing Scenarios f.1 and u.2. Even for very limited forms of dynamic pricing strategies such as the stopping-time ones  with only two prices, the revenue gains relative to the uniform pricing in the incomplete information setup are already remarkable (around $12\%$). 

\medskip
Besides, this demand learning can compensate almost all revenue and welfare loss due to ex ante uncertainty on demand. The difference in revenue under optimal uniform pricing and the corresponding social welfare between incomplete and complete information is around \EUR{2,000} and \EUR{2,500}, respectively (comparing u.4 and u.2 in Table \ref{tab:counter_sw_cs}), while these differences decrease to around \EUR{30} and \EUR{40} only under optimal dynamic pricing (see f.2 and f.1).\footnote{This finding is in line with \cite{lin2006dynamic}, who reports a similar near-optimality of demand learning in a simulation study.} Given that observed revenues are well below those estimated under f.1 or even s.1, this suggests a potential lack of effective demand learning in the actual pricing strategy that leads to the revenue and welfare losses relative to optimal dynamic pricing strategies. 

\medskip
The reason of the very modest loss due to incomplete information in dynamic pricing compared to the complete information setup is that information accumulates quickly. To illustrate this point, we simulate expected revenues under a class of intermediate stopping-time pricing strategies, where the firm is only allowed to dynamically price the first $x\%$ seats, turning to uniform pricing for the remaining seats. Thus, $x=0$ and $x=100$ correspond respectively to the optimal uniform and stopping-time pricing strategies. By quantifying the revenue gain from $x$ to $x+1$, we can characterize how much can be marginally gained from being able to extract information on demand from additional purchases ($1\%$ of total seats) and optimally adjust its pricing. 

\begin{center}
	\begin{figure}[H]
		\caption{Revenues (lower bound) under intermediate pricing strategies}\label{fig:intermediate}
		\vspace{-7cm}
		\begin{center}
				\includegraphics[width=\textwidth]{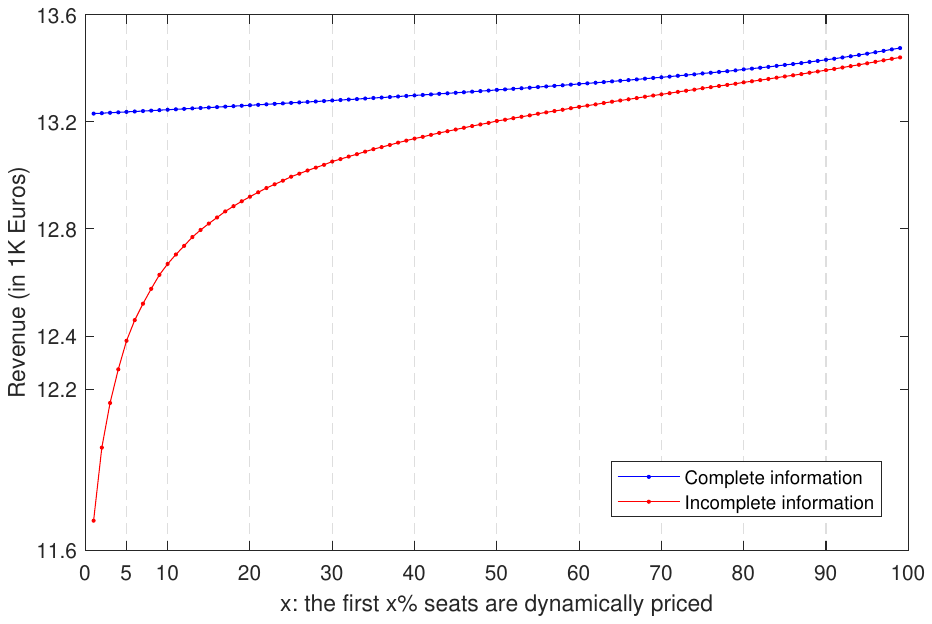}
		\end{center}
				\vspace{-7cm}
	\end{figure}
\end{center}

\medskip
Figure \ref{fig:intermediate} displays the lower bounds of the optimal revenues under these intermediate pricing strategies under complete (blue) and incomplete (red) information for $x=1,...,100$.\footnote{We also simulate the upper bounds of these revenues. The obtained curve is very similar.} Under incomplete information, demand learning is rather quick, as we can see from the important concavity of the red line. With just $x=5$, the firm already achieves a revenue equal to the observed one; by learning from  $50\%$ of the seats, it obtains a revenue around $3\%$ lower than that of the complete information.\footnote{It may be that if the information set of revenue managers is coarser than we assume (with, e.g., uncertainty on  price elasticity and $s_m(\cdot)$), information acquisition is slower. On the other hand, we do not account for the possibility that $\eta_{dm}$ is serially correlated across adjacent departure dates, rather than independent. While our estimation of demand is robust to such correlation \citep[e.g.,][]{levine1983remark}, this would create information spillover in demand learning and accelerate demand learning. We leave these important questions for future research.} On the other hand, the blue line shows that the revenue gains under complete information are small. The incremental revenue from $x$ to $x+1$ is almost constant and barely reaches \EUR{3}. This latter result could be expected, given that the difference between uniform pricing and the full stopping-time pricing is small under complete information. 

\medskip
The striking difference in the pattern of marginal gain between complete and incomplete information settings is also in line with our previous findings: in terms of revenue improvement by dynamic pricing, the effect of learning overall demand $(\xi_{am},\xi_{dm})$ is much more important than that of pinning down the uncertainty in demand process when $(\xi_{am},\xi_{dm})$ is fixed. 

\subsection{Robustness checks}\label{sec:robustness_checks}

\paragraph{Time-varying price elasticities.} One could expect that consumers purchasing their tickets earlier would be more price elastic than those buying their tickets late. For instance, the latter could include more business travelers. If so, our price elasticity estimate would be biased. Also, consumers with different price sensitivities can be screened to increase revenues based on their purchasing time, and our counterfactuals would fail to include such an intertemporal price discrimination. 

\medskip
We entertain this possibility by replacing $\eps$ in Assumption \ref{hyp:cons_demand}.1 with $\eps_{\text{early}} 1\{k\leq S \} + \eps_{\text{late}} 1\{k> S \}$, for some threshold $S$, so that $V_{dm}$ is now a Poisson process with intensity
\begin{equation}
I_{dm}(t,p) = \xi_{dm} s_m(t) \left[\eps_\text{early} p^{-1-\eps_{\text{early}}}\ind{t\le t_S} + \eps_\text{late} p^{-1-\eps_{\text{late}}}\ind{t> t_S}\right],	
	\label{eq:I_two_elast}
\end{equation}
where $t_S\in(0,1)$ is the closing time of fare class $S$ and is assumed to be a function of $W_m$. This last restriction holds for instance if the moment when fare class $S$ is closed is fixed initially depending on $W_m$, and not adjusted afterwards. Then, Equation \eqref{eq:binomial} still holds, with $\eps$ simply replaced by $\eps_{\text{early}} 1\{k\leq S \} + \eps_{\text{late}} 1\{k> S \}$. Thus, we can identify both elasticities provided there is enough variation in relative prices.

\medskip
Table \ref{tab:demand_two_elas} in Appendix \ref{app:res_TV_elast} reports demand estimates using thresholds $S = 9$, $10$, and $11$. Across all cases, early purchasers appear more price elastic than late purchasers, with elasticities greater in magnitude than the baseline ($-4.04$) but still within its $95\%$ confidence interval. Other parameters, such as $\beta_0$ and $\lambda_0$, remain close to the baseline results.

\medskip
Next, we assess the robustness of the time-invariance elasticity condition on counterfactual revenues and social welfare. We check the lower bound in \eqref{eq:revenue_gamma_ave} under two elasticities with $S=10$. We refer to Section  2 
of the Online Appendix for computational details. Table \ref{tab:counter_rev_two_elas} in Appendix \ref{app:additional_tab_fig} summarizes the results on revenues and social welfare with revenue-maximizing uniform, stopping-time and ``full'' dynamic pricing, under both complete and incomplete information. In all scenarios, the estimated lower bounds of revenues are within around \EUR{500} (or 4\%) of those in Table \ref{tab:counter_sw_cs}. The revenue losses in the observed scenario  remain above 4.9\% relative to optimal dynamic pricing strategies (Scenarios s.3 and f.1).  Relative differences in the estimated lower bounds across pricing strategies also align closely with those in Table \ref{tab:counter_sw_cs}. The lower bounds on social welfare become smaller when we relax the assumption of time-invariant price elasticity. However, they are still informative and imply important welfare losses of the observed pricing strategy relative to revenue-maximizing dynamic pricing strategies.

\medskip
Overall, the results  suggest that despite the difference in price elasticity of early and late purchasers and the possibility of intertemporal price discrimination, our results are robust to the assumption of a time-invariant price elasticity.

\paragraph{Alternative parametric specifications.} We conduct two robustness checks. First, we simulate counterfactual revenues with a lognormal specification on $\eta_{dm}$ instead of a gamma distribution (Assumption \ref{hyp:gamma}(ii)). The drawback of a lognormal specification is that it is not conjugate with the Poisson distribution. Then, the updated distribution of $\eta_{dm}$ in the incomplete information setup takes a complicated form, making it  difficult to compute counterfactual scenarios. Nevertheless, this issue does not appear for uniform pricing and complete information. Table \ref{tab:supply2_benforroni} in Appendix \ref{app:additional_tab_fig} shows the results for Scenarios u.2, u.4, s.6 and f.2. Even if the bounds are wider than in the baseline specification, the results are similar. Second, we have focused so far on the demand model corresponding to Column I in Table \ref{tab:binomial}. Counterfactual revenues are harder to compute under the richer specification associated to Column II in that table. Nevertheless, we can still compute such revenues for a few scenarios. The results hardly change compared to those in Table \ref{tab:supply_main}, see Table \ref{tab:rev_column2} in Appendix \ref{app:alt_specif}.

\paragraph{Effect of specific routes.} Table \ref{DES2} show that the routes to Marseille and C\^ote basque have unusually high and low loads, so one may worry that revenue management was very different for these lines. We recompute the counterfactual revenues by excluding these two routes. The results are hardly affected, with changes in the bounds by at most 1\% over all scenarios.

\section{Conclusion}

We have developed a new, nonlinear difference-in-difference identification strategy for demand estimation when standard approaches, based in particular on instrumental variables, are not available. Applying it to railway transportation, we obtain price elasticity estimates that are larger in magnitude than those obtained in the literature, which may be due to aggregation biases in previous studies. We further show how these demand estimates can be used to assess the optimality of iDTGV's pricing strategy. The results indicate that the observed pricing practice entails significant revenue and welfare losses relative to a fully flexible dynamic benchmark. An important part of such losses may come from insufficient demand learning from consumer purchases.

\medskip
We leave two important questions for future research. First, by imposing independence between $V_{am}$ and $V_{bm}$, we have assumed that the two goods target different segments of consumers. Relaxing this assumption would broaden the scope of applications of our method. Second, consumers' search and the timing of each purchase was unobserved in our application, but in some cases these data are accessible. Presumably, it would be possible to weaken some of our assumptions with such additional data.

\newpage
\bibliographystyle{chicago}
\bibliography{biblio}

\appendix
\section*{Appendix}

\section{Microfoundations of the demand model}\label{app:microfoundation} 

This section aims at showing that Assumption \ref{hyp:cons_demand} may be derived from a model of consumer's choice. First, we suppose that the number of travelers who search at time $t\in[0,1]$ for traveling to destination $d$ (e.g., from Paris to Marseille) at time $1$ follows a non-homogeneous Poisson process with rate $\kappa_{dm}\lambda_{mt}$. These consumers then choose between $\overline{L}$ modes (iDTGV, airline, another train, etc.), with mode 1 corresponding to iDTGV.  We assume that traveler $i$ searching at time $t$ for route $r=(d,m)$ has the following indirect utility when choosing mode $\ell$:
\[
U_{itr\ell}={v_{itr}}-\alpha \ln p_{tr\ell}-{c_{itr\ell}},
\]
where $v_{itr}$ is $i$'s expected utility at time $t$ to take route $r$ (e.g., their utility of traveling from Paris to Toulouse to attend a rugby match in Toulouse), $p_{tr\ell}$ is the price at time $t$ of mode $\ell$ for route $r$,  and $c_{itr}$ is $i$'s expected cost incurred when choosing mode $\ell$ that is not captured by price (e.g., their traveling time to Gare Montparnasse in Paris). The probability of purchasing an iDTGV ticket at time $t$ for route $r$ is then:
\begin{equation}
\begin{aligned}
s_{tr1}=&\Pr\left(U_{itr1}\geq  U_{itr\ell}, \forall \ell \ge 2\right)=\Pr\left(C_{itr\ell}\geq \frac{p_{tr1}^\alpha}{p_{tr\ell}^\alpha}C_{itr1}, \forall \ell \ge 2\right),\\
\end{aligned}
\end{equation}
where $C_{itr\ell}=\exp(c_{itr\ell})$. Suppose that $\{C_{itr\ell}/A_{tr\ell}\}_{\ell=1}^{\overline{L}}$ are independent and follow Pareto distributions with parameters $(1,\delta_{m\ell})$, respectively, so that $
\Pr\left(C_{itr\ell}>c \right)=\left(1\wedge \frac{A_{tr\ell}}{c} \right)^{\delta_{m\ell}}$). The parameter $A_{tr\ell}$ can be interpreted as the minimal cost of taking mode $\ell$ for route $r$ among travelers who search at $t$. Suppose also that $A_{rt\ell}/A_{rt1}=a_{r\ell}$ does not depend on $t$, i.e., the minimal cost of taking mode $\ell$ relative to iDTGV does not depend on the time of search (the cost of traveling to airport relative to train station, or the relative cost of taking another train departing at a different hour within $m$). Then, as long as $a_{r\ell}\geq p^\alpha_{tr\ell}/p^\alpha_{tr1}$ for all $\ell=2,...,\overline{L}$, we obtain:
\begin{equation}\label{eq:purchase_prob}
\begin{aligned}
s_{tr1}&=\mathbb{E}\left[\Pr\left(C_{itr\ell}\geq \frac{p_{tr1}^\alpha}{p_{tr\ell}^\alpha}C_{itr1}, \forall \ell \ge 2\bigg|\frac{C_{itr1}}{A_{tr1}}\right)\right]\\
&=\mathbb{E}\left[\Pr\left(\frac{C_{itr\ell}}{A_{tr\ell}}\geq a^{-1}_{r\ell}\frac{p_{tr1}^\alpha}{p_{tr\ell}^\alpha}\frac{C_{itr1}}{A_{tr1}}, \forall \ell \ge 2\bigg|\frac{C_{itr1}}{A_{tr1}}\right)\right]\\
&=\prod_{\ell=2}^{\overline{L}}(p_{tr\ell}^{\alpha \delta_{m\ell}} a_{r\ell}^{\delta_{m\ell}})\mathbb{E}\left[p_{tr1}^{-\alpha\sum_{\ell=2}^{\overline{L}} \delta_{m\ell}}\left(\frac{C_{itr1}}{A_{tr1}}\right)^{-\sum_{\ell=2}^{\overline{L}}{\delta_{m\ell}}} \right]\\
&=\frac{\delta_{m1}}{\sum_{\ell=1}^{\overline{L}}{\delta_{m\ell}}}\prod_{\ell=2}^{\overline{L}}\left(p_{tr\ell}^\alpha a_{r\ell}\right)^{\delta_{m\ell}}p_{tr1}^{-\alpha\sum_{\ell=2}^{\overline{L}} \delta_{m\ell}}.\\
\end{aligned}
\end{equation}
Then, given $(p_{tr\ell},a_{r\ell})_{\ell=2}^{\overline{L}}$, the intensity of purchases of iDTGV tickets at time $t$ with price $p_{tr1}$ is
\begin{equation}\label{eq:microfoundation}
I_{dm}(t,p)=-\kappa_{dm}\lambda_{mt}\frac{\partial s_{rt1}}{\partial p_{tr1}}=\kappa_{dm}\lambda_{mt}\alpha\frac{\delta_{m1}\sum_{\ell=2}^{\overline{L}}{\delta_{m\ell}}}{\sum_{\ell=1}^{\overline{L}}{\delta_{m\ell}}} \prod_{\ell=2}^{\overline{L}}\left({p_{tr\ell}^\alpha a_{r\ell}}\right)^{\delta_{m\ell}}p^{-1-\alpha\sum_{\ell=2}^{\overline{L}} \delta_{m\ell}}.
\end{equation}
To obtain our demand model from \eqref{eq:microfoundation}, we finally assume that $p_{tr\ell}=p_{tm\ell}b_{dm}$, a separability restriction in the pricing dynamics of mode $\ell$. Then, we obtain:
\begin{equation}\label{eq:microfoundation2}
	\begin{aligned}
		I_{dm}(t,p)&=\underbrace{\frac{\delta_{m1}\kappa_{dm}b_{dm}^{\alpha\sum_{\ell=2}^{\overline{L}}\delta_{m\ell}}\prod_{\ell=2}^{\overline{L}}a_{dm\ell}^{\delta_{m\ell}}}{\sum_{\ell=1}^{\overline{L}}\delta_{m\ell}}\int_{0}^1\lambda_{mu} \prod_{\ell=2}^{\overline{L}} p_{um\ell}^{\alpha\delta_{m\ell}}du}_{\xi_{dm}}\\
		&\times \underbrace{\frac{\lambda_{mt} \prod_{\ell=2}^{\overline{L}} p_{tm\ell}^{\alpha\delta_{m\ell}}}{\int_{0}^1\lambda_{mu} \prod_{\ell=2}^{\overline{L}} p_{um\ell}^{\alpha\delta_{m\ell}}du}}_{s_m(t)} \underbrace{\left(\alpha\sum_{\ell=2}^{\overline{L}}\delta_{m\ell}\right)}_{\eps}p^{-1-\alpha\sum_{\ell=2}^{\overline{L}} \delta_{m\ell}}.
	\end{aligned}
\end{equation}
The price-elasticity $\eps$ is related to both travelers' disutility of price ($\alpha$) and non-price costs for alternative modes of transportation, through the term $\sum_{\ell=2}^{\overline{L}}\delta_{m\ell}$. Note that Equation \eqref{eq:microfoundation2} allows $\eps$ to depend on $m$. The term $s_m(t)$  captures both overall demand through the term $\lambda_{mt}$ and competition effect, through the term $\prod_{\ell=2}^{\overline{L}} p_{tm\ell}^{\alpha\delta_{m\ell}}$.

\paragraph{Welfare.} Start with the traveler's probability of purchasing a ticket at iDTGV ($\ell=1$):
\[
\Pr\left(C_{itr\ell}\geq \frac{p_{tr1}^\alpha}{p_{tr\ell}^\alpha, }C_{itr1},\ \forall \ell \ge 2\right) = \Pr\left(\max_{\ell \ge 2}\left\{\frac{C^{1/\alpha}_{itr\ell}}{C^{1/\alpha}_{itr1}}p_{tr\ell} \right\}\geq {p_{tr1}}\right).
\]
In other words, given $p_{tr\ell}$, $\ell \ge 2$, traveler's willingness-to-pay (WTP) is $\max_{\ell \ge 2}\left\{\frac{C^{1/\alpha}_{itr\ell}}{C^{1/\alpha}_{itr1}}p_{tr\ell} \right\}$. From \eqref{eq:purchase_prob}, we know that the WTP follows a Pareto distribution and 
\[
\mathbb{E}\left[\max_{\ell \ge 2}\left\{\frac{C^{1/\alpha}_{itr\ell}}{C^{1/\alpha}_{itr1}}p_{tr\ell} \right\}\Big|\max_{\ell \ge 2}\left\{\frac{C^{1/\alpha}_{itr\ell}}{C^{1/\alpha}_{itr1}}p_{tr\ell} \right\}\geq p_{tr1} \right]= \frac{\alpha\sum_{\ell=2}^{\overline{L}}\delta_{m\ell}}{\underbrace{\alpha\sum_{\ell=2}^{\overline{L}}\delta_{m\ell}}_{\eps}-1} p_{tr1}.
\]
We now consider a case in which $\overline{L}=2$ and the only alternative is not traveling whose monetized indirect utility is $0$. Note that for any (potentially stochastic) pricing strategy at the iDTGV, $(p_{tr1})_{t\in[0,1]}$, the induced social welfare at time $t$ in case of a sale (i.e., the event that the WTP is no less than $p_{tr1}$) is $(\frac{\eps}{\eps-1}p_{tr1})_{t\in[0,1]}$, and $0$ otherwise. The flow of social welfare is then equal to the revenue flow scaled by $\frac{\eps}{\eps-1}$. As a result, the social welfare corresponding to $(p_{tr1})_{t\in[0,1]}$, which is an integral of the social welfare flow, is written as a scaled quantity of the corresponding revenue:
\begin{equation}\label{eq:social_welfare}
    SW = R\times \frac{\eps}{\eps-1},
\end{equation}
which resembles the welfare conclusion in \cite{mcafee2008dynamic} that the optimal revenue pricing strategy also maximizes social welfare and that the social welfare is the optimal revenue times $\frac{\eps}{\eps-1}$.

This win-win property depends crucially on the specification of constant elasticity. To see this, suppose that we have early and late arrivals with different elasticities, $\eps_{\text{early}}$ and $\eps_{\text{late}}$ respectively. Denote by $R_{\text{early}}$ and $R_{\text{late}}$ the revenues corresponding to the early and late elasticities, respectively. Given the total capacity $C=C_{\text{early}}+C_{\text{late}}$, we then have:
\[
SW(C)=\frac{\eps_{\text{early}}}{\eps_{\text{early}-1}}R_{\text{early}}+\frac{\eps_{\text{late}}}{\eps_{\text{late}}-1}R_{\text{late}}.
\]
When $\eps_{\text{early}}>\eps_{\text{late}}$, we would expect that the optimal allocation of capacity to late arrivals in terms of total social welfare should be more than the optimal allocation in terms of revenue. This is because $R_{\text{late}}$ in $SW(C)$ has a higher weight. Then, the pricing strategy that maximizes $SW(C)$ differs from the one that maximizes revenue.

The formula above also implies a way to compare social welfare in the observed scenario between the counterfactual ones. For instance, if we define the cutoff of early and late as fare class nine. Then, an approximate for the social welfare in the observed scenario would be 
\[
\frac{\eps_{\text{early}}}{\eps_{\text{early}-1}}\sum_{k\leq 9}R_{k}+\frac{\eps_{\text{late}}}{\eps_{\text{late}}-1}\sum_{k\geq 10}R_{k}.
\]
The analysis above also applies to consumer surplus. Under constant elasticity $\eps$, the consumer surplus (CS) of a price path is the corresponding revenue times $\frac{1}{\eps-1}$. With $(\eps_{\text{early}},\eps_{\text{late}})$, we have:
\[
CS(C)=\frac{1}{\eps_{\text{early}-1}}R_{\text{early}}+\frac{1}{\eps_{\text{late}}-1}R_{\text{late}}.
\]
An approximate for the consumer surplus in the observed scenario would be 
\[
\frac{1}{\eps_{\text{early}-1}}\sum_{k\leq 9}R_{k}+\frac{1}{\eps_{\text{late}}-1}\sum_{k\geq 10}R_{k}.
\]

\paragraph{Compensating variation in prices.} Combining \eqref{eq:microfoundation2} and revenue formula \eqref{eq:revenue_gamma_ave}, we can compute the compensating variation in prices of competing transportation modes, $p_{tm\ell}$, that offsets the loss due to the observed revenue management relative to counterfactual pricing strategies. Suppose that $p_{tm\ell}$ decreases to $(1-\Delta)p_{tm\ell}$  for all $\ell=2,...,\overline{L}$ and $t\in[0,1]$, i.e., an overall relative decrease in competitors' prices by $\Delta$. According to  \eqref{eq:microfoundation2}, $\xi_{dm}$ will decrease to $(1-\Delta)^{\alpha\sum_{\ell=2}^{\overline{L}}\delta_{m\ell}}\xi_{dm}=(1-\Delta)^{\eps}\xi_{dm}$. Then, because of  \eqref{eq:revenue_gamma_ave}, $R_{r}^I$ decreases by $1-[(1-\Delta)^{\eps}]^{1/\eps}=\Delta$. For instance,  consider Scenario s.3 in Table \ref{tab:supply_main} and its lower bound \EUR{13,100}. To offset the gain of the counterfactual scenario relative to observed scenario relative to this counterfactual one, competitors' prices need to decrease by $1-12.21/13.10=6.8\%$ on average.

\section{Estimation and inference: details}\label{sub:estimation_and_inference}

We estimate $\theta_0:=(\eps,\beta_0,\lambda_{a0},\lambda_{b0})$ as follows. Let $Y_{jkm}=1$ if seat $j$ in fare class $k$ for train $m$ is sold to $a$, $Y_{jkm}=0$ otherwise. By \eqref{eq:binomial}, we have
$$\Pr(Y_{jkm}=1|\xi_{am},\xi_{bm}) = \Lambda\left(\ln(\xi_{bm}/\xi_{am}) - \eps \ln(p_{bkm}/p_{akm})\right),$$
and the $(Y_{jkm})_{j=1,...,n_{km}}$ (with $n_{km}:=n_{akm}+n_{bkm}$) are independent. Thus, we can estimate $\eps$ and $\ln(\xi_{bm}/\xi_{am})$ by maximizing the likelihood of a logit model including train fixed effects. Because the number of sales for each train is large (usually above 250), the bias related to the estimation of these fixed effects is expected to be negligible. 

\begin{rmk}\label{rmk:aggregation_dest}
Note that we made in Assumption \ref{hyp:gamma}(i) the simplifying assumption that trains only had two destinations, an intermediate $a$ and a final one $b$. But recall from Table \ref{DES1} that most of them serve more than just two cities, so $a$ or $b$ actually correspond to more than one city. If so, we modify (i) by assuming that
\begin{equation}
\xi_{dm}=\left[\sum_{c\in d} \exp(X'_{cm}\beta_0)\right]g_0(W_m)\eta_{dm},
\label{eq:aggreg_cities}
\end{equation}
where $c$ is an index for cities belonging to either $a$ or $b$. For instance, in a train to C\^ote d'Azur, $c$ corresponds to Avignon for destination $a$ whereas $b$ includes Cannes, Saint-Rapha\"el and Nice, see again Table \ref{DES1}. Note, on the other hand, that all cities $c\in d$ are priced equally, so we do not need to take into account price variations between cities.
\end{rmk}

Second, under Assumption \ref{hyp:gamma} (with the equality in (i) replaced by \eqref{eq:aggreg_cities} to account for multiple cities in each $d\in\{a,b\}$),
$$\ln(\xi_{bm}/\xi_{am})=\ln\left[\frac{\sum_{c \in b}\exp(X_{cm}'\beta_0)}{\sum_{c \in a}\exp(X_{cm}'\beta_0)}\right] +\ln\left(\frac{\eta_{bm}}{\eta_{am}}\right), \quad \eta_{bm}/\eta_{am} \indep (X_{cm})_c.$$
Then, we estimate $\beta_0$ by nonlinear least squares, replacing $\ln(\xi_{bm}/\xi_{am})$ by its estimator. Finally, we estimate $(\lambda_{a0},\lambda_{b0})$ by maximum likelihood on the sample $(\widehat{\ln(\eta_{bm}/\eta_{am})})_m$, with
$$\widehat{\ln(\eta_{bm}/\eta_{am})} = \widehat{\ln(\xi_{bm}/\xi_{am})}-\ln\left[\frac{\sum_{c \in b}\exp(X_{cm}'\widehat{\beta})}{\sum_{c \in a}\exp(X_{cm}'\widehat{\beta})}\right].$$

\begin{rmk}
	We could directly estimate $\theta_0$ by maximum likelihood, as under Assumptions \ref{hyp:cons_demand}-\ref{hyp:gamma}, the distribution of $(Y_{jkm})_{j=1,...,n_{km}, k=1,...,K}$ is fully parametric. We do not follow this path for numerical reasons. In fact, the corresponding estimator is much more complicated to compute, something turning out to be important when considering inference based on subsampling.
\end{rmk}

Next, we estimate the lower and upper bounds on $g_0(W_m)$ by the empirical counterparts of \eqref{eq:lower_bound} and \eqref{eq:upper_bound}, where the conditional expectations $\E(.|W_m)$ are replaced by empirical means (as $W_m$ is discrete here).
We then estimate bounds on $R_r^I$ by the empirical counterpart of \eqref{eq:revenue_gamma_ave}.

\medskip
Turning to inference, we consider below statistical tests of whether $R_r^I$ improves upon the observed revenue  $R_{\text{obs}}$. Specifically, we test for
\begin{equation}\label{eq:test}
	\text{H}_0:\ \Delta=R_r^I- R_{\text{obs}}\leq 0 \text{ vs } \text{H}_1:\ \Delta>0.
\end{equation}
To this end, we consider critical regions of the form $\{\sqrt{N}\hat{\Delta}_l>\delta \}$ with $N$ the number of trains, $\delta$ specified below,  $\hat{\Delta}_l:=\hat{R}_{\text{lower}}-\hat{R}_{\text{obs}}$ and $\hat{R}_{\text{lower}}$ and $\hat{R}_{\text{obs}}$  estimators of the lower bound of $R_{r}^I$ and $R_{\text{obs}}$, respectively. By definition, $\Delta_l\leq \Delta$. Then, under H$_0$,
\begin{equation}\label{eq:inference}
	\begin{aligned}
		\Pr\left(\hat{\Delta}_l>\delta \right)&=	\Pr\left(\sqrt{N}(\hat{\Delta}_l-\Delta_l)>\delta -\sqrt{N}\Delta_l\right)\\
		&\leq\Pr\left(\sqrt{N}(\hat{\Delta}_l-\Delta_l)>\delta -\sqrt{N}\Delta\right) \\
		&\leq\Pr\left(\sqrt{N}(\hat{\Delta}_l-\Delta_l)>\delta\right).\\
	\end{aligned}
\end{equation}
Then, by using for $\delta$ a consistent estimator of the quantile of order $1-\alpha$ of the asymptotic distribution of $\sqrt{N}(\hat{\Delta}_l-\Delta_l)$ ($q_{l,1-\alpha}$, say), we ensure control of the asymptotic level of our test. Because of the maximum in \eqref{eq:lower_bound}, the asymptotic distribution of $\sqrt{N}(\hat{\Delta}_l-\Delta_l)$ may not be Gaussian and the bootstrap may be invalid. We thus rely instead on subsampling \citep{politis1999subsampling} and estimate $q_{l,1-\alpha}$ by the $(1-\alpha)$-quantile of the subsampling distribution of $\sqrt{N}(\hat{\Delta}_l-\Delta_l)$. We also use \eqref{eq:inference} and this subsampling distribution to compute an upper bound of the p-value corresponding to \eqref{eq:test}, which we will report in the next section. 

\paragraph{A computational issue.} In all the counterfactual scenarios, we consider a separate pricing strategy for destinations $a$ and $b$, more flexible than the actual practice. On the other hand, without further restrictions, the state space is large: the optimal strategy at any time $t$ depends on the remaining seats for both destinations, a typical numerical challenge in multiple-resource quantity-based revenue management \citep{van2005introduction}. To reduce the state space, we approximate the original multiple-resource problem by multiple independent single-resource ones.\footnote{Such techniques are frequently used in quantity-based revenue management practices, especially in the presence of multiple resources, i.e., network revenue management. See Section 2 of \cite{van2005introduction}.} Concretely, we fix ex ante the total number of seats available for stops $a$ ($C_{am}$, say) and thus to $b$ ($C_{bm}=C_m-C_{am}$, with $C_m$ the total number of seats in train $m$). Then, depending on the scenario we consider, we either consider the optimal pre-allocation $C_{am}$, or fix it so that $C_{am}$ matches the observed average sales for $a$. In either case, fixing $C_{am}$ allows us to solve the optimization problem separately for each destination (given the independence of $\eta_{am}$ and $\eta_{bm}$ imposed in Assumption \ref{hyp:gamma}(i)) rather than jointly, greatly reducing the computational burden of the optimization problem. Compared to the fully optimal counterfactual revenues without pre-allocation,  our results may thus be seen as lower bounds. But this only reinforces some of our conclusions. 

\medskip
To get a sense on the quantitative effect of these pre-allocations, we simulate counterfactual revenues under unconstrained uniform pricing without pre-allocating capacities among intermediate and final destinations. The corresponding formulas are in Appendices \ref{ssub:proofs} and \ref{proof_thm_limited_rev}, see the sections ``uniform pricing'' therein. In the complete information setup, we obtain a set estimate of $[13.45, 14.68]$, corresponding to an increase in between 1.4\% and 1.8\% compared to Scenario u.4. In the incomplete information setup, we obtain a higher gain of around 6\%, with a set estimate of $[11.96, 14.68]$. This 6\% might be the upper bound on possible gains from not imposing any pre-allocation, as one could expect that the effects of pre-allocation can be more easily mitigated with more flexible pricing strategies. This suggests that the other counterfactual revenues would not change much either without pre-allocation.

\section{Additional results on the application}\label{app:additional_tab_fig}

\subsection{Results with time-varying elasticities}
\label{app:res_TV_elast}

\begin{table}[H]
	\caption{Binomial model of demand with $(\eps_{\text{early}},\eps_{\text{late}})$}\label{tab:demand_two_elas}
~\vspace{-0.5cm}
\begin{center}		
	\begin{threeparttable}
		\begin{tabular}{p{6.5cm}ccc}
			\toprule
			Specification&S=9&S=10&S=11\\ 
            \midrule
			Price elasticity &  &&\\                  
			$\eps_{\text{early}}$&$4.76$    &${4.55}$&$4.53$\\
			$\eps_{\text{late}}$&$3.01$&${3.17}$&$2.88$\\
			\midrule          
			Destination effects &&&\\
			\quad    Population (in M. inhabitants) &$2.11$&${2.14}$&$2.15$\\
			\quad    Regional capital&$0.25$&${0.24}$&$0.24  $ \\
			\quad    Travel time by train (in hours)&$-1.90$&${-1.94}$ &$ -1.93   $\\
			\quad     Travel time by train, squared&$0.33 $&${0.33}$&$0.33$\\
			\hline
			Gamma distributions&&&\\
			\quad $\lambda_{a0}$ (intermediate)&$3.63$&${3.63}$&$3.63    $\\	
			\quad $\lambda_{b0}$ (final)&$2.62$&${2.62}$&$2.62$\\			
			\hline
			Control for $X_d\times W_m$&Yes&Yes&Yes\\
			$R^2$ of the reg. of $\ln(\xi_{bm}/\xi_{am})$&$0.500$&$0.501$&$0.501$\\
			\bottomrule
		\end{tabular}
		\begin{tablenotes}
			\footnotesize
			\item \emph{Notes:} The total number of trains is 2,909 and the total number of observations (fare classes $\times$ trains) is 21,988.
		\end{tablenotes}
	\end{threeparttable}
\end{center} 
\end{table}
\begin{table}[H]
	\begin{center}
	\caption{Counterfactual revenues and social welfare with $(\eps_{\text{early}},\eps_{\text{late}})$: lower bounds}\label{tab:counter_rev_two_elas}
	\begin{threeparttable}
		\begin{tabular}{lcccc}
			\toprule
			Scenario &  \multicolumn{2}{c}{Estimate (in K\EUR)}  & \multicolumn{2}{c}{Table \ref{tab:counter_sw_cs}}\\
			&Revenue &SW&Revenue&SW\\
			\hline
			\textbf{Observed}&$12.21$&$15.88$&$12.21$&$16.23$\\
			\textbf{Uniform pricing strategy} &&&&\\
			u.2 Incomplete information, unconstrained prices 
            & $[11.03,11.54]$&$14.13$&$11.29$&$15.00$\\
			u.4 Complete information, unconstrained prices
            & 
            $[12.81,13.39]$ &$16.42$&$13.23$&$17.58$\\
			\textbf{Stopping-time pricing strategy} &&&&\\
			s.3 Incomplete information &${[12.84,13.96]}$
			&$16.45$&$13.44$&$17.86$\\
			s.6 Complete information &$[12.85,14.00]$ 
			&$16.47$&$13.48$&$17.91$\\
			\textbf{``Full'' dynamic pricing strategy} &&&&\\
			f.1 Incomplete information &$[12.89,13.98]$
			&$16.50$&$13.47$&$17.90$\\
			f.2 Complete information &$[12.91,14.01]$
			&$16.55$&$13.50$&$17.94$\\
			\hline
		\end{tabular}
	\begin{tablenotes}\footnotesize
		\item \emph{Notes}: The row ``Observed'' corresponds to the point estimates of revenues and social welfare  in the observed scenario. See Appendix \ref{app:microfoundation} for their derivations. 
		Other rows report the lower bounds estimates for counterfactual revenues (second and forth columns) and social welfare (third and fifth columns). 
		For the second and third columns, we set $S=10$ and estimate the demand with $(\eps_{\text{early}},\eps_{\text{late}})$ in the second column of Table \ref{tab:demand_two_elas}.  The second column reports the set estimate of revenue lower bounds. The method of deriving the lower bound estimates in these two columns is detailed in Section  \ref{app:constant_elas_details} of the Online Appendix. 
		\end{tablenotes}
	\end{threeparttable}
\end{center}
\end{table}

\subsection{Robustness to alternative assumptions and specifications}
\label{app:alt_specif}

\begin{table}[H]
		\caption{Counterfactual revenues with log-normally distributed $\eta_{dT}$}\label{tab:supply2_benforroni}
		\vspace{-0.5cm}
		
	\setlength{\tabcolsep}{5pt}
		\begin{center}
			\small{
			\begin{threeparttable}
				\begin{tabular}{lcc}
					\toprule
					Scenario& \multicolumn{2}{c}{Point or set estimate (in K\EUR)} \\
					\midrule
					{\textbf{Observed pricing strategy}} &\multicolumn{2}{c}{$12.21$}\\
					\midrule
					\textbf{Uniform pricing strategy} &Log-normal specification&Baseline\\
					u.2\ Incomplete inf., uncons. prices &$[10.03,12.30]$&$[    11.29,   12.31]$\\
					u.4\ Complete inf., uncons. prices &$[12.67,    15.36]$&$[13.23,14.42]$\\
					\midrule
					\textbf{Complete information}&&\\
					s.6\ Stopping-time pricing strategy &$[12.90   ,15.66 ]$&$[13.48,14.70]$\\
					f.2\ ``Full'' dynamic pricing strategy&$[12.92   ,15.68]$&$[13.50, 14.72]$\\
					\bottomrule
				\end{tabular}
				\begin{tablenotes}
					\footnotesize
					\item \emph{Notes:} The baseline results correspond to those in Table \ref{tab:supply_main}. See the notes of that table for more details.
				\end{tablenotes}
			\end{threeparttable}
		}
		\end{center}
\end{table}
\vspace{-0.5cm}
\begin{table}[H]
			\caption{Counterfactual revenues based on Column II in Table \ref{tab:binomial}}\label{tab:rev_column2}
			\vspace{-0.5cm}
			
\small{
	\begin{center}
	\begin{threeparttable}
		\begin{tabular}{lcc}
			\toprule
			Scenario& \multicolumn{2}{c}{Point or set estimate (in K\EUR)} \\ \midrule
{\textbf{Observed pricing strategy}} &\multicolumn{2}{c}{$12.21$}\\
\midrule
			\textbf{Uniform pricing strategy} &Specification II&Baseline\\
			u.1\ Incomplete information, constrained prices &$[11.51,12.21]$&$[11.15,12.21]$ \\
			u.2\ Incomplete information, unconstrained prices &$[11.77    ,12.37]$&$[11.29,   12.31]$\\
			u.3\ Complete information, constrained prices &$[12.85,13.73]$&$[12.56,13.86]$\\
			u.4\ Complete information, unconstrained prices &$[13.63,  14.35]$&$[13.23,14.42]$\\
			\midrule
			\textbf{Stopping-time pricing strategy} &&\\
			s.3\ Incomplete information  &$[13.83, 14.57]$&$[13.44, 14.66]$ \\
			s.6\ Complete information &$[13.87, 14.60]$&$[13.48,14.70]$ \\
			\midrule
			\textbf{``Full'' dynamic pricing strategy} & &\\
			f.1\ Incomplete information  & $[13.86,  14.59]$&$[13.47,    14.68]$\\
			f.2\ Complete information &$[13.89, 14.62]$&$[13.50, 14.72]$\\
			\bottomrule
	\end{tabular}
		\begin{tablenotes}
			\footnotesize
			\item \emph{Notes: } The baseline results are those of Table \ref{tab:supply_main}. See the notes of that table for more details.
		\end{tablenotes}
	\end{threeparttable}
\end{center}
}
\end{table}

\section{Proofs}

\subsection{ Theorem \ref{thm:ident_eps}} 
\label{sub:proof_of_theorem_ref_thm_ident_eps}

\subsubsection{Main proof}

First, note that because the realization of $\tau_k$ is determined by the Poisson process before $\tau_k$ and is independent of $V_{dm}([\tau_k,\tau_{k+1}),[p_{dkm},\infty))$ for $d\in\{a,b\}$, it suffices to show \eqref{eq:binomial} if $\tau_k$ is replaced by any fixed number that we suppose equal to 0 without loss of generality. To ease the exposition, we often omit the index $m$ hereafter and define $\mu_d=\xi_d p_{dk}^{-\eps}$ and $\rho=\mu_a/(\mu_a+\mu_b)$. We also introduce $D_{d,\tau_n}=V_{dm}([0,\tau_n),[p_{dkm},\infty))$ for $d\in\{a,b\}$, $D_{\tau_n}=D_{a,\tau_n}+D_{b,\tau_n}$ and $\tau_n=\inf\{t>0:D_t\geq n\}\wedge1$. We will show that for all $n\geq 1$,
\begin{equation}\label{eq:binomial_proof}
D_{a,\tau_n}|D_{\tau_n}, s_m(\cdot), \xi_{a},\xi_b \sim \text{Binomial}\left(D_{\tau_n},\rho\right).
\end{equation}
Given the previous discussion and because the right-hand side of \eqref{eq:binomial_proof} does not depend on $s_m(\cdot)$, \eqref{eq:binomial} will follow from \eqref{eq:binomial_proof}.

\medskip
To prove \eqref{eq:binomial_proof}, we introduce,  for any $n\geq 1$, the hitting times $\sigma_n=\inf\{t\in [0,1]: D_t\geq n\}$, with $\sigma_n =2$ if $D_1 < n$. Let us also fix $\underline{t}\in (0,1)$ and let us
partition the interval $I=[\underline{t},1]$ into $m$ intervals $I_1,...,I_m$ of equal length $\Delta t=(1-\underline{t})/m$. Finally, for all  $c\leq n$, let
\begin{equation}\label{cell_prob}
q_{c,n;k}=\Pr[D_{a,\sigma_n}=c | D_{\sigma_n}=n, \sigma_n\in I_k].
\end{equation}
By Lemma \ref{lem:approx_p}, there exists $(c_l,c_r)$, independent of $k$ and $m$, such that for all $k=1,...,m$,
\begin{equation*}
	-c_{l}(1+n)\Delta t\leq q_{c,n;k}- \binom{n}{c} \rho^c(1-\rho)^{n-c}\leq c_{r}\Delta t.
\end{equation*}
Moreover, we have
\begin{equation*}
\begin{aligned}
\Pr[D_{a,\sigma_n}=c | D_{\sigma_n}=n, \sigma_n\in I]&=\frac{\sum_{k=1}^m\Pr[D_{a,\sigma_n}=c, D_{\sigma_n}=n, \sigma_n\in I_k]}{\sum_{k=1}^m\Pr[D_{\sigma_n}=n, \sigma_n\in I_k]}\\
&\in\left[\min_{k=1,...,m}q_{c,n;k}, \max_{k=1,...,m}q_{c,n;k}\right].
\end{aligned}
\end{equation*}
Consequently,
\begin{equation*}
-c_{l}(1+n)\Delta t\leq \Pr[D_{a,\sigma_n}=c| D_{\sigma_n}=n, \sigma_n\in I] - \binom{n}{c}  \rho^c(1-\rho)^{n-c}\leq c_{r}\Delta t.	
\end{equation*}
By letting $m\to\infty$ and then let $\underline{t}\to 0$, we obtain
\begin{equation}\label{original_result}
\Pr[D_{a,\sigma_n}=c | D_{\sigma_n}=n, \sigma_n\leq 1]=\binom{n}{c} \rho^c(1-\rho)^{n-c}.
\end{equation}
Now, because $D_{\tau_n}=n$ if and only if $\sigma_n\leq1$, we obtain \eqref{eq:binomial_proof} in this case. Further, because $D_{\tau_n}=n'<n$ if and only if $D_1=n'$ and $\sigma_n=2$, we have
\begin{equation*}
\Pr[D_{a,\tau_n}=c|D_{\tau_n}=n']=\Pr[D_{a,1}=c|D_1=n',\sigma_n=2]=\Pr[D_{a,1}=c|D_1=n']=\binom{n'}{c} \rho^c(1-\rho)^{n'-c}.
\end{equation*}
Thus, \eqref{eq:binomial_proof} also holds when $D_{\tau_n}=n'$, $n'<n$. The result follows.

\subsubsection{A key lemma} 
\label{ssub:key_lemma}

The proof crucially relies on the following lemma, which we prove below. 

\begin{lem}\label{lem:approx_p}
Suppose that Assumption \ref{hyp:cons_demand} holds. Then, there exists $c_l$ and $c_r$, independent of $k$ and $m$, such that for all $k=1,...,m$,
	\begin{equation}
		\label{eq:lemma_for_binom}
	-c_{l}(1+n)\Delta t\leq q_{c,n;k}-\binom{n}{c} \rho^c(1-\rho)^{n-c}\leq c_{r}\Delta t.
	\end{equation}
\end{lem}

\textbf{Proof:} First, observe that $\{\sigma_n\in I_k\}=\{D_{\underline{t}+(k-1)\Delta t}<n,\  D_{\underline{t}+k\Delta t}\geq n\}$. Then
\begin{align}
& \Pr[D_{a,\sigma_n}=c, D_{\sigma_n}=n, \sigma_n\in I_k] \nonumber \\
=&\Pr[D_{a,\sigma_n}=c, D_{\sigma_n}=n, D_{\underline{t}+(k-1)\Delta t}<n,\  D_{\underline{t}+k\Delta t}\geq n] \nonumber \\
=&\Pr[D_{a,\sigma_n}=c, D_{\sigma_n}=n, D_{\underline{t}+(k-1)\Delta t}=n-1,\  D_{\underline{t}+k\Delta t}\geq n] \nonumber \\
&+\Pr[D_{a,\sigma_n}=c, D_{\sigma_n}=n, D_{\underline{t}+(k-1)\Delta t}<n-1,\  D_{\underline{t}+k\Delta t}\geq n] \label{eq:decomp1_1em}
\end{align}
We first show that the second term in \eqref{eq:decomp1_1em} is negligible, as being of order $(\Delta t)^2$. Simple algebra shows that if $U\sim\mathcal{P}(\mu)$, then $\Pr(U\geq 2)\leq \mu^2$. Hence,
\begin{align*}
&\Pr[D_{a,\sigma_n}=c, D_{\sigma_n}=n, D_{\underline{t}+(k-1)\Delta t}<n-1,\  D_{\underline{t}+k\Delta t}\geq n]\\
\leq &\Pr[D_{\underline{t}+k\Delta t}-D_{\underline{t}+(k-1)\Delta t}\geq2] \\
\leq & \left[(\mu_a+\mu_b)\int_{\underline{t}+(k-1)\Delta t}^{\underline{t}+k\Delta t}s(t)dt\right]^{2} \\
\leq & \left[(\mu_a+\mu_b)\bar{s}\Delta t\right]^2,
\end{align*}
where $\overline{s}=\sup_{t\in[0,1]} s(t)$. Now, the first term in \eqref{eq:decomp1_1em} satisfies:
\begin{align*}
&\Pr[D_{a,\sigma_n}=c, D_{\sigma_n}=n, D_{\underline{t}+(k-1)\Delta t}=n-1,\  D_{\underline{t}+k\Delta t}\geq n]\\
=&\Pr[D_{a,\sigma_n}=n, D_{\sigma_n}=n, D_{\underline{t}+(k-1)\Delta t}=n-1,\  D_{\underline{t}+k\Delta t}=n]\\
+&\Pr[D_{a,\sigma_n}=c, D_{\sigma_n}=n, D_{\underline{t}+(k-1)\Delta t}=n-1,\  D_{\underline{t}+k\Delta t}>n],
\end{align*}
where the second term can be similarly controlled as above:
\begin{align*}
\Pr[D_{a,\sigma_n}=c, D_{\sigma_n}=n, D_{\underline{t}+(k-1)\Delta t}=n-1,\  D_{\underline{t}+k\Delta t}>n]&\leq\Pr[D_{\underline{t}+k\Delta t}-D_{\underline{t}+(k-1)\Delta t}\geq2]\\
&\leq \left[(\mu_a+\mu_b)\bar{s}\Delta t\right]^2.
\end{align*}
As a consequence,
\begin{align}
&\Pr[D_{a,\sigma_n}=c, D_{\sigma_n}=n, D_{\underline{t}+(k-1)\Delta t}=n-1,\  D_{\underline{t}+k\Delta t}=n]\nonumber \\
<& \Pr[D_{a,\sigma_n}=c, D_{\sigma_n}=n, \sigma_n\in I_k]\nonumber \\
\leq & \Pr[D_{a,\sigma_n}=c, D_{\sigma_n}=n, D_{\underline{t}+(k-1)\Delta t}=n-1,\  D_{\underline{t}+k\Delta t}=n]\nonumber \\
&+2[(\mu_a+\mu_b)\bar{s}]^{2}(\Delta t)^2. \label{jump2_2}
\end{align}
Now, we have
\begin{align}
& \Pr[D_{a,\sigma_n}=c, D_{\sigma_n}=n, D_{\underline{t}+(k-1)\Delta t}=n-1,\  D_{\underline{t}+k\Delta t}=n] \nonumber \\
= & \Pr[D_{a,\sigma_n}=c,D_{\underline{t}+k\Delta t}=n,\ D_{\underline{t}+(k-1)\Delta t}=n-1] \nonumber\\
&\Pr[D_{a,\sigma_n}=c, \  D_{\underline{t}+k\Delta t}=n,\ D_{\underline{t}+(k-1)\Delta t}=n-1]\nonumber \\
=&\Pr[D_{a,\sigma_n}=c, \  D_{\underline{t}+k\Delta t}=n,\ D_{\underline{t}+(k-1)\Delta t}=n-1, D_{a,\underline{t}+(k-1)\Delta t}=c-1]\nonumber \\
&+\Pr[D_{a,\sigma_n}=c, \  D_{\underline{t}+k\Delta t}=n,\ D_{\underline{t}+(k-1)\Delta t}=n-1, D_{a,\underline{t}+(k-1)\Delta t}=c]\nonumber \\
=& \Pr[D_{a,\underline{t}+k\Delta t}=c, \  D_{\underline{t}+k\Delta t}=n,\ D_{\underline{t}+(k-1)\Delta t}=n-1, D_{a,\underline{t}+(k-1)\Delta t}=c-1]\nonumber \\
&+\Pr[D_{a,\underline{t}+k\Delta t}=c, \  D_{\underline{t}+k\Delta t}=n,\ D_{\underline{t}+(k-1)\Delta t}=n-1, D_{a,\underline{t}+(k-1)\Delta t}=c].
\label{second_decomposition}
\end{align}
Now, by independence between $(D_{a,t})_{t\geq 0}$ and $(D_{b,t})_{t\geq 0}$, and independence between $D_{d,t+s}-D_{d,t}$ and $D_{d,t}$ for all $s>0$ and $d\in\{a,b\}$,
\begin{align*}
&\Pr[D_{a,\underline{t}+k\Delta t}=c, \  D_{\underline{t}+k\Delta t}=n,\ D_{\underline{t}+(k-1)\Delta t}=n-1, D_{a,\underline{t}+(k-1)\Delta t}=c-1]\\
= &  \Pr[D_{a,\underline{t}+(k-1)\Delta t}=c-1]\Pr[D_{a,\underline{t}+k\Delta t}=c|D_{a,\underline{t}+(k-1)\Delta t}=c-1] \\
& \times \Pr[D_{b,\underline{t}+(k-1)\Delta t}=n-c]\Pr[D_{b,\underline{t}+k\Delta t}=n-c|D_{b,\underline{t}+(k-1)\Delta t}=n-c] \\
= & \frac{\mu_a^c\mu_b^{n-c}\left(\int_0^{\underline{t}+(k-1)\Delta t}s(t)dt\right)^{n-1}}{(n-c)!(c-1)!}\exp\left\{-(\mu_a+\mu_b)\int_0^{\underline{t}+k\Delta t}s(t)dt\right\}\int_{\underline{t}+(k-1)\Delta t}^{\underline{t}+k\Delta t}s(t)dt.
\end{align*}
Similarly,
\begin{align*}
	& \Pr[D_{a,\underline{t}+k\Delta t}=c, \  D_{\underline{t}+k\Delta t}=n,\ D_{\underline{t}+(k-1)\Delta t}=n-1, D_{a,\underline{t}+(k-1)\Delta t}=c] \\
= & \frac{\mu_a^c\mu_b^{n-c}\left(\int_0^{\underline{t}+(k-1)\Delta t}s(t)dt\right)^{n-1}}{(n-c-1)!c!}\exp\left\{-(\mu_a+\mu_b)\int_0^{\underline{t}+k\Delta t}s(t)dt\right\}\int_{\underline{t}+(k-1)\Delta t}^{\underline{t}+k\Delta t}s(t)dt.
\end{align*}
By plugging the last two equalities  into \eqref{second_decomposition}, we obtain
\begin{equation*}
\begin{aligned}
&\Pr[D_{a,\sigma_n}=c, \  D_{\underline{t}+k\Delta t}=n,\ D_{\underline{t}+(k-1)\Delta t}=n-1]\\
=&\frac{n\mu_a^c\mu_b^{n-c}\left(\int_0^{\underline{t}+(k-1)\Delta t}s(t)dt\right)^{n-1}}{(n-c)!c!}\exp\left\{-(\mu_a+\mu_b)\int_0^{\underline{t}+k\Delta t}s(t)dt\right\}(\mu_a+\mu_b)\int_{\underline{t}+(k-1)\Delta t}^{\underline{t}+k\Delta t}s(t)dt.\\
\end{aligned}
\end{equation*}
Inequality \eqref{jump2_2} then becomes
\begin{align}
&\frac{n\mu_a^c\mu_b^{n-c}(\int_0^{\underline{t}+(k-1)\Delta t}s(t)dt)^{n-1}}{(n-c)!c!}\exp\{-(\mu_a+\mu_b)\int_0^{\underline{t}+k\Delta t}s(t)dt\}(\mu_a+\mu_b)\int_{\underline{t}+(k-1)\Delta t}^{\underline{t}+k\Delta t}s(t)dt\nonumber \\
< & P(D_{a,\sigma_n}=c, D_{\sigma_n}=n, \sigma_n\in I_k)\nonumber \\
\leq & \frac{n\mu_a^c\mu_b^{n-c}(\int_0^{\underline{t}+(k-1)\Delta t}s(t)dt)^{n-1}}{(n-c)!c!}\exp\{-(\mu_a+\mu_b)\int_0^{\underline{t}+k\Delta t}s(t)dt\}(\mu_a+\mu_b)\int_{\underline{t}+(k-1)\Delta t}^{\underline{t}+k\Delta t}s(t)dt\nonumber \\
&+2[(\mu_a+\mu_b)\bar{s}]^2(\Delta t)^2.
\label{two_side_control}
\end{align}
By summing (\ref{two_side_control}) over $c=0,1,...,n$, we obtain
\begin{align}
&\frac{n(\mu_a+\mu_b)^n(\int_0^{\underline{t}+(k-1)\Delta t}s(t)dt)^{n-1}}{n!}\exp\{-(\mu_a+\mu_b)\int_0^{\underline{t}+k\Delta t}s(t)dt\}(\mu_a+\mu_b)\int_{\underline{t}+(k-1)\Delta t}^{\underline{t}+k\Delta t}s(t)dt\nonumber \\
<&P(D_{\sigma_n}=n, \sigma_n\in I_k)\nonumber \\
\leq & \frac{n(\mu_a+\mu_b)^n(\int_0^{\underline{t}+(k-1)\Delta t}s(t)dt)^{n-1}}{n!}\exp\{-(\mu_a+\mu_b)\int_0^{\underline{t}+k\Delta t}s(t)dt\}(\mu_a+\mu_b)\int_{\underline{t}+(k-1)\Delta t}^{\underline{t}+k\Delta t}s(t)dt.\nonumber \\
&+2(n+1)[(\mu_a+\mu_b)\bar{s}]^2(\Delta t)^2.
\label{two_side_control_2}
\end{align}
By combining (\ref{two_side_control}), (\ref{two_side_control_2}), and (\ref{cell_prob}), we obtain the following inequalities:
\begin{equation*}
-c_{l,k}(1+n)(\Delta t)^2\leq q_{c,n;k}-\binom{n}{c}\rho^c(1-\rho)^{n-c}\leq c_{r,k}(\Delta t)^2,
\end{equation*}
where
\begin{align*}
c_{r,k} & = \frac{2(n+1)[(\mu_a+\mu_b)\bar{s}]^2}{\frac{n(\mu_a+\mu_b)^n(\int_0^{\underline{t}+(k-1)\Delta t}s(t)dt)^{n-1}}{n!}\exp\{-(\mu_a+\mu_b)\int_0^{\underline{t}+k\Delta t}s(t)dt\}(\mu_a+\mu_b)\int_{\underline{t}+(k-1)\Delta t}^{\underline{t}+k\Delta t}s(t)dt},\\
c_{l,k} &= c_{r,k}\binom{n}{c}\rho^c(1-\rho)^{n-c}.
\end{align*}
Finally, note that $c_{r,k}\Delta t \leq c_r$ where
$$c_r = \frac{2(n+1)[(\mu_a+\mu_b)\bar{s}]^2\exp\{(\mu_a+\mu_b) \int_0^1 s(t)dt\}}{\frac{n(\mu_a+\mu_b)^n(\int_0^{\underline{t}}s(t)dt)^{n-1}}{n!} (\mu_a+\mu_b)\inf_{t\in I}s(t)}.$$
Moreover, $c_r$ does not depend on $k$ and $m$. Finally, defining $c_l=c_{r}\binom{n}{c} \rho^c(1-\rho)^{n-c}$, $c_l$ does not depend on $k$ and $m$ either, and  \eqref{eq:lemma_for_binom} holds for all $k=1,...,m$.


\subsection{ Theorem \ref{thm:ident_xi}} 
\label{sub:proof_of_theorem_ref_thm_ident_xi}

\subsubsection{An additional assumption}

Hereafter, $d_X$ denotes the dimension of $X_{dm}$, $\Supp(A)$ denotes the support of any random variable $A$,  and $\Delta X_m:=X_{bm}-X_{am}$. We first state Assumption \ref{hyp:supp_X} appearing in the statement of the theorem.

\begin{hyp}
  There exists a component $X_{djm}$ of $X_{dm}$ such that $\Supp(\Delta X_{jm}|\Delta X_{-jm})=\R$, where $X_{d-jm}$ is the vector stacking all the components of $X_{dm}$ except $X_{djm}$. Also, there exists $x^1_{-j}, x^2_{-j},...,x^{d_X}_{-j} \in \Supp(X_{b-jm}-X_{a-jm})$ such that the matrix
$$M:=\begin{pmatrix}
	x^1_{-j}{}' - x^2_{-j}{}' \\
	\vdots \\
	x^1_{-j}{}' -  x^{d_X}_{-j}{}'	
\end{pmatrix} $$
is nonsingular. Finally, $\sup \Supp\left(\sum_{k=1}^K n_{akm}+n_{bkm}\right)\geq K+1$.
  \label{hyp:supp_X}
\end{hyp}
The most restrictive condition is $\Supp(\Delta X_{jm}|\Delta X_{-jm})=\R$. However, the proof below reveals that identification is not achieved at infinity. The large support condition simply ensures that we can produce the compensating variations used in the proof. Also, the nonparametric identification of the distribution of $\eta_{bm}/\eta_{am}$ can be obtained without large support, using the analyticity of the density of the logistic distribution. The last condition ($\sup \Supp\left(\sum_{k=1}^K n_{akm}+n_{bkm}\right)\geq K+1$) easily holds in our application. It shows that our result does not require a large number of units of the perishable good (namely, tickets for a given train in our application) to be applicable.

\subsubsection{Proof}

Consider two fare classes $k,k'$ such that  $p_{ak}/p_{bk}\ne p_{ak'}/p_{bk'}$ (hereafter, we implicitly reason conditional on prices). Fix $x\in\R$ and let
\begin{equation}
\widetilde{x}=x - \frac{\eps}{\beta_{0j}} \ln\left(\frac{p_{ak}}{p_{bk}}\frac{p_{bk'}}{p_{ak'}}\right).
	\label{eq:def_xpr}
\end{equation}
Then, $x\beta_{0j}-\eps \ln(p_{ak}/p_{bk}) = \widetilde{x}\beta_{0j}-\eps \ln(p_{ak'}/p_{bk'})$. In turn, given the index structure,
\begin{align*}
& \Pr(n_{bkm}=n_b|n_{akm}+n_{bkm}=n, \Delta X_{-jm}, \Delta X_{jm}=x) \\
= & \Pr(n_{bk'm}=n_b|n_{ak'm}+n_{bk'm}=n, \Delta X_{-jm}, \Delta X_{jm}=\widetilde{x}).	
\end{align*}
Conversely, there is a single solution $\widetilde{x}$ to this equation, given by \eqref{eq:def_xpr}. Hence, $\widetilde{x}$ and thus $\beta_{0j}$ are identified. Similarly, for any two $x_{-j}\ne \widetilde{x}_{-j}$ in the support of $\Delta X_{-jm}$,
\begin{align*}
& \Pr(n_{akm}=n_b|n_{akm}+n_{bkm}=n, \Delta X_{-jm}=x_{-j}, \Delta X_{jm}=x) \\
= & \Pr(n_{akm}=n_b|n_{akm}+n_{bkm}=n, \Delta X_{-jm}=\widetilde{x}_{-j}, \Delta X_{jm}=\widetilde{x}).	
\end{align*}
if and only if $x\beta_{0j}+x_{-j}'\beta_{0-j}=\widetilde{x}\beta_{0j}+\widetilde{x}_{-j}'\beta_{0-j}$. By considering $x^1_{-j},...,x^{d_X}_{-j}$ as in Assumption \ref{hyp:supp_X}, we obtain $M\beta_{0j}=y$ for some identified vector $y$. Since $M$ is nonsingular, $\beta_{0-j}$ is identified.

\medskip
We now show the nonparametric identification of the cdf $F$ of $\ln(\eta_{am}/\eta_{bm})$. Since $\sup \Supp\left(\sum_{k=1}^K n_{akm}+n_{bkm}\right)\geq K+1$, there exists a fare class $k$ for which $\sup\Supp(n_{akm}+n_{bkm})\ge 2$. Fix $n\ge 2$. The distribution of $n_{bkm}|n_{akm}+n_{bkm}=n, \Delta X_{m}=x$ is a binomial mixture, with mixture distribution $G_x$, say. Then \citep[see, e.g.][]{d2017measuring}, the first $n$ moments of $G_x$ are identified. In particular, we identify $\int_0^1 p(1-p)dG_x(p)$. Now, given the structure of the problem,
$$\int_0^1 p(1-p)dG_x(p) = \int \Lambda'(x'\beta_0 - u) dF(u),$$
with $\Lambda'=\Lambda(1-\Lambda)$ the density of the logistic distribution. By varying $x_j$ over $\R$, we thus identify the distribution of $U+V$, where $U$ and $V$ are independent, $U$ is logistic and $V\sim F$. Taking the Fourier transform, we thus identify $\Psi_U \times \Psi_V$, where $\Psi_U$ and $\Psi_V$ (resp. $\Psi_V$) denotes the characteristic function of $U$ (resp. $V$). Since $\Psi_U(t)=\pi t/\sinh(\pi t)\ne 0$, $\Psi_V$ is identified. Hence, $F$ is identified as well.

\medskip
Finally, under Assumption \ref{hyp:gamma}(ii), $\eta_{bm}/\eta_{am}$ follows a beta prime distribution with parameters $(\lambda_{b0},\lambda_{a0})$. Because this beta prime distribution is identified by what precedes, so are $(\lambda_{b0},\lambda_{a0})$.


\subsection{Theorem \ref{thm:counterf_rev}} 
\label{sub:theorem_ref_thm_counterf_rev}

The idea is to define a new time variable and new Poisson processes such that (i) $s_m(\cdot)$ does not appear anymore in this new setup; (ii) the corresponding counterfactual revenues are equal to the initial ones. The result then follows. The strategy also applies to counterfactual loads.

First, fix $(I,r)\in \{c,i\}\times\{u,f,s,sM,sM+\}$ and let $S_m(t)=\int_0^t s_m(u)du$. By assumption, $S_m(.)$ is continuous and strictly increasing. We now define our new setup, using tildes. First, let $(\widetilde{X}_{am},\widetilde{X}_{bm},\widetilde{W}_m)=(X_{am},X_{bm},W_m)$,  $\widetilde{t}=S_m(t)$ and for $d\in\{a,b\}$, $t\in [0,1]$, $A\subset [0,\infty)$ and $p>0$, let us define $\widetilde{V}_{dm}$ as
$$\widetilde{V}_{dm}([0, \widetilde{t}], A) = V_{dm}([0, t], A).$$
By Assumption \ref{hyp:cons_demand} and definition of $\widetilde{t}$, $\widetilde{V}_{dm}$ is a Poisson process with intensity $(t,p)\mapsto \xi_{dm} \eps p^{-\eps-1}$. In other words, $\widetilde{\xi}_{dm}=\xi_{dm}$, $\widetilde{\eps}=\eps$ and $\widetilde{s}_m(t)=1$. In particular, $\widetilde{V}_{dm}$ does not depend on $s_m(\cdot)$.

\medskip
We now prove that $\widetilde{R}^I_r$, the optimal revenue associated with $(\widetilde{V}_{am}, \widetilde{V}_{bm})$, satisfies $\widetilde{R}^I_r=R^I_r$. Let us consider a pricing strategy $(p^I_{r,a}(.),p^I_{r,b}(.))$ associated with the processes $(V_{am},V_{bm})$, satisfying the constraints associated with $r$ and $I$ and leading, on expectation, to the optimal revenue $R^I_r$.\footnote{There may be no such pricing strategy, but only sequences of pricing strategies with corresponding expected revenue tending to $R^I$. If so, we just replace $p^I_r(.)$ by the corresponding sequence in the rest of the proof.} As feasible pricing strategies, $p^I_{r,a}(t)$ and $p^I_{r,b}(t)$ only depend on purchases up to $t$, on $\eps$, $s_m(\cdot)$ and on $(\xi_{am},\xi_{bm})$ (if $I=c$) or on $f_{\xi_{am},\xi_{bm}|X_{am},X_{bm},W_m}$ (if $I=i$). Now, let us define, for $d\in\{a,b\}$ and $t\in[0,1]$,
$$\widetilde{p}^I_{r,d}(\widetilde{t}) =p^I_{r,d}(t).$$
By construction, $\widetilde{p}^I_{r,d}(\widetilde{t})$ only depends on purchases (associated with $\widetilde{V}_{dm}$) up to $\widetilde{t}$, on $\widetilde{\eps}=\eps$, $s_m(\cdot)$ and on $(\widetilde{\xi}_{am},(\widetilde{\xi}_{bm})=(\xi_{am},\xi_{bm})$ (if $I=c$) or on $f_{\widetilde{\xi}_{am},\xi_{bm}|\widetilde{X}_{am},\widetilde{X}_{bm},\widetilde{W}_m}=f_{\xi_{am},\xi_{bm}|X_{am},X_{bm},W_m}$ (if $I=i$).
 Also, as $p^I_{r,d}$, it satisfies all the constraints associated with $r$, since no constraints are related to time. Hence, $(\widetilde{p}^{I}_{r,a},\widetilde{p}^I_{r,b})$ is a feasible pricing strategy, up to one point: it depends on $s_m(\cdot)$. This dependence is nevertheless useless, since the Poisson processes $(\widetilde{V}_{am}, \widetilde{V}_{bm})$ do not depend on $s_m(\cdot)$. Hence, the expected revenue ($\widetilde{\widetilde{R}}{}^I_r$, say) associated with $(\widetilde{p}^{I}_{r,a},\widetilde{p}^I_{r,b})$ satisfies $\widetilde{\widetilde{R}}{}^I_r\le \widetilde{R}^I_r$. Also, by construction, we obtain with $(\widetilde{p}^I_{r,a},\widetilde{p}^I_{r,b})$, associated with $(\widetilde{V}_{am},\widetilde{V}_{bm})$, the same purchases at the same prices as with the pricing strategy $(p^I_{r,a},p^I_{r,b})$ associated to $(V_{am},V_{bm})$. In other words, $\widetilde{\widetilde{R}}{}^I_r=R^I_r$. Therefore, $R^I_r\leq \widetilde{R}^I_r$.

 \medskip
Finally, consider a pricing strategy $(\widetilde{p}^{I,o}_{r,a}(\cdot),\widetilde{p}^{I,o}_{r,b}(\cdot))$ associated with $(\widetilde{V}_{am}, \widetilde{V}_{bm})$ , satisfying the constraints associated with $r$ and $I$ and leading, on expectation, to the optimal revenue $\widetilde{R}^I_r$. Then let $p^{I,o}_{r,a}(t) = \widetilde{p}^{I,o}_{r,d}(\widetilde{t})$. By construction, $p^{I,o}_{r,a}(t)$ only depends on purchases (associated with $V_{dm}$) up to $t$, on $\eps$, $s_m(\cdot)$ (through $\widetilde{t}=S_m(t)$) and on $(\xi_{am},\xi_{bm})$ (if $I=c$) or on $f_{\xi_{am},\xi_{bm}|X_{am},X_{bm},W_m}$ (if $I=i$). Thus, it is a feasible pricing strategy. Reasoning as above, this yields $\widetilde{R}^I_r \le R^I_r$. Hence, $\widetilde{R}^I_r=R^I_r$ and the result follows.


\newpage
\begin{center}
{\Large Instrument-Free Demand Estimation Using Relative Prices Variation, with an Application to Railway Transportation\\[2mm] Online Appendix}	
\end{center}

\section{Expressions for the counterfactual quantities}\label{app:counter_rev}

In this appendix, we first introduce some notations for the revenues under counterfactual pricing strategies. We then provide their formula, as well as those of counterfactual loads and social welfare, under Assumption \ref{hyp:gamma}(ii). The proofs of these formulas are given in Section \ref{app:proof_formulas} below. The formulas are given conditional on $(X_{am},X_{bm},W_m)$ and for simplicity, we assume here that $C_{am}$ and $C_{bm}$ are constant; if not, the results should just be seen conditional on $(C_{am},C_{bm})$. Due to the independence between $\eta_{am}$ and $\eta_{bm}$ in Assumption \ref{hyp:gamma}(i) and the pre-allocation of capacity, we can separably simulate the counterfactual revenue for each destination, and sum them up to obtain the revenue for train $m$. Consequently, we focus on pricing for destination $d$ served by train $m$ to simplify the exposition. We both consider arbitrary distributions for $\xi_{dm}$ and the gamma distribution in Assumption \ref{hyp:gamma}(ii).

\medskip
We use subscripts $u$, $f$, and $s$ to refer to uniform, full dynamic, and stopping-time pricing, respectively. For stopping-time pricing strategies, $sM$ (resp. $sM+$) denotes those for which only $M$ fares (resp. $M$ increasing fares), are allowed. Finally, superscripts $c$ and $i$ are employed to denote respectively complete and incomplete information settings. For instance, $R_f^c$ refers to the counterfactual optimal revenue under uniform pricing and complete information. In addition to the scenarios described in Section \ref{sec:mode}, we consider intermediate  pricing strategies described in Section \ref{sec:counterfactuals}, see Figure \ref{fig:intermediate} and the discussion above it. The corresponding revenues are denoted by $R^I_{iK}$, with $I\in\{c,i\}$ and where $K\in[0,100]$ indexes the proportion of seats that are dynamically priced.

\medskip
As shown in Appendix \ref{app:proof_formulas}, under Assumption \ref{hyp:gamma}(i), the revenue formulas of price strategies of interest under both complete and incomplete information have the following form: for $I\in\{c,i \}$ and $r\in\{u,f,s,sM,sM+,iK \}$,
\begin{equation}\label{eq:revenue_gamma_ave}
    R_{r}^I=\alpha_r^I(\eps,C_{dm},f)\exp\{X_{dm}'\beta_0/\eps\}g^{1/\eps}_0(W_m),
\end{equation}
where $f(.)$ denotes the distribution of $\eta_{dm}$. In each scenario, we will display the result for a general $f(.)$, and then apply it to the case where $f$ is the density of a gamma distribution $\Gamma(\lambda_{d0},1)$. We specify $\alpha_r^I(\eps,C_{dm},f)$ in each case. Finally,  $D(q)$ denotes hereafter a random variable satisfying $D(q)\sim \mathcal{P}(q)$.

\medskip
Finally, we focus on revenues hereafter. Under the microfoundation in Appendix \ref{app:microfoundation}, above, we can obtain similar expressions  for the social welfare using \eqref{eq:revenue_gamma_ave}.

\subsection{Complete information}

\paragraph{Uniform pricing}
${\alpha}^c_{u}=\max_{q>0}\left\{q^{-\frac{1}{\eps}}\mathbb{E}[D(q)\wedge{C_{dm}}]\right\}\mathbb{E}\left[\eta_{dm}^{1/\eps} \right]$.
The corresponding load is $\mathbb{E}\left[D(q^*)\wedge C_{dm} \right]$ where $q^*=\text{argmax}_{q>0}\left\{q^{-\frac{1}{\eps}}\mathbb{E}[D(q)\wedge{C_{dm}}]\right\}$.
\paragraph{Full-dynamic pricing}
$\alpha^c_f=\alpha^c_{C_{dm},f}\mathbb{E}\left[\eta_{dm}^{1/\eps} \right]$, where $\alpha^c_{0,f}=0$ and for all $k\geq 1$, $\alpha^c_{k,f}=(\alpha^c_{k,f}-\alpha_{k-1,f}^{c})^{1-\eps}\left(1-1/\eps\right)^{\eps-1}$. The corresponding load is $1$.
\paragraph{Stopping-time pricing}
$\alpha^c_s=\alpha^c_{C_{dm},s}\mathbb{E}\left[\eta_{dm}^{1/\eps} \right]$, where $\alpha^c_{0,s}=0$ and for all $k\geq 1$,
\begin{equation*}
\begin{aligned}
&\alpha_{k,s}^c=\max_{q>0}\left\{q^{-\frac{1}{\eps}}(1-e^{-q})+\alpha_{k-1,s}^c\int_0^1qe^{-sq}(1-s)^{\frac{1}{\eps}}ds\right\}.
\end{aligned}
\end{equation*}
The corresponding load $D_k$ satisfies: for $r=1,...,k$ and $D_0=0$,
\[
D_r=(1-e^{-\alpha_{r,s}^c})(1+D_{r-1}).
\]
\paragraph{Stopping-time pricing with $M$ fares}
$
R^c_{sM}=\alpha^c_{C_{dm},sM}\mathbb{E}\left[\eta_{dm}^{1/\eps} \right],
$	
where $\alpha^c_{C_{dm}, sM}=\max_{q>0}\alpha_{C_{dm},M}(q)$, $\alpha_{k,0}(q)= q^{-\frac{1}{\eps}}\mathbb{E}[D(q)\wedge k]$ and for all $k\in\{1,..,C_{dm}\}$ and $j\in \{1,...,k\}$,
\begin{equation*}
\begin{aligned}
\alpha_{k,j}(q)&=\max\Big\{\int_0^1qe^{-qz}\left[q^{-\frac{1}{\eps}}+\alpha_{k-1,j\wedge (k-1)}(q(1-z))(1-z)^{\frac{1}{\eps}}\right]dz,\\
&\hspace{1.7cm} \max_{q>0}\int_0^1qe^{-qz}\left[q^{-\frac{1}{\eps}}+\alpha_{k-1,j-1}(q(1-z))(1-z)^{\frac{1}{\eps}} \right]dz\Big\}.\\
\end{aligned}
\end{equation*}

\paragraph{Stopping-time pricing with $M$ increasing fares}
$\alpha^c_{sM+}=\alpha^c_{C_{dm},sM+}\mathbb{E}\left[\eta_{dm}^{1/\eps} \right]$,
where $\alpha^c_{C_{dm},sM+}=\max_{q>0}\alpha^+_{C_{dm},M}(q)$ with $\alpha^+_{k,0}(q)=\alpha_{k,0}(q)$, 
\begin{align*}
\alpha^+_{k,j}(q)&=\max\Big\{q \int_0^1e^{-qz}\left[q^{-\frac{1}{\eps}}+\alpha^+_{k-1,j\wedge (k-1)}(q(1-z))(1-z)^{\frac{1}{\eps}}\right]dz,\\
&\hspace{2cm} \max_{q'\in (0, q]} q' \int_0^1e^{-q'z}\left[q'^{-\frac{1}{\eps}}+\alpha^+_{k-1,j-1}(q'(1-z))(1-z)^{\frac{1}{\eps}}\right]dz.\Big\}.
\end{align*}

\paragraph{Intermediate-$K$ stopping-time pricing} $\alpha^c_{iK}=\alpha^c_{C_{dm},iK} \mathbb{E}\left[\eta_{dm}^{1/\eps} \right]$, where
$\alpha^c_{C_{dm}(1-K\%),iK}=\max_{q>0}q^{-\frac{1}{\eps}}\mathbb{E}[D(q)\wedge(C_{dm}(1-K\%))]$ and for $k>C(1-K\%)$,
$$\alpha^c_{k,iK}=\max_{q>0} \left\{q^{-\frac{1}{\eps}}(1-e^{-q})+\alpha^c_{k-1,iK}\int_0^1q e^{-q s}(1-s)^{\frac{1}{\eps}}ds\right\}.$$
\medskip

Under Assumption \ref{hyp:gamma}(ii), $\eta_{dm}\sim\Gamma(\lambda_{d0},1)$. We have $\mathbb{E}\left[\eta_{dm}^{1/\eps} \right]=\Gamma(\lambda_{d0}+1/\eps)/\Gamma(\lambda_{d0})$.

\subsection{Incomplete Information}

Hereafter, we denote by $\gamma_{\lambda,\mu}$ the density of the $\Gamma(\lambda,\mu)$ distribution.

\paragraph{Uniform pricing} $\alpha^i_{u}=\max_{q>0}\left\{\int_{\mathbb{R}^+}q^{-\frac{1}{\eps}}\mathbb{E}[D(qz)\wedge C_{dm}]f(z)dz\right\}$. Under Assumption \ref{hyp:gamma}(ii),
$\alpha^i_{u}=\max_{q>0}\left\{\int_{\mathbb{R}^+}q^{-\frac{1}{\eps}}\mathbb{E}[D(qz)\wedge C_{dm}]\gamma_{\lambda_0,1}(z)dz\right\}.$ The corresponding load is $\int_{\mathbb{R}^+}\mathbb{E}[D(q^*z)\wedge C_{dm}]\gamma_{\lambda_0,1}(z)dz$ where $q^*=\text{argmax}_{q>0}\left\{\int_{\mathbb{R}^+}q^{-\frac{1}{\eps}}\mathbb{E}[D(qz)\wedge C_{dm}]\gamma_{\lambda_0,1}(z)dz\right\}$.

\paragraph{Full-dynamic pricing} Under Assumption \ref{hyp:gamma},
$\alpha^i_f=\alpha^i_{C_{dm},f}(\lambda_{d0})$, where $\alpha^i_{0,f}(\lambda)=0$ for any $\lambda>0$ and for all $k\in\{1,...,C_{dm}\}$,
$$\alpha^i_{k,f}(\lambda)=\lambda\left(1-\frac{1}{\eps}\right)^{\eps-1}\left[-\alpha^i_{k-1,f}(\lambda+1)+(1+\frac{1}{\lambda\eps})\alpha^i_{k,f}(\lambda)\right]^{1-\eps}.$$

\paragraph{Stopping-time pricing}
$\alpha^i_s=\alpha^i_{C_{dm},s}(f)$, where $\alpha^i_{0,s}(f)=0$ and for any $k\in\{1,..,C_{dm}\}$, $$\alpha^i_{k,s}(f)=\max_{q>0}q\int_0^1\left[q^{-1/\eps}+(1-u)^{\frac{1}{\eps}}\alpha^i_{k-1,s}(T(f;qu))\right]\left[\int_0^\infty ze^{-quz}f(z)dz\right]du$$
and $T(f;q)$ is a transformation of density function $f$ defined in Lemma \ref{ConjugatePrior_general} below. Under Assumption \ref{hyp:gamma}(ii), $\alpha^i_s=\alpha^i_{C_{dm},s}(\lambda_{d0})$,
where $\alpha^i_{0,s}(\lambda)=0$ for $\lambda>0$, and for all $k\in\{1,...,C_{dm}\}$,
\[\alpha^i_{k,s}(\lambda)=\max_{q>0}q\int^1_0\frac{\lambda}{(1+qs)^{\lambda+1}}\left[q^{-\frac{1}{\eps}}+\left(\frac{1-s}{1+qs}\right)^{\frac{1}{\eps}}\alpha^i_{k-1,s}(\lambda+1)\right]ds.
\]
The corresponding load $D_k$ satisfies: for $r=1,...,k$ and $D_0(\lambda,\mu)=0$ for any $\lambda,\mu>0$,
\[
D_r(\lambda,\mu)=(1-e^{-\alpha^i_{r,s}(\lambda)\mu})(1+D_{r-1}(\lambda+1,[1+\alpha^i_{r,s}(\lambda)]\mu)).
\]

\paragraph{Stopping-time pricing with $M$ fares}
$
\alpha^i_{M,s}(M,f)=\alpha^i_{sM}(f)
$
where $\alpha^i_{sM}(f)=\max_{q>0} c_{C_{dm},M}(q,f)$ and for all $k$, $c_{k,0}(q,f)=q^{-\frac{1}{\eps}}\int \mathbb{E}[D(qz)\wedge k]f(z)dz$ and
\begin{equation*}
\begin{aligned}
c_{k,j}(q,f)=\max\Big\{& q \int_0^1\int ze^{-qzu}f(z)dz\Big[q^{-1/\eps}+c_{k-1,j\wedge (k-1)}(q(1-u),T(f;qu)) \\
& (1-u)^{\frac{1}{\eps}}\Big]du, \;\max_{q'>0}q'\int_0^1\int ze^{-q'zu}f(z)dz\Big[q'{}^{-1/\eps} +c_{k-1,j-1}(q'(1-u), \\
& T(f;q'u))  (1-u)^{\frac{1}{\eps}}\Big]du \Big\}\\
\end{aligned}
\end{equation*}
for any $j\in \{1,...,k\}$, $T$ being the same transform as in the case of stopping-time pricing. Further under Assumption \ref{hyp:gamma}(ii), $\alpha^i_{sM}(M,\lambda_{d0})=\alpha^i_{sM}(\lambda_0),$
where $\alpha^i_{sM}(\lambda)=\max_{q>0}c_{C_{dm},M}(q,\lambda)$ with, for all $k$, $c_{k,0}(q,\lambda)=q^{-\frac{1}{\eps}}\int \mathbb{E}[D(qz)\wedge k]\gamma_{\lambda,1}(z)dz$ and for all $k\in\{1,...,C_{dm}\}$ and all $j\in\{1,...,k\}$,
{\small
\begin{equation*}
\begin{aligned}
c_{k,j}(q,\lambda)&=\max\left\{q \int_0^1\frac{\lambda}{(1+qu)^{\lambda+1}}\left[q^{-\frac{1}{\eps}}+c_{k-1,j\wedge(k-1)}\left(\frac{q(1-u)}{1+qu},\lambda+1\right)\left(\frac{1-u}{1+qu}\right)^{\frac{1}{\eps}}\right]du, \right. \\
&\left. \quad\max_{q'>0} q' \int_0^1\frac{\lambda}{(1+q'u)^{\lambda+1}}\left[q'^{-\frac{1}{\eps}}+c_{k-1,j-1}\left(\frac{q'(1-u)}{1+q'u},\lambda+1\right)\left(\frac{1-u}{1+q'u}\right)^{\frac{1}{\eps}}\right]du\right\}.\\
\end{aligned}
\end{equation*}}

\paragraph{Stopping-time pricing with $M$ increasing fares}
$
\alpha^i_{sM+}(M,f)=\alpha^i_{C_{dm},sM+}(f),
$
where $\alpha^i_{C_{dm},sM+}(f)=\max_{q>0}c^+_{C_{dm},M}(q,f)$ with, for any $k\in\{0,...,C_{dm}\}$,  $c^+_{k,0}(q,f)=c_{k,0}(q,f)$ and for any $j\geq 1$,
\begin{equation*}
\begin{aligned}
c^+_{k,j}(q,f)=\max\Big\{& q \int_0^1\int ze^{-qzu}f(z)dz\Big[q^{-1/\eps}+c^+_{k-1,j\wedge (k-1)}(q(1-u),T(f;qu)) \\
& (1-u)^{\frac{1}{\eps}}\Big]du, \;\max_{q'\in(0,q]}q'\int_0^1\int ze^{-q'zu}f(z)dz\Big[q'{}^{-1/\eps} +c^+_{k-1,j-1}(q'(1-u), \\
& T(f;q'u))  (1-u)^{\frac{1}{\eps}}\Big]du \Big\}\\
\end{aligned}
\end{equation*}
Under Assumption \ref{hyp:gamma}(ii), we have
$R^i_{sM+}(M,\lambda_{d0})=\alpha^i_{sM+}(\lambda_0)$,
where $\alpha^i_{sM+}(\lambda)=\max_{q>0}c^+_{C_{dm},M}(q,\lambda)$ with  $c^+_{k,j}(q,\lambda)=c^+_{k,j}(q,\gamma_{\lambda,1})$ as defined above. Further, we have the following simplifications:
{\small
	\begin{equation*}
	\begin{aligned}
	c^+_{k,j}(q,\lambda)&=\max\left\{q \int_0^1\frac{\lambda}{(1+qu)^{\lambda+1}}\left[q^{-\frac{1}{\eps}}+c^+_{k-1,j\wedge(k-1)}\left(\frac{q(1-u)}{1+qu},\lambda+1\right)\left(\frac{1-u}{1+qu}\right)^{\frac{1}{\eps}}\right]du, \right. \\
	&\left. \quad\max_{q'\in(0,q]} q' \int_0^1\frac{\lambda}{(1+q'u)^{\lambda+1}}\left[q'^{-\frac{1}{\eps}}+c^+_{k-1,j-1}\left(\frac{q'(1-u)}{1+q'u},\lambda+1\right)\left(\frac{1-u}{1+q'u}\right)^{\frac{1}{\eps}}\right]du\right\}.\\
	\end{aligned}
\end{equation*}}

\paragraph{Intermediate-$K$ stopping-time pricing} $\alpha^i_{iK}=\alpha^i_{C_{dm},iK}(f)$, where $\alpha^i_{C_{dm}(1-K\%),iK}(f)=\max_{q>0}\left\{\int_{\mathbb{R}^+}q^{-\frac{1}{\eps}}\mathbb{E}[D(qz)\wedge (C_{dm}(1-K\%))]f(z)dz\right\}$ and for any $k>C_{dm}(1-K\%)$,
$$\alpha^i_{k,iK}(f)=\max_{q>0}q\int_0^1\left[q^{-1/\eps}+(1-u)^{\frac{1}{\eps}}\alpha^i_{k-1,iK}(T(f;qu))\right]\left[\int_0^\infty ze^{-quz}f(z)dz\right]du.$$
Under Assumption \ref{hyp:gamma}(ii), $\alpha^i_s=\alpha^i_{C_{dm},iK}(\lambda_0)$,
where $$\alpha^i_{C_{dm}(1-K\%),iK}(\lambda)=\max_{q>0}\left\{\int_{\mathbb{R}^+}q^{-\frac{1}{\eps}}\mathbb{E}[D(qz)\wedge (C_{dm}(1-K\%))]\gamma_{\lambda,1}(z)dz\right\}$$
 for $\lambda>0$, and for all $k>C_{dm}(1-K\%)$,
$$\alpha^i_{k,iK}(\lambda)=\max_{q>0}\left\{q\int^1_0\frac{\lambda}{(1+qs)^{\lambda+1}}\left[q^{-\frac{1}{\eps}}+\left(\frac{1-s}{1+qs}\right)^{\frac{1}{\eps}}\alpha^i_{k-1,iK}(\lambda+1)\right]ds.\right\}.$$



\section{Bounding counterfactual revenues with a time-varying elasticity}\label{app:constant_elas_details}
In this appendix, we give details on the partial identification of the lower bound in \eqref{eq:revenue_gamma_ave} with a time-varying elasticity, namely when the intensity of $V_{dm}$ satisfies \eqref{eq:I_two_elast}. We first show how to partially identify $g_0$ in this context and then turn to counterfactual revenues, social welfare, and consumer surplus.

\paragraph{Demand estimation: $g_0(.)$.} To partially identify $g_0(.)$, we still rely on consumers' rationality described in section \ref{sec:partial_id} and weak optimality condition in \eqref{eq:upper_bound_rev}.  Because $\eps_{\text{early}}\neq\eps_{\text{late}}$, the resulting moment inequalities differ from those under the assumption of constant price elasticity. The inequalities originating from consumers' rationality are modified as follows. First, for $k\leq S$ (recalling that $t_S$ is a function of $W_m$),
\begin{equation}\label{eq:rationality_two_elas1}
\resizebox{0.9\textwidth}{!}{$
\begin{aligned}
	\E\Big[&\sum_{j=k}^K n_{djm}-C_m \wedge D\left(\exp\{X'_{dm}\beta_0\}\eta_{dm}\left[p^{-\eps_{\text{early}}}_{dkm} g^\text{early}_0(W_m)+p^{-\eps_{\text{late}}}_{d(S+1)m}g^\text{late}_0(W_m)\right]\right)\big|W_m\Big]\leq 0,\\
\end{aligned}
$}
\end{equation}
and for $k> S$,
\begin{equation}\label{eq:rationality_two_elas2}
\resizebox{0.9\textwidth}{!}{$
\begin{aligned}
\E\Big[&\sum_{j=k}^K n_{djm}-C_m \wedge D\left(\exp\{X'_{dm}\beta_0\}\eta_{dm}p^{-\eps_{\text{late}}}_{dKm}\left[g^\text{early}_0(W_m)+g^\text{late}_0(W_m)\right]\right)\big|W_m\Big]\leq 0,\\
\end{aligned}
$}
\end{equation}
where $g^\text{early}_0(W_m)=g_0(W_m)\int_0^{t_S}s_m(t)dt$ and $g^\text{late}_0(W_m)=g_0(W_m)\int_{t_S}^1s_m(t)dt$. Note that the left-hand side of \eqref{eq:rationality_two_elas1} is strictly decreasing in $\overline{g}(W_t):=p^{-\eps_{\text{early}}}_{dkm} g^\text{early}_0(W_m)+p^{-\eps_{\text{late}}}_{dSm}g^\text{late}_0(W_m)$. Then, given $d$, $k\leq S$, and $W_m$, we obtain a lower bound on $\overline{g}(W_t)$. Similarly, we obtain a lower bound on $g^\text{early}_0(W_m)+g^\text{late}_0(W_m)$ from \eqref{eq:rationality_two_elas2} given $d$, $k\leq S$, and $W_m$. As a result, for a given $W_m$, we obtain $24$ linear constraints on $(g^\text{early}_0(W_m),g^\text{late}_0(W_m))$.

\medskip
Turning to the weak optimality condition, we have
\begin{align*}
\mathbb{E}[R_m(p_a,p_b)|W_m] = \max_{C_{am}+C_{bm}=C_m}\bigg\{&\sum_{d=a,b}p_d\int_0^\infty \mathbb{E}\big[ D\left(\exp\{X_{dm}'\beta_0 \}\big(p_d^{-\eps_\text{early}}g^\text{early}_0(W_m) \right. \\
& \left. +p_d^{-\eps_\text{late}}g^\text{late}_0(W_m)\big)z\right) \wedge C_{dm}\big]\gamma_{\lambda_{d0},1}(z)dz\bigg\}.
\end{align*}
Let $R(g_0^\text{early}(W_m),g_0^\text{late}(W_m );\eps_\text{early},\eps_\text{late},\beta_0,\lambda_0):= \underset{k=1,...,K}{\max}\mathbb{E}[R_m(p_{akm},p_{bkm})|W_m]$. Then, we obtain:
\begin{equation}\label{eq:weak_optimality_two_elas}
R(g_0^\text{early}(W_m),g_0^\text{late}(W_m);\eps_\text{early},\eps_\text{late},\beta_0,\lambda_0)\leq \mathbb{E}\left[R^\text{obs}_m|W_t\right].
\end{equation}
\noindent Unlike the moment inequalities built on the consumers' rationality, the weak optimality condition does not deliver a linear constraint on $(g^\text{early}_0(W_m),g^\text{late}_0(W_m))$ because of the maximization over $k$. The estimation of the identified set  of the counterfactual revenue is complicated by this non-linearity. To circumvent this numerical challenge, we exploit a looser weak optimality inequality, namely $$R(g_0^\text{early},g_0^\text{late};\eps_\text{early},\eps_\text{late},\beta_0,\lambda_0)\geq R(g_0^\text{early},g_0^\text{late};\eps_\text{early},\eps_\text{early},\beta_0,\lambda_0).$$
This inequality holds because $\eps_{\text{early}}<\eps_{\text{late}}$ and $p_{dkm}\ge 1$ for all $m$, $d\in\{a,b\}$ and $k\in\{1,...,12\}$. Then, \eqref{eq:weak_optimality_two_elas} implies
\begin{equation*}
R(g_0^\text{early}(W_m),g_0^\text{late}(W_m);\eps_\text{early},\eps_\text{early},\beta_0,\lambda_0)\leq \mathbb{E}\left[R^\text{obs}_m|W_m\right].
\end{equation*}
In other words, the actual revenue management yields a higher revenue than the optimal uniform pricing strategy with a price chosen from the price grid and early purchasers's price elasticity. This looser inequality leads to an upper bound on $g^\text{early}_0(W_m)+g^\text{late}_0(W_m)$ and  delivers a linear constraint.

\medskip
To summarize, for a given $W_m$, consumers' rationality and the loosened version of the weak optimality condition imply linear constraints on $(g^\text{early}_0(W_m),g^\text{late}_0(W_m))$. The computation of the estimator of the identified set of counterfactual revenues then reduces to an optimization problem with linear constraints.

\paragraph{Counterfactual revenues, uniform pricing.} We give details for the optimal uniform pricing under complete information. The derivation for the case under limited information is similar. Given $W_m$ and capacity $C_{dm}$ for destination $d$, the expected revenue $R^c_{udm}$ of the optimal uniform pricing strategy under complete information for destination $d$ in train $m$ is
\begin{align*}
R^c_{udm}(\eps_{\text{early}},\eps_{\text{late}},g^\text{early}_0,g^\text{late}_0,C_{dm})
=\int_0^\infty & \max_{p_d>0}\left\{p_d \mathbb{E}\big[D\big(\exp\{X_{dm}'\beta_0 \} \left(p_d^{-\eps_\text{early}} g^\text{early}_0(W_m) \right.\right.  \\
& \left. \left.\left. +p_d^{-\eps_\text{late}} g^\text{late}_0(W_m)\right)z\right)\wedge C_{dm}\big]\right\}\gamma_{\lambda_{d0},1}(z)dz.
\end{align*}
Similarly to the case of time-invariant elasticity, $R^c_{udm}(\eps_{\text{early}},\eps_{\text{late}},g^\text{early}_0,g^\text{late}_0,C_{dm})$ will not depend on the pattern of $s_m(\cdot)$. Differently, it is a function of the average demand corresponding to different elasticity levels, $g^\text{early}_0$ and $g^\text{late}_0$.

The expected optimal revenue at the train level is:
\[
R^c_{um}(\eps_{\text{early}},\eps_{\text{late}},g^\text{early}_0,g^\text{late}_0,C_m)=\underset{C_{am}+C_{bm}=C_m}{\max}\sum_{d=a,b}R^c_{udm}(\eps_{\text{early}},\eps_{\text{late}},g^\text{early}_0,g^\text{late}_0,C_{dm}).
\]
Denote by $\mathcal{G}$ the set of linear constraints on $(g^\text{early}_0,g^\text{late}_0)$ derived in the previous section. Then, the lower bound of the set estimate of $R^c_{um}(\eps_{\text{early}},\eps_{\text{late}},g^\text{early}_0,g^\text{late}_0,C_m)$ is expressed as:
\[
\underline{R}^c_{um}(\eps_{\text{early}},\eps_{\text{late}},C_m)=\underset{(g^\text{early}_0,g^\text{late}_0)\in\mathcal{G}}{\min}\underset{C_{am}+C_{bm}=C_m}{\max}\sum_{d=a,b}R^c_{udm}(\eps_{\text{early}},\eps_{\text{late}},g^\text{early}_0,g^\text{late}_0,C_{dm}).
\]
Solving this program exactly is difficult. Rather, we obtain lower and upper bounds on $\underline{R}^c_{um}(\eps_{\text{early}},\eps_{\text{late}},C_m$. First, note that using Jensen's inequality, we have
\[
p^{-\eps_{\text{early}}}_{d} g^\text{early}_0+p^{-\eps_{\text{late}}}_{d}g^\text{late}_0\geq (g^\text{early}_0+g^\text{late}_0)p^{-\eps(g^\text{early}_0,g^\text{late}_0)}_{d},
\]
where $\eps(g^\text{early}_0,g^\text{late}_0)=\frac{g^\text{early}_0\eps_{\text{early}}+g^\text{late}_0\eps_{\text{late}}}{g^\text{early}_0+g^\text{late}_0}$. Then,
\[
R^c_{udm}(\eps_{\text{early}},\eps_{\text{late}},g^\text{early}_0,g^\text{late}_0,C_{dm})\geq R^c_{udm}(\eps(g^\text{early}_0,g^\text{late}_0),\eps(g^\text{early}_0,g^\text{late}_0),g^\text{early}_0,g^\text{late}_0,C_{dm}),
\]
with $R^c_{udm}(\eps(g^\text{early}_0,g^\text{late}_0),\eps(g^\text{early}_0,g^\text{late}_0), g^\text{early}_0,g^\text{late}_0,C_{dm})$ the expected revenue of the optimal uniform pricing under constant price elasticity $\eps(g^\text{early}_0,g^\text{late}_0)$. Then,
\begin{equation*}
\begin{aligned}
&\underline{R}^c_{um}(\eps_{\text{early}},\eps_{\text{late}},C_m)\\
&\geq \underset{(g^\text{early}_0,g^\text{late}_0)\in\mathcal{G}}{\min}\underset{C_{am}+C_{bm}=C_m}{\max}\sum_{d=a,b}R^c_{udm}(\eps(g^\text{early}_0,g^\text{late}_0),\eps(g^\text{early}_0,g^\text{late}_0),g^\text{early}_0,g^\text{late}_0,C_{dm}).\\
\end{aligned}
\end{equation*}
Note that $R^c_{udm}(\eps(g^\text{early}_0,g^\text{late}_0),\eps(g^\text{early}_0,g^\text{late}_0),g^\text{early}_0,g^\text{late}_0,C_{dm})$ is a function of $g^\text{early}_0+g^\text{late}_0$ and $\eps(g^\text{early}_0,g^\text{late}_0)$. Moreover, for $d=a,b$ and any $C_{dm}$, it is increasing with respect to $g^\text{early}_0+g^\text{late}_0$ and decreasing with respect to $\eps(g^\text{early}_0,g^\text{late}_0)$. Then
$$\underset{C_{am}+C_{bm}=C_m}{\max}\sum_{d=a,b}R^c_{udm}(\eps(g^\text{early}_0,g^\text{late}_0),\eps(g^\text{early}_0,g^\text{late}_0),g^\text{early}_0,g^\text{late}_0,C_{dm})$$
is also increasing with respect to $g^\text{early}_0+g^\text{late}_0$ and decreasing with respect to $\eps(g^\text{early}_0,g^\text{late}_0)$. As a result, its minimization with respect to $(g^\text{early}_0,g^\text{late}_0)\in\mathcal{G}$ is achieved either  when $g^\text{early}_0+g^\text{late}_0$ is minimized or $\eps(g^\text{early}_0,g^\text{late}_0)$ is maximized. Denote the minimizer of $g^\text{early}_0+g^\text{late}_0$ by $\underline{g}_0$ and the maximizer of  $\eps(g^\text{early}_0,g^\text{late}_0)$ by $\overline{\eps}$. Then,
\begin{equation*}
	\resizebox{\textwidth}{!}{$
	\begin{aligned}
	&\underset{(g^\text{early}_0,g^\text{late}_0)\in\mathcal{G}}{\min}\underset{C_{am}+C_{bm}=C_m}{\max}\sum_{d=a,b}R^c_{udm}(\eps(g^\text{early}_0,g^\text{late}_0),\eps(g^\text{early}_0,g^\text{late}_0),g^\text{early}_0,g^\text{late}_0,C_{dm})\\
	&=\min\Big\{\underset{(g^\text{early}_0,g^\text{late}_0)\in\mathcal{G},g^\text{early}_0+g^\text{late}_0=\underline{g}_0}{\min}\underset{C_{am}+C_{bm}=C_m}{\max}\sum_{d=a,b}R^c_{udm}(\eps(g^\text{early}_0,g^\text{late}_0),\eps(g^\text{early}_0,g^\text{late}_0),g^\text{early}_0,g^\text{late}_0,C_{dm}),  \\
&\underset{(g^\text{early}_0,g^\text{late}_0)\in\mathcal{G},\eps(g^\text{early}_0,g^\text{late}_0)=\overline{\eps}}{\min}\underset{C_{am}+C_{bm}=C_m}{\max}\sum_{d=a,b}R^c_{udm}(\overline{\eps},\overline{\eps},g^\text{early}_0,g^\text{late}_0,C_{dm})\Big\}.\\
	\end{aligned}
$}
\end{equation*}
To perform the first minimization, it suffices to compute the upper bound of $\eps(g^\text{early}_0,g^\text{late}_0)$ in $\mathcal{G}$ subject to $g^\text{early}_0+g^\text{late}_0=\underline{g}_0$, and compute the corresponding optimal revenue. Similarly, to perform the second minimization, it suffices to compute the lower bound of $g^\text{early}_0+g^\text{late}_0$ in $\mathcal{G}$ subject to $\eps(g^\text{early}_0,g^\text{late}_0)=\overline{\eps}$, and compute the corresponding optimal revenue with constant price elasticity $\overline{\eps}$. The minimum of the two simulated revenues then bounds $\underline{R}^c_{um}(\eps_{\text{early}},\eps_{\text{late}},C_m)$ from below. Second, note that
\begin{equation*}
	\resizebox{\textwidth}{!}{$
	\begin{aligned}
	\underline{R}^c_{um}(\eps_{\text{early}},\eps_{\text{late}},C_m)&\leq \underset{(g^\text{early}_0,g^\text{late}_0)\in\mathcal{G},g^\text{late}_0=0 }{\min}\underset{C_{am}+C_{bm}=C_m}{\max}\sum_{d=a,b}R^c_{udm}(\eps_{\text{early}},\eps_{\text{late}},g^\text{early}_0,g^\text{late}_0,C_{dm})\\
	&=\underset{(g^\text{early}_0,g^\text{late}_0)\in\mathcal{G},g^\text{late}_0=0 }{\min}\underset{C_{am}+C_{bm}=C_m}{\max}\sum_{d=a,b}R^c_{udm}(\eps_{\text{early}},\eps_{\text{early}},g^\text{early}_0,0,C_{dm}).
	\end{aligned}
$}
\end{equation*}
The revenue $R^c_{udm}(\eps_{\text{early}},\eps_{\text{early}},g^\text{early}_0,0,C_{dm})$ is the expected revenue of the optimal uniform pricing under constant price elasticity $\eps_{\text{early}}$ and only depends on $g_0^{\text{early}}$. Then it suffices to compute the lower bound of $g_0^{\text{early}}$ in $\mathcal{G}\cap\{(g_0^{\text{early}},0):g_0^{\text{early}}\in\mathbb{R}^+ \}$ and simulate the corresponding optimal revenue with constant price elasticity $\eps_{\text{early}}$  to bound $\underline{R}^c_{um}(\eps_{\text{early}},\eps_{\text{late}},C_m)$ from above.

\paragraph{Counterfactual revenues, dynamic pricing.} Given an allocation   $(C_{am},C_{bm})$ between destinations $a$ and $b$ served by train $m$, we can consider the optimal dynamic pricing (stopping-time or the full one) separably for either destination. Take the full dynamic pricing under complete information as example. Using the notations in Appendix \ref{ssub:proofs}, the Bellman equation is written as:  for  $t< t_S$,
\begin{equation}\label{eq:bellman_two_elas_late}
	\resizebox{0.91\textwidth}{!}{$
	V^*_k{}'(t)=\partial_1 S_m(1-t,1)\frac{\xi_{dm}}{\eps_{\text{late}}-1}\left(1-\frac{1}{\eps_{\text{late}}}\right)^{\eps_{\text{late}}}\left[V^*_k(t)-V^*_{k-1}(t)\right]^{1-\eps_{\text{late}}}, \text{ with } V^*_{k}(0)=0 \ \forall k\geq 0,
	$}
\end{equation}
and for $t\geq t_S$,
\begin{equation}\label{eq:bellman_two_elas_early}
V^*_k{}'(t)=\partial_1 S_m(1-t,1)\frac{\xi_{dm}}{\eps_{\text{early}}-1}\left(1-\frac{1}{\eps_{\text{early}}}\right)^{\eps_{\text{early}}}\left[V^*_k(t)-V^*_{k-1}(t)\right]^{1-\eps_{\text{early}}},
\end{equation}
where the initial conditions $V^*_{k}(t_S)$ are defined by \eqref{eq:bellman_two_elas_late}. We are interested in $V^*_k{}'(1)$. 

If the elasticity in \eqref{eq:bellman_two_elas_late} were $\eps_{\text{early}}$, we would then obtain the same explicit solution in Appendix \ref{app:counter_rev}. However, because $\eps_{\text{late}}<\eps_{\text{early}}$,  the initial conditions for \eqref{eq:bellman_two_elas_early} at $t=t_S$ are no longer the same. Such jumps in the initial conditions introduce substantial challenges in deriving explicit formula for $V^*_k{}(1)$. 

Nevertheless, similarly to the case of uniform pricing above, we can show that $V^*_k{}(1)$ depends on average demand $g^\text{early}_0$ and $g^\text{late}_0$, rather than the pattern of $s_m(\cdot)$. Note that following the technique of deriving the formula for the case of time-invariant elasticity, we obtain from \eqref{eq:bellman_two_elas_late} that for $t<t_S$, i.e., $S_m(1-t,1)< S_m(1-t_S,1)$, 
\[
V_k^*(t) = \alpha^c_{k,f} [S_m(1-t,1)\xi_{dm}]^{\frac{1}{\eps_{\text{late}}}}=\alpha^c_{k,f} (\exp\{X'_{dm}\beta_0\}\eta_{dm}g_0(W_m)S_m(1-t,1))^{\frac{1}{\eps_{\text{late}}}},
\]
where $\alpha^c_{0,f}=0$ and for all $k\geq 1$, $\alpha^c_{k,f}=(\alpha^c_{k,f}-\alpha_{k-1,f}^{c})^{1-\eps_{\text{late}}}\left(1-1/\eps_{\text{late}}\right)^{\eps_{\text{late}}-1}$. Moreover, from \eqref{eq:bellman_two_elas_early}, we can derive that for any $t\geq t_S$, i.e., $S_m(1-t,1)\geq S_m(1-t_S,1)$, we can express $V_k^*(t)=\xi_{dm}^{\frac{1}{\eps_{\text{early}}}} f_k(S_m(1-t,1))$ where function $f_k$ satisfies : for any $s\geq S_m(1-t_S,1)$,
\[
f_k'(s) = \frac{1}{\eps_{\text{early}}-1}\left(1-\frac{1}{\eps_{\text{early}}}\right)^{\eps_{\text{early}}}\left[f_k(s)-f_{k-1}(s)\right]^{1-\eps_{\text{early}}}
\]
with 
\[
f_k(S_m(1-t_S,1))=\alpha^c_{k,f} [g_{0}^{\text{late}}]^{\frac{1}{\eps_{\text{late}}}} [\exp\{X'_{dm}\beta_0\}\eta_{dm}]^{\frac{1}{\eps_{\text{late}}}-\frac{1}{\eps_{\text{early}}}}g_0(W_m)^{-\frac{1}{\eps_{\text{early}}}}
\]
for any $k$, where $g_0(W_m)=g_{0}^{\text{early}}+g_{0}^{\text{late}}$. Then, $f_k(S_m(1-t,1))$ is a function of $S_m(1-t,1)$ and $(g_0^{\text{early}},g_0^{\text{late}})$, so is  $V_k^*(t)$ for $t\geq t_S$. When setting $t=1$, $S_m(0,1)=1$ and $V_k^*(t)$ will depend on $(g_0^{\text{early}},g_0^{\text{late}})$, rather than the pattern of $s_m(\cdot)$. 
\smallskip

We propose lower and upper bounds for $V^*_k{}(1)$. To obtain the lower bound, we consider a preallocation of $C_{dm}$ into $C_{dm\text{early}}$ and $C_{dm\text{late}}$, each corresponding to the capacity during early and late periods for destination $d$,  respectively. Any seat allocated to the early period cannot be re-used during the late one. Intuitively, this preallocation is suboptimal because any unsold seats during the early period could have been sold later. Therefore, the optimal dynamic pricing with such preallocation constraints  provides a lower bound for the unconstrained one. Technically, this constrained dynamic pricing corresponds to setting $V_k^*(t_S)=0$ in \eqref{eq:bellman_two_elas_early} for any $k>0$. Then, we can derive the revenues before and after $t_S$ by using the formulas in Appendix \ref{app:counter_rev}, and  explicitly express the total revenue as a nonlinear function of $(g_0^{\text{early}},g_0^{\text{late}})$.
\smallskip

Denote by $V_k^*(t;\eps_{\text{early}})$ the value function corresponding to the Bellman equation with constant elasticity $\eps_{\text{early}}$ for $t\in[0,1]$.  To obtain the upper bound, we use the following property:
$$
V^{*}_k(t)-V^{*}_{k-1}(t)>V_k^{*}(t;\eps_{\text{early}})-V_{k-1}^{*}(t;\eps_{\text{early}})\text{ for }t>t_S,
$$
i.e., the opportunity cost of selling a seat during the early period is greater when consumers in the late period are less price elastic. Using this property and formulas in Appendix \ref{app:counter_rev}, we have:
$$
V^*_k(1)-V^*_k(t_S)<V_k^*(1;\eps_{\text{early}})-V_k^*(t_S;\eps_{\text{early}})=\alpha_{k,f}^c\xi_{dm}^{1/\eps_{\text{early}}}[(g_{0}^{\text{early}}+g_{0}^{\text{late}})^{1/\eps_{\text{early}}}-(g_{0}^{\text{early}})^{1/\eps_{\text{early}}}].
$$
Then, using \eqref{eq:bellman_two_elas_late}, we obtain the upper bound for $V_k(1)$:
\[
V_k^*(1)< \alpha_{k,f}^c\left[\xi_{dm}^{1/\eps_{\text{early}}}[(g_{0}^{\text{early}}+g_{0}^{\text{late}})^{1/\eps_{\text{early}}}-(g_{0}^{\text{early}})^{1/\eps_{\text{early}}}]+\xi_{dm}^{1/\eps_{\text{late}}}(g_{0}^{\text{late}})^{1/\eps_{\text{late}}} \right].
\]
We can then minimize the lower/upper bound with respect to $(g_0^{\text{early}},g_0^{\text{late}})$ to obtain the lower/upper bound for the lower bound of the counterfactual revenue.
\paragraph{Social welfare and consumer surplus.} For any pricing strategy and information environment, the corresponding social welfare is always equal to $\frac{\eps_{\text{early}}}{\eps_{\text{early}}-1}R_{\text{early}}+\frac{\eps_{\text{late}}}{\eps_{\text{late}}-1}R_{\text{late}}$, where $R_{\text{early}}$ and $R_{\text{late}}$ are the revenues generated by early and late arrivals (see Appendix \ref{app:microfoundation}). Besides, the corresponding consumer surplus is $\frac{1}{\eps_{\text{early}}-1}R_{\text{early}}+\frac{1}{\eps_{\text{late}}-1}R_{\text{late}}$. Because $\eps_{\text{early}}>\eps_{\text{late}}>1$, we obtain lower bounds for social welfare and consumer surplus as $\frac{\eps_{\text{early}}}{\eps_{\text{early}}-1} (R_{\text{early}}+R_{\text{late}})$ and $\frac{1}{\eps_{\text{early}}-1} (R_{\text{early}}+R_{\text{late}})$, respectively. We then use  the lower bounds on the revenues corresponding to optimal uniform and dynamic pricing strategies  in Table \ref{tab:counter_rev_two_elas} to derive the lower bounds for social welfare. Despite the simplicity, these lower bounds may be conservative. 

Concretely, under  the optimal uniform pricing, we obtain  $\frac{\eps_{\text{early}}}{\eps_{\text{early}}-1}\underline{R}^c_{um}(\eps_{\text{early}},\eps_{\text{late}},C_m)$ as a lower bound for the social welfare and $\frac{\underline{R}^c_{um}(\eps_{\text{early}},\eps_{\text{late}},C_m)}{\eps_{\text{early}}-1}$ as a lower bound for the corresponding consumer surplus.  Similar formula apply to the social welfare and consumer surplus under the optimal dynamic pricing strategy.

\section{Proof of the formulas for counterfactual revenues}
\label{app:proof_formulas}

\subsection{Complete information}
\label{ssub:proofs}
\subsubsection{Uniform pricing}

Given $\xi_{dm}$, the revenue under uniform price $p_d$ is
\begin{equation*}
R^c_{u}(p_d,\xi_{dm})=\mathbb{E}[p_dV_{dm}([0,\tau_{C_{dm}}\wedge1),[p_d,\infty))|\xi_{dm}],
\end{equation*}
where $\tau_{C}=\inf\{t:V_{dm}([0,t),[p_d,\infty))\geq{C_{dm}}\}$ is the stopping time of selling out all $C_{dm}$ seats. Then,
	\begin{equation*}
	\begin{aligned}
	R^c_u(\xi_{dm})&=\max_{p>0}R^c_u(p,\xi_{dm})=\max_{p>0}p\mathbb{E}[D(\xi_{dm}p^{-\eps})\wedge{C_{dm}}|\xi_{dm}].
	\end{aligned}
	\end{equation*}
	We obtain the result by defining $q=\xi_{dm}p^{-\eps}$ and integrate over $\xi_{dm}$.

Without pre-allocation of capacities among intermediate and final destinations, given $(p_a,p_b)$, $(\xi_{am},\xi_{bm})$, and using Theorem \ref{thm:ident_eps}, we have:
\begin{equation*}\label{eq:uniform_without_allocation}
\begin{aligned}
	R^c_u(p_a,p_b,\xi_{am},\xi_{bm})&=\mathbb{E}\left[\mathbb{E}\left[p_aD(\xi_{am}p_a^{-\eps})+p_bD(\xi_{bm}p_b^{-\eps}) \big|\left(D(\xi_{am}p_a^{-\eps}\xi_{bm}p_b^{-\eps})\wedge C_m\right)\right]\right]\\
	&=\frac{\xi_{am}p_a^{1-\eps}+\xi_{bm}p_b^{1-\eps}}{\xi_{am}p_a^{-\eps}+\xi_{bm}p_b^{-\eps}}\mathbb{E}\left[D(\xi_{am}p_a^{-\eps}+\xi_{bm}p_b^{-\eps})\wedge C_m\right].\\
\end{aligned}
\end{equation*}
Then, the optimal revenue under uniform pricing without pre-allocation is achieved when $p_a=p_b$ and therefore:
\[
R^c_u(\xi_{am},\xi_{bm})=\max_{p>0} p\mathbb{E}\left[D((\xi_{am}+\xi_{bm})p^{-\eps})\wedge C_m\right]
\]

\subsubsection{Full dynamic pricing.}
Denote by $V_k(t,p_d)$ the expected revenue when there remains $k$ vacant seats before the departure and the current seat is priced at $p_d$ at time $1-t$. From $1-t$ to $1-t+\Delta t$, the probability of selling one seat is $s_m(1-t)\xi_{dm}p_{d}^{-\eps}\Delta t+o(\Delta t)$ and generates $p_d$ revenue if one seat is sold. With probability $o(\Delta t)$, more than one seats are sold. Then, following \cite{Gallego_vanRyzin_94} (Section 2.2.1 on page 1004), we have:
	\begin{align}
	V^*_k(t) = &\max_{p_d>0}\Big\{s_m(1-t)\xi_{dm}p_{d}^{-\eps}\Delta t\left(p_d+V_{k-1}^*(t-\Delta t)\right) \nonumber  \\
	& \hspace{1.7cm} + [1-s_m(1-t)\xi_{dm}p_{d}^{-\eps}\Delta t] V_k^*(t-\Delta t)
	+ o(\Delta t)
	\Big\}. \label{full_yield}
	\end{align}
Letting $\Delta t\to0$, this equation shows that $V^*_k$ is continuous. Further, by considering $(V^*_k(t) - V^*_k(t-\Delta t))/\Delta t$ and letting $\Delta t\to 0$, $V_k^*$ is differentiable, with\footnote{For conditions that enable to interchange $\underset{\Delta t\to 0}{\lim}$ and $\max$, we refer to \cite{bremaud1981point} for details.}
\begin{equation}\label{full_yield_K}
	V^*_k{}'(t)  	= \max_{p_d>0}s_m(1-t)\xi_{dm}p_{d}^{-\eps} \left[p_{d}+V^*_{k-1}(t)-V^*_k(t)\right]
\end{equation}
with boundary conditions $V^*_k(0)=0$ for any $k=1,...,C_{dm}$ and $V^*(t,0)=0$ for any $t\in[0,1]$. As a consequence, the optimal price $p^*_{tk}$ can be obtained from the first-order condition of the right-hand side of \eqref{full_yield_K}:
	\begin{equation}\label{opt_price:full_full}
	p^*_{tk}=\frac{\eps}{\eps-1}\left[V^*_k(t)-V^*_{k-1}(t)\right].
	\end{equation}
	By plugging $p^*_{tk}$ into \eqref{full_yield_K} and using $S_m(t,1)=\int_t^{1}s_m(s)ds$ (where we let $S_m(t,t'):=\int_t^{1}s_m(s)ds$), we obtain:
	\begin{equation}\label{full_yield_K_bis}
	V^*_k{}'(t)=\partial_1 S_m(1-t,1)\frac{\xi_{dm}}{\eps-1}\left(1-\frac{1}{\eps}\right)^{\eps}\left[V^*_k(t)-V^*_{k-1}(t)\right]^{1-\eps},
	\end{equation}
where $\partial_j S_m$ denotes the derivative of $S_m$ with respect to its $j$-th argument. We now prove by induction on $k$ that
\begin{equation}
V^*_k(t)=\alpha^c_{k,f} [\xi_{dm}S_m(1-t,1)]^{\frac{1}{\eps}}	
	\label{eq:Vstar_full_ci}
\end{equation}
for all $k\in\{0,...,C_{dm}\}$, with $\alpha^c_f(0)=0$ and $\alpha^c_{k,f}=(\alpha^c_{k,f}-\alpha^c_{k-1,f})^{1-\eps}\left(1-\frac{1}{\eps}\right)^{\eps-1}$.
	
\medskip
The result holds for $k=0$ since $V^*_0(t)=0$. Next, suppose that \eqref{eq:Vstar_full_ci} holds for $k-1\geq 0$ and let us show that the result holds for $k$. By plugging this solution for $k-1$ into the differential equation \eqref{full_yield_K}, we obtain:
	\begin{equation}
	 V^*_k{}'(t) = \partial_1 S_m(1-t,1)\frac{\xi_{dm}}{\eps-1}\left(1-\frac{1}{\eps}\right)^{\eps}   \left[V^*_k(t)-\alpha^c_{k-1,f}[\xi_{dm}S_m(1-t,1)]^{\frac{1}{\eps}}\right]^{1-\eps}, \label{full_yield_K_induction}
	\end{equation}
	with $V^*_k(0)=0$. We can check that $V^*_k(t)=\alpha^c_{k,f}[\xi_{dm}S_m(1-t,1)]^{1/\eps}$ is a solution to \eqref{full_yield_K_induction}. To show uniqueness, let $\phi(v,z)=\frac{1}{\eps-1}\left(1-\frac{1}{\eps}\right)^{\eps}\left[v-\alpha^c_{k-1,f}z^{1/\eps}\right]^{1-\eps}$. Consider the diffeomorphism $z(t)=\xi_{dm}S_m(1-t,1)$ and define $\bar{V}^*_k(z)=V_k^*(t(z))$. Then, \eqref{full_yield_K_induction} can be written as
	\begin{equation}\label{full_yield_K_induction2}
	\bar{V}_k^*{}'(z)=\phi(\bar{V}^*_k(z),z),
	\end{equation}
	with $\bar{V}^*_k(0)=0$. It is enough to prove that $\bar{V}^*_k$ is the unique solution of \eqref{full_yield_K_induction2} and we prove this by contradiction. Suppose that there is another differentiable solution $\tilde{V}_k(.)$ different from $\bar{V}^*_k(z)=\alpha^c_{k,f}z^{1/\eps}$. Without loss of generality, $\tilde{V}_k(z_0)> \bar{V}^*_k(z_0)$ for some $z_0>0$. Because $\tilde{V}_k(0)=\bar{V}^*_k(0)=0$, then $z_m=\sup\{z\leq z_0:\tilde{V}_k(z_0)\leq\bar{V}^*_k(z_0) \}$ exists and $z_m<z_0$. Moreover, $\tilde{V}_k(z_m)=\bar{V}^*_k(z_m)$. Then, \eqref{full_yield_K_induction2} implies the contradiction
	\[
	0<\tilde{V}_k(z_0)-\bar{V}^*_k(z_0)=\int_{z_m}^{z_0}[\phi(\tilde{V}_k(z),z)-\phi(\bar{V}^*_k(z),z)]dz\leq 0,
	\]
	where the second inequality follows from the fact that $\phi$ is a decreasing function of $z$ and
	$\tilde{V}_k(s)>\bar{V}^*_k(s)$ for all $s\in (z_m,z_0]$. Finally, we conclude that $\bar{V}_k^*(.)$ is the unique solution. Hence, the result holds for $k$, and \eqref{eq:Vstar_full_ci} holds. By taking $t=1, k={C_{dm}}$ and integrating over $\xi_{dm}$, we obtain the desired formula.

\textbf{\emph{Load.}} We now derive the formula for load under the optimal dynamic pricing \eqref{opt_price:full_full} which denote by $D_k(t,t')$ for any $0\leq t< t'\leq 1$. First, note that $V_{dm}$ becomes an one-dimensional non-homogeneous Poisson process with intensity $\frac{s_m(t)}{S_m(t,1)}(\alpha^c_{K,f})^{\frac{1-\eps}{\eps}}$. 
When $k=1$, for any $t,t'$, 
\[
\Pr(D_1(t,t')=0)=\exp\left\{-\int_{t}^{t'}\frac{s_m(s)}{S_m(s,1)}(\alpha_{1,f}^c)^{\frac{1-\eps}{\eps}}ds\right\}=\left(\frac{S_m(t',1)}{S_m(t,1)}\right)^{(\alpha_{1,f}^c)^{\frac{1-\eps}{\eps}}}.
\]
As $t'\to 1$, we have $\Pr(D_1(t,t')=0)\to0$. As a result, we have $D_1(t,1)=1$ for any $t$.

Now suppose that $\Pr(D_r(t,1)=r)=1$ for $r=1,2,...,k-1$ and any $t$. We prove  $\Pr(D_k(t,1)=k)=1$. Denote by $\tau>t$ the hitting time of the first sale after $t$. Then, 
\begin{equation*}
\begin{aligned}
\Pr(D_k(t,1)=k)&=\mathbb{E}_\tau\left[\Pr(D_k(t,1)=k|\tau<1) \right]\Pr(\tau<1)\\
&=\mathbb{E}_\tau[\Pr(D_{k-1}(\tau,1)=k-1|\tau<1)]\Pr(\tau<1)\\
&=\Pr(\tau<1).\\
\end{aligned}
\end{equation*}
Note that for $t<t'<1$,
\[
\Pr(D_k(t,t')=0)=\exp\left\{-\int_{t}^{t'}\frac{s_m(s)}{S_m(s,1)}(\alpha_{k,f}^c)^{\frac{1-\eps}{\eps}}ds\right\}=\left(\frac{S_m(t',1)}{S_m(t,1)}\right)^{(\alpha_{k,f}^c)^{\frac{1-\eps}{\eps}}}.
\]
Then, $\Pr(\tau<1)\geq 1-\Pr(D_k(t,t')=0)$. Let $t'\to1$, we have $\Pr(\tau<1)\geq 1$ and therefore $\Pr(D_k(t,1)=k)=\Pr(\tau<1)= 1$. Setting $t=0$, we obtain the load formula.
\subsubsection{Stopping-time pricing.}

Denote by $V_k(t,p_d)$ the expected optimal revenue at time $1-t$ when pricing the next seat at $p_d$ and with $k$ remaining seats. In this scenario, prices do not change until the next seat is sold. Define $\tau_{1-t,p_d}=\inf\{s>0:V_{dm}([1-t,1-t+s),[p_d,\infty))\geq 1\}$. Then,
\begin{equation*}
\begin{aligned}
\Pr[\tau_{1-t;p_d}>s]&=\Pr[V_{dm}([1-t,1-t+s),[p_d,\infty))=0]=\exp\{-S_m(1-t,1-t+s)\xi_dp_d^{-\eps}\},\\
\end{aligned}
\end{equation*}
and the density of $\tau_{1-t,{p}_d}$ is
\begin{equation}\label{stopping-time}
f_{\tau_{1-t,{p}_d}}(s)=\xi_dp_d^{-\eps}\partial_2 S_m(1-t,1-t+s) e^{-S_m(1-t,1-t+s)\xi_dp_d^{-\eps}}.
\end{equation}
Then, the Bellman equation is
\begin{align}
V_k(t,p_d)&=\E\left[\mathbf{1}_{\tau_{1-t,p_d}<t}\left(p_d+V^*_{k-1}({t-\tau_{1-t,\bm{p}}})\right)\right]\nonumber \\
&=\int_0^tf_{\tau_{1-t,p_d}}(s)\left(p_d+V^*_{k-1}({t-s})\right)ds\nonumber \\
&=\int_0^t\xi_dp_d^{-\eps}\partial_2 S_m(1-t,1-t+s)e^{-S_m(1-t,1-t+s)\xi_dp_d^{-\eps}} \times \left(p_d+V^*_{k-1}( {t-s})\right)ds. \label{eq:bellman1}
\end{align}
Let $V^*_k(t)=\max_{p>0} V_k(t,p)$. We now show by induction that
\begin{equation}\label{eq:value_function_stopping_time_f}
	V^*_k(t)=\alpha^c_{k,s} [\xi_{dm}S_m(1-t,1)]^{\frac{1}{\eps}},
\end{equation}
where $\alpha^c_{0,s}=0$ and
\begin{equation*}
\alpha^c_{k,s}=\max_{q>0}\left\{q^{-\frac{1}{\eps}}(1-e^{-q})+\alpha^c_{k-1,s}\int_0^1qe^{-sq}(1-s)^{\frac{1}{\eps}}ds\right\}.
\end{equation*}
The result holds for $k=0$ since $V_0^*(1-t)=0$. Now, suppose that \eqref{eq:value_function_stopping_time_f} is true for $k-1\geq 0$. By using the change of variable $z=S_m(1-t,1-t+s)/S_m(1-t,1)$ and applying \eqref{eq:value_function_stopping_time_f} for $V^*_{k-1}(t)$ in Equation \eqref{eq:bellman1}, we get
\begin{align*}
V_k(t,p)  = & \int_0^1\xi_{dm}S_m(1-t,1)p^{-\eps}e^{-S_m(1-t, 1)\xi_{dm}p^{-\eps}z}\left(p+[\xi_{dm}S_m(1-t,1)(1-z)]^{\frac{1}{\eps}}\alpha^c_{k-1,s}\right)dz\\
=&[\xi_{dm}S_m(1-t,1)]^{\frac{1}{\eps}}\left(q^{-\frac{1}{\eps}}(1-e^{-q})+\alpha^c_{k-1,s}\int^1_0qe^{-qz}(1-z)^{\frac{1}{\eps}}dz\right),
\end{align*}
where $q=\xi_{dm}S_m(1-t,1)p^{-\eps}$. As a consequence,
\begin{equation*}
\begin{aligned}
V^*_k(t)&=\max_{p>0}V_k(t,p)\\
&=[\xi_{dm}S_m(1-t,1)]^{\frac{1}{\eps}} \max_{q>0}\left\{q^{-\frac{1}{\eps}}(1-e^{-q})+\alpha^c_{k-1,s} \int^1_0qe^{-qz}(1-z)^{\frac{1}{\eps}}dz\right\}\\
&=\alpha^c_{k,s}[\xi_{dm}S_m(1-t,1)]^{\frac{1}{\eps}},
\end{aligned}
\end{equation*}
and \eqref{eq:value_function_stopping_time_f} is true for $k$. Thus, \eqref{eq:value_function_stopping_time_f} holds for all $k\in\{0,...,C_{dm}\}$. Finally, by taking $t=1,k={C_{dm}}$ and the expectation with respect to $\xi_{dm}|(X_{dm},W_m)$, we obtain the desired expression.

To obtain the load formula, we first derive the optimal pricing from the Bellman equation. Then, we express the load under the optimal pricing at time $t$ similarly to \eqref{eq:bellman1} and prove the desired formula by induction.
\subsubsection{Stopping-time pricing with $M$ fares.}
Denote by $V_k(0;t,p,j)$ (resp. $V_k(1;t,p,j)$) the expected revenue of the firm at time $1-t$, with a current price $p$, a remaining capacity $k$ and a remaining number of fares $j$, if it decides to keep the same price $p$ (resp. to choose a new price). Then, we have the following Bellman equations:
\begin{equation}\label{limited_fares_stopping_time_1}
\left\{\begin{aligned}
&V_k(1;t,p,j)=\max_{p'>0}\int_{0}^{t}f_{\tau_{1-t,p'}}(s)\left[p'+V^*_{k-1}(t-s,p',j-1)\right]ds,\\
&V_k(0;t,p,j)=\int_{0}^{t}f_{\tau_{1-t,p}}(s)\left[p+V^*_{k-1}(t-s,p,j)\right]ds,\\
&V^*_k(t,p,j)=\max_{d\in\{0,1\}}V_k(d;t,p,j),\\
\end{aligned}\right.
\end{equation}
with initial conditions $V^*_0(t,p,j)=0$. We show by induction on $k$ that for all $(k,j)\in\{0,...,C_{dm}\}\times \N$,
\begin{equation}\label{eq:stopping_time_M_f}
	V^*_k(t,p,j)=\alpha_{k,j}(q(t,p))\left[\xi_{dm}S_m(1-t,1)\right]^{\frac{1}{\eps}},
\end{equation}
where $q(t,p)=p^{-\eps}\xi_{dm}S_m(1-t,1)$, $\alpha_{k,0}(q) = q^{-\frac{1}{\eps}}\mathbb{E}[D(q)\wedge k]$ and for $j\geq 1$,
\begin{align*}
	\alpha_{k,j}(q)&=\max\Big\{q \int_0^1e^{-qz}\left[q^{-\frac{1}{\eps}}+\alpha_{k-1,j\wedge (k-1)}(q(1-z))(1-z)^{\frac{1}{\eps}}\right]dz,\\
	&\hspace{2cm} \max_{q'>0} q' \int_0^1e^{-q'z}\left[q'^{-\frac{1}{\eps}}+\alpha_{k-1,j-1}(q'(1-z))(1-z)^{\frac{1}{\eps}}\right]dz.\Big\}.
\end{align*}
Because for any $j\geq k$ and $d\in\{0,1\}$, we have $V_k(d;t,p,j)=V_k(d;t,p,k)$, it suffices to prove the result for $j\leq k$. The result holds for $k=j=0$ since $V^*_0(t,p,j)=0$. Now, suppose that \eqref{eq:stopping_time_M_f} holds for $k-1\geq 0$ and all $j\leq k-1$. If $j=0$, the price cannot be changed anymore, so $V^*_k(t,p,j)$ is simply the revenue with price $p$ from $1-t$ to $1$, and \eqref{eq:stopping_time_M_f} holds.

\medskip
If $j\geq 1$, we have, by Equations \eqref{stopping-time}, \eqref{limited_fares_stopping_time_1}, the change of variable $z=S_m(1-t,1-t+s)/S_m(1-t,1)$ and the induction hypothesis,
\begin{align}
	& V_k(0;t,p,j) \nonumber \\
	=&\int_{0}^{t}f_{\tau_{1-t,p}}(s)\left[p+V^*_{k-1}(t-s,p,j\wedge (k-1))\right]ds\nonumber \\
	=&\int_{0}^{t}\xi_{dm}p^{-\eps}\partial_2 S_m(1-t,1-t+s)e^{-\xi_{dm}p^{-\eps}S_m(1-t,1-t+s)}\nonumber \\
	&\left[p+\alpha_{k-1,j\wedge (k-1)}(\xi_{dm}S_m(1-t+s,1)p^{-\eps})[\xi_{dm}S_m(1-t+s,1)]^{\frac{1}{\eps}}\right]ds\nonumber \\
	=&\int_0^1\xi_{dm}p^{-\eps}S_m(1-t,1)e^{-\xi_{dm}p^{-\eps}S_m(1-t,1)z}\nonumber \\
	&\left[p+\alpha_{k-1,j\wedge (k-1)}(\xi_{dm}S_m(1-t,1)p^{-\eps}(1-z))[\xi_{dm}S_m(1-t,1)]^{\frac{1}{\eps}}(1-z)^{\frac{1}{\eps}}\right]dz\nonumber \\
	=&[\xi_{dm}S_m(1-t,1)]^{\frac{1}{\eps}}  \int_0^1q(t,p)e^{-q(t,p)z}\left[q(t,p)^{-\frac{1}{\eps}}+\alpha_{k-1,j\wedge (k-1)}(q(t,p)(1-z))(1-z)^{\frac{1}{\eps}}\right]dz, \label{eq:induction_M+}
\end{align}
With the same reasoning, we also obtain
\begin{equation*}
	\begin{aligned}
	V_k(1;t,p,j)=&\max_{p'>0}\int_{0}^{t}f_{\tau_{1-t,p'}}(s)\left[p'+V^*_{k-1}(t-s,p',j-1)\right]ds\\
	=&[\xi_{dm}S_m(1-t,1)]^{\frac{1}{\eps}}\max_{q>0}\int_0^1qe^{-qz}\left[q^{-\frac{1}{\eps}}+\alpha_{k-1,j-1}(q(1-z))(1-z)^{\frac{1}{\eps}}\right]dz.\\
	\end{aligned}
\end{equation*}
Then, $V^*_k(t,p,j)=\max_{d\in\{0,1\}}V_k(d;t,p,j)=\alpha_{k,j}(q(t,p))[\xi_{dm}S_m(1-t,1)]^{\frac{1}{\eps}}$.
Thus, \eqref{eq:stopping_time_M_f} holds for $k$, and hence for all $k\in\{0,...,C_{dm}\}$. By setting $t=0$ and optimizing $V^*_k(t,p,j)$ over $p$ (or equivalently over $q(t,p)$) and taking the expectation with respect to $\xi_{dm}|(X_{dm},W_m)$, we obtain the desired expression.

\subsubsection{Stopping-time pricing with $M$ increasing prices.}

The reasoning is very similar to the previous case. The only change in \eqref{limited_fares_stopping_time_1} is in the formula of $V_k(1;t,p,j)$: the maximization is now over $p'\geq p$ rather than $p'\geq 0$, since the new price has to be higher than the current one. Then, following a similar strategy by induction, we get
$$V^*_k(t,p,j)=\alpha^+_{k,j}(q(t,p))\left[\xi_{dm}S_m(1-t,1)\right]^{\frac{1}{\eps}},$$
where $\alpha^+_{k,0}(q)=\alpha_{k,0}(q)$ and
\begin{align*}
	\alpha^+_{k,j}(q)&=\max\Big\{q \int_0^1e^{-qz}\left[q^{-\frac{1}{\eps}}+\alpha^+_{k-1,j\wedge (k-1)}(q(1-z))(1-z)^{\frac{1}{\eps}}\right]dz,\\
	&\hspace{2cm} \max_{q'\in (0, q]} q' \int_0^1e^{-q'z}\left[q'^{-\frac{1}{\eps}}+\alpha^+_{k-1,j-1}(q'(1-z))(1-z)^{\frac{1}{\eps}}\right]dz.\Big\}.
\end{align*}
We obtain the result by taking $t=0$, $k=C_{dm}$ and defining $\alpha^c_{C_{dm}, sM+}=\max_{q>0} \alpha^+_{C_{dm},M}(q)$.

\subsubsection{Intermediate-$K$ stopping-time pricing}

The proof is the same as that for \eqref{eq:value_function_stopping_time_f} except for the initial value because the firm must apply uniform pricing whenever there remain $C_{dm}(1-K\%)$ seats. Thus, the Bellman equation and the updating of the constants $\alpha^c_{iK,k}$ take the same form as under the stopping-time pricing strategy in \eqref{eq:value_function_stopping_time_f} for $k\geq C_{dm}(1-K\%)$. The initial value becomes $\alpha^c_{iK,C_{dm}(1-K\%)}$, which comes from the optimal uniform pricing with $C_{dm}(1-K\%)$ seats.


\subsection{Incomplete information}
\label{proof_thm_limited_rev}

Denote $Y_{dm}=\exp\{X_{dm}'\beta_0\} g_0(W_m)$. Denote the density function of $\xi_{dm}$ by $f$. Under Assumption \ref{hyp:gamma}(ii), $f$ is a gamma density  $\Gamma(\lambda_{d0},Y_{dm}^{-1})$.

\subsubsection{Uniform pricing.}  We have:
	\begin{equation*}
	\begin{aligned}
	R^i_u&=\max_{p>0}R^i_u(p;\eps,f)=\max_{p>0}p\int_{z>0}\mathbb{E}[D(p^{-\eps}z)\wedge C_{dm}]f(z)dz.\\
	\end{aligned}
	\end{equation*}
By the change of variable $q=Y_{dm}p^{-\eps}$, we obtain the desired formula.

\noindent\textbf{Proof of \eqref{eq:uniform_rev}.} Given $(p_{a},p_{b})$ and $(C_{am},C_{bm})$,  the total revenue generated by $m$ is:
\begin{align}
& \mathbb{E}\left[R_m(p_a,p_b;C_{am},C_{bm})|W_m \right] \notag \\
= & \sum_{d=a,b}p_{d}\int_{z>0} \mathbb{E}\left[D(p^{-\eps}_d\exp\{X'_{dm} \beta_0 \}g_0(W_m)z)\wedge C_{dm}\right] \gamma_{\lambda_{d0},1}(z)dz, \label{eq:rev_unif_WT}
\end{align}
where $\gamma_{\lambda_{d0},1}(z)$ is the density of a $\Gamma(\lambda_{d0},1)$. Then, we obtain \eqref{eq:uniform_rev} by maximizing the revenue in \eqref{eq:rev_unif_WT} over all possible allocations $(C_{am},C_{bm})$ subject to $C_{am}+C_{bm}=C_m$.

Without capacity pre-allocation among intermediate and final destinations, we obtain:
\begin{equation*}
\resizebox{\textwidth}{!}{$
	\begin{aligned}
	\mathbb{E}\left[R_m(p_a,p_b;C_{m})|W_m \right]&=\mathbb{E}\left[\mathbb{E}\left[\sum_{d=a,b}p_{d}D_{dm}\big| (D_{am}+D_{bm})\wedge C_m,z_a,z_b\right]\right]\\
	&=\mathbb{E}\left[\left(D\left(\sum_{d=a,b}p^{-\eps}_d\exp\{X'_{dm} \beta_0 \}g_0(W_m)z_d\right)\wedge C_m\right)\frac{\sum_{d=a,b}p^{1-\eps}_d\exp\{X'_{dm} \beta_0 \}g_0(W_m)z_d}{\sum_{d=a,b}p^{-\eps}_d\exp\{X'_{dm} \beta_0 \}g_0(W_m)z_d} \right]\\
	\end{aligned}
	$}
\end{equation*}
where $z_a$ and $z_b$ follows $\Gamma(\lambda_{a0},1)$ and $\Gamma(\lambda_{b0},1)$, respectively, and are independent. Then, the optimal revenue is obtained by maximizing $	\mathbb{E}\left[R_m(p_a,p_b;C_{m})|W_m \right]$ over $(p_a,p_b)\in\mathbb{R}^2_+$.

\subsubsection{Full dynamic pricing.} Define $V_k(t,p,f)$ as the expected revenue at time $1-t$ when there remains $k$ vacant seats before the departure, the current seat is priced at $p$ and the density of $\xi_{dm}$, given the current information, is $f$. Let also $V^*_k(t,f)=\max_{p>0}V_k(t,p,f)$. When $\eta_m\sim \Gamma(\lambda,\mu)$, we use respectively $V_k(t,p,\lambda,\mu)$
and $V^*_k(t,\lambda,\mu)$ instead of $V_k(t,p,\gamma_{\lambda,\mu})$ and  $V^*_k(t,\gamma_{\lambda,\mu})$.

\medskip
Between $1-t$ and $1-t+\Delta t$, if one seat is sold, which occurs with probability $\xi_{dm}p^{-\eps}\partial_1 S_m(1-t,1)\Delta t+o(\Delta t)$, the posterior cdf of $\xi_{dm}$, $F_1(.;\Delta t)$ satisfies
$$F_1(\xi;\Delta t)\propto[p^{-\eps}\partial_1 S_m(1-t,1)\xi\Delta t+o(\Delta t)]\xi^{\lambda-1} e^{-\mu \xi},$$
and the corresponding density is
\begin{align*}
	f_1(\xi;\Delta t)=& \xi^{\lambda}e^{-\mu \xi}\frac{\mu^{\lambda+1}}{\Gamma(\lambda+1)}+o(\Delta t).
\end{align*}
As $\Delta t\to 0$, the posterior density converges to $\gamma_{\lambda+1,\mu}$. If the seat is not sold between $1-t$ and $1-t+\Delta t$, then the posterior cdf of $\eta_m$ is
$$F_0(\xi;\Delta t) \propto \xi^{\lambda-1} \exp(-\mu(t,\Delta t,p) \xi),$$
where $\mu(t,\Delta t,p)=\mu+p^{-\eps}S_m(1-t,1-t+\Delta t)$. Therefore, the posterior density  is $\gamma_{\lambda,\mu(t,\Delta t,p)}$. Then, the Bellman equation can be written as:
\begin{equation*}
\begin{aligned}
V_k(t,p,\lambda,\mu) = & \int \Big\{[p^{-\eps}\xi \partial_1 S_m(1-t,1)\Delta t+o(\Delta t)] \times \left[p+V_{k-1}^*(t-\Delta t,f_1(.;\Delta t))\right]\\
&\; +\left[1-p^{-\eps}\xi \partial_1 S_m(1-t,1)\Delta t-o(\Delta t)\right]\times V_k^*(t-\Delta t, \lambda,\mu(p, t, \Delta))\Big\} \gamma_{\lambda,\mu}(\xi)d\xi.\\
\end{aligned}
\end{equation*}
Then,
{\small
\begin{align*}
V_k(t,p,\lambda,\mu) = & V^*_k(t-\Delta t,\lambda,\mu)
+ \int \Big\{[p^{-\eps}\xi\partial_1 S_m(1-t,1)\Delta t+o(\Delta t)] \\
 \times  & \left[p+V_{k-1}^*(t-\Delta t,f_1(.;\Delta t))\right] +\left[V_k^*(t-\Delta t,\lambda,\mu(p, t, \Delta t))-V_k^*(t-\Delta t,\lambda,\mu)\right]\\
-& V^*_k(t-\Delta t,\lambda,\mu( t, \Delta,p)) [p^{-\eps}\xi\partial_1 S_m(1-t,1)\Delta t+o(\Delta t)]\Big\} \gamma_{\lambda,\mu}(\xi)d\xi.
\end{align*}}
Then, using $V^*_k(t,\lambda,\mu)=\max_{p>0} V_k(t,p,\lambda,\mu)$ and letting  $\Delta t\to0$, we obtain:\footnote{For conditions that enable to interchange $\underset{\Delta t\to 0}{\lim}$ and $\max$, we refer to \cite{bremaud1981point} for details.}

\begin{align*}
& \partial_1 V^*_k(t,\lambda,\mu) \\
=&\max_{p>0}\int \Big\{p^{-\eps}\xi\partial_1 S_m(1-t,1)\left[p+V_{k-1}^*(t,\lambda+1,\mu)-V^*_k(t,\lambda,\mu)\right]\nonumber \\
&\hspace{1.5cm}+\lim_{\Delta t\to0}\frac{V^*_k(t-\Delta t,\lambda,\mu(t,\Delta t,p))-V^*_k(t-\Delta t,\lambda,\mu)}{\Delta t}\Big\} \gamma_{\lambda,\mu}(\xi)d\xi \nonumber \\
=&\partial_1 S_m(1-t,1)\max_{p>0}\int \Big\{p^{-\eps}\xi \left[p+V_{k-1}^*(t,\lambda+1,\mu)-V^*_k(t,\lambda,\mu)\right]+\partial_3 V^*_k(t,\lambda,\mu)p^{-\eps}\Big\} \gamma_{\lambda,\mu}(\xi)d\xi\nonumber \\
=&\partial_1 B_{m}(1-t,1)\max_{p>0} \Big\{p^{-\eps}\frac{\lambda}{\mu}\left[p+V_{k-1}^*(t,\lambda+1,\mu)-V^*_k(t,\lambda,\mu)\right]+\partial_3 V^*_k(t,\lambda,\mu)p^{-\eps}\Big\}.
\end{align*}
Solving for the optimal price, we then obtain:
\begin{align*}
\partial_1 V^*_k(t,\lambda,\mu)=&\left[\frac{\eps}{\eps-1}\right]^{-\eps}\frac{\lambda}{\mu(\eps-1)} \partial_1 S_m(1-t,1)\nonumber \\
&\times \left[-V_{k-1}^*(t,\lambda+1,\mu)+V^*_k(t,\lambda,\mu)-\frac{\mu}{\lambda}\partial_3 V^*_k(t,\lambda,\mu)\right]^{1-\eps}.
\end{align*}
Letting $z(t)=S_m(1-t,1)$ and $\bar{V}^*(z(t),\lambda,\mu)=V^*(t,\lambda,\mu)$, we obtain:
\begin{align}
\partial_1 \bar{V}^*_k(z,\lambda,\mu)=\left[\frac{\eps}{\eps-1}\right]^{-\eps}\frac{\lambda}{\mu(\eps-1)}\bigg[& -\bar{V}_{k-1}^*(z,\lambda+1,\mu)+\bar{V}^*_k(z,\lambda,\mu) \nonumber \\
& -\frac{\mu}{\lambda}\partial_3 \bar{V}^*_k(z,\lambda,\mu)\bigg]^{1-\eps}.
\label{bellman_limited_rev_5}
\end{align}	
We prove by induction on $k$ that for all $k\in\{0,...,C_{dm}\}$.
\begin{equation}\label{eq:full_l}
	\bar{V}^*_k(z,\lambda,\mu)=\left(\frac{z}{\mu}\right)^{\frac{1}{\eps}}\alpha^i_{k,f}(\lambda),
\end{equation}
where $\alpha^i_f(0,\lambda)=0$ and for $k\geq 1$,
$$\alpha^i_{k,f}(\lambda)=\lambda\left(1-\frac{1}{\eps}\right)^{\eps-1}\left[-\alpha^i_{k-1,f}(\lambda+1)+\left(1+\frac{1}{\lambda\eps}\right)\alpha^i_{k,f}(\lambda)\right]^{1-\eps}.$$

The result holds for $k=0$ since $V_0^*(z,\lambda,\mu)=0$. Suppose that \eqref{eq:full_l} holds for $k-1$. Then,  \eqref{bellman_limited_rev_5} and the induction hypothesis yield
\begin{equation}\label{bellman_limited_rev_6}
\resizebox{0.92\textwidth}{!}{$
\partial_1 \bar{V}^*_k(z,\lambda,\mu)=\left[\frac{\eps}{\eps-1}\right]^{1-\eps}\frac{\lambda}{\mu(\eps-1)}\left[ -\left(\frac{z}{\mu}\right)^{\frac{1}{\eps}}\alpha^i_{k-1,f}(\lambda+1)+\bar{V}^*_k(z,\lambda,\mu) -\frac{\mu}{\lambda}\partial_3 \bar{V}^*_k(z,\lambda,\mu)\right]^{1-\eps}.
$}
\end{equation}	
The function $(z,\lambda,\mu)\mapsto \alpha^i_{k,f}(\lambda)\left(z/\mu\right)^{1/\eps}$ is a solution to \eqref{bellman_limited_rev_6}. We now show that $\bar{V}^*_k(z,\lambda,\mu)$ is equal to this solution.
First, note that $V^*_k(t,\lambda,\mu)$ remains unchanged if the distribution of $S_m(t,t')\xi$ remains unchanged. Now,
\[
S_m(t,t')\xi= (S_m(t,t')/\delta) \times (\delta\xi),
\]
with $\delta\xi\sim\Gamma(\lambda,\mu/\delta)$. Hence, $V^*_k(t,\lambda,\mu)$ remains unchanged if we replace $\mu$ by $\mu/\delta$ and $z(t)$  by $z(t)/\delta$. Given the definition of $\bar{V}^*_k(z,\lambda,\mu)$, this implies $\bar{V}^*_k(z/\delta,\lambda,\mu/\delta)=\bar{V}^*_k(z,\lambda,\mu)$ for all $\delta>0$.  Then, to prove the induction step, we only need to show that $V(x):=V_k^*(x,\lambda,1)$ satisfies $V(x)=\alpha^i_{k,f}(\lambda)x^{1/\eps}$.  By Equation \eqref{bellman_limited_rev_6},
\begin{equation}\label{ode:pde}
V'(x) = \left[\frac{\eps}{\eps-1}\right]^{1-\eps}\frac{\lambda}{\eps-1}\left[-x^{\frac{1}{\eps}}\alpha^i_{k-1,f}(\lambda+1) + V(x) +\frac{x}{\lambda}V'(x)
\right]^{1-\eps},
\end{equation}
with initial condition $V(0)=0$. Suppose that \eqref{ode:pde} has two distinct solutions $V_1, V_2$ and let $x_0$ be such that $V_1(x_0)\neq V_2(x_0)$, say $V_1(x_0)>V_2(x_0)$. Define $x_m=\sup\{x\leq x_0:V_1(x)\leq V_2(x) \}$. Because $V_1(0)=V_2(0)$ and $V_1(x_0)>V_2(x_0)$,  we have $0\leq x_m<x_0$ and $V_1(x)>V_2(x)$ for $x\in(x_m,x_0]$. Moreover, because both solutions are continuous, $V_1(x_m)=V_2(x_m)$. According to \eqref{ode:pde}, because $\eps>1$, as long as $V_1(x)>V_2(x)$, we have $V_1'(x)<V_2'(x)$. Then,
\[
V_1(x_0)-V_2(x_0) = \int_{x_m}^{x_0}\left[V_1'(x)-V_2'(x)\right]dx<0,
\]
which contradicts $V_1(x_0)>V_2(x_0)$. Hence, $V(x)=\alpha^i_{k,f}(\lambda)x^{1/\eps}$, and the induction step holds. Thus, \eqref{eq:full_l} is satisfied for $k\in\{0,...,C_{dm}\}$. Finally, we obtain the desired result by taking $t=0$ and $k=C_{dm}$.

\subsubsection{Stopping-time pricing}

The difference from the stopping-time pricing under complete information is that the firm updates in a Bayesian way its belief on the distribution of $\xi_{dm}$. Even if the firm continuously updates its belief, only moments where a sale occurs matter, since this is the time where it can decide to change its prices. Thus, starting at time $1-t$, we can focus on time $1-t+\tau_{t,p}$. The next lemma characterizes the corresponding posterior distribution of $\xi_{dm}$.

\begin{lem}\label{ConjugatePrior_general}
	Suppose that the density function of $\xi_{dm}$ at time $1-t$ is $f$ and the firm prices the next seat at ${p}$. Then, the posterior distribution of $\xi_{dm}|\tau_{1-t;{p}}=s$ is $T(f;q(t,s,{p}))$, with $q(t,s,{p})=p^{-\eps}S_m(1-t,1-t+s)$ and
	$$T(f;u)(z)=\frac{ze^{-uz}f(z)}{\int ze^{-uz}f(z)dz}.$$
\end{lem}

\textbf{Proof:} As Equation \eqref{stopping-time} shows, given $\xi_{dm}=z$, the density function of $\tau_{1-t;{p}}$ is
\begin{equation}
	\label{eq:f_tau_B}	
f_{\tau_{1-t,{p}}|\xi_{dm}}(s|z)=p^{-\eps}z\partial_2 S_m(1-t,1-t+s)e^{-zq(t,s,{p})}.
\end{equation}
Then, the joint distribution of $(\tau_{1-t,{p}},\xi_{dm})$ is
	\begin{align*}
	f_{\tau_{1-t,{p}},\xi_{dm}}(s,z)=&p^{-\eps}z\partial_2 S_m(1-t,1-t+s)e^{-q(t,s,{p})z}f(z)
	\end{align*}
	The result follows.

\bigskip
Now, using the same notation as in the full dynamic pricing case above and the same arguments as in proof of \eqref{eq:bellman1}, we have
$$V_k(t,{p},f)=\int_{0}^t f_{\tau_{1-t,{p}}}(s)\Big
	[p +V^*_{k-1}(t-s;T(f;q(t,s,{p})))\Big]ds.$$
and
\begin{equation}
	V^*_k(t,f)= \max_{p>0}\int_{0}^t f_{\tau_{1-t,p}}(s)\left[p+V^*_{k-1}(t-s; T(f;q(t,s,p)))\right]ds. \label{eq:for_induct_ST}
\end{equation}
We now prove by induction on $k$ that for all $k\in\{0,...,C_{dm}\}$,
\begin{equation}\label{eq:stop_l}
	V_k^*(t;f)=\left[{S_m(1-t,1)}\right]^{\frac{1}{\eps}}\alpha^i_{k,s}(f).
\end{equation}
where $\alpha^i_s(0,f)=0$ and for all $k\in\{1,..,C_{dm}\}$, $$\alpha^i_{k,s}(f)=\max_{q>0} q \int_0^1 \left[q^{-1/\eps}+(1-u)^{\frac{1}{\eps}}\alpha^i_{k-1,s}(T(f;qu))\right]\int_0^\infty ze^{-quz}f(z)dzdu.$$

\medskip	
The result holds for $k=0$ since $V_0^*(t;f)=0$. Suppose that it holds for $k-1\geq 0$.  First, by \eqref{eq:f_tau_B}, we have
\begin{equation}
	\label{eq:f_tau_marg}
f_{\tau_{1-t,{p}},S_m}(s,z)=\int_0^\infty p^{-\eps}z\partial_2 S_m(1-t,1-t+s)e^{-q(t,s,{p})z}f(z)dz. 	
\end{equation}
Using \eqref{eq:for_induct_ST}, we obtain
	\begin{align*}
	V^*_k(t,f)	=&\max_{p>0}\int_{0}^t f_{\tau_{1-t,p}}(s)\bigg\{p+\left[S_m(1-t+s,1)\right]^{\frac{1}{\eps}} \times \alpha^i_{k-1,s}(T(f;q(t,s,p)))\bigg\}ds\\
	=&\max_{p>0}\int_0^1 \bigg[\int_0^\infty q(t,p)z  e^{-q(t,p)uz}f(z)dz\bigg] \\
	 &  \hspace{1cm} \times \bigg\{p+\left[S_m(1-t,1) (1-u)\right]^{\frac{1}{\eps}}\alpha^i_{k-1,s}(T(f;q(t,p) u))\bigg\}  du\\
	=&\left[S_m(1-t,1)\right]^{\frac{1}{\eps}}\max_{q>0}q \int_0^1 \left[\int_0^\infty ze^{-quz}f(z)dz\right] \left[q^{-1/\eps}+(1-u)^{\frac{1}{\eps}}\alpha^i_{k-1,s}(T(f;qu))\right] du.
\end{align*}
The second equality follows using  the change of variable $u=S_m(1-t,1-t+s)/S_m(1-t,1)$ and the third by the change of variable $q=q(t,p)$. Hence, the induction step holds, and  \eqref{eq:stop_l} is satisfied for all $k\in\{0,...,C_{dm}\}$. We obtain the desired expression by taking $t=0$ and $k=C_{dm}$.

\medskip
If Assumption \ref{hyp:gamma}(ii) further holds, we obtain by Lemma \ref{ConjugatePrior_general} that if $f=\gamma_{\lambda,\mu}$, then $T(f;u) =\gamma_{\lambda+1,\mu+u}$. Let $V_k(t,{p};\lambda,\mu)$ and $V^*_k(t;\lambda,\mu)$ be defined as in the full dynamic pricing case. Then, by the same induction as above, we have, for all $k\in\{0,...,C_{dm}\}$,
\begin{equation}\label{eq:stopping_gamma_l}
V_k^*(t;\lambda,\mu)=\left[\frac{S_m(1-t,1)}{\mu}\right]^{\frac{1}{\eps}}\alpha^i_{k,s}(\lambda),
\end{equation}
where $\alpha^i_s(0,\lambda)=0$ for $\lambda>0$, and
$$\alpha^i_{k,s}(\lambda)=\max_{q>0} q \int^1_0\frac{\lambda}{(1+qs)^{\lambda+1}}\left[q^{-\frac{1}{\eps}}+\left(\frac{1-s}{1+qs}\right)^{\frac{1}{\eps}}\alpha^i_{k-1,s}(\lambda+1)\right]ds.$$
The result follows by taking $t=0$, $\mu=Y_{dm}^{-1}$, and $k=C_{dm }$, we obtain the desired expression. To obtain the load formula, we follow the same techniques to that under complete information.

\subsubsection{Stopping-time pricing with $M$ fares.}
 As in the complete information case, let $V_k(0;t,p,j)$ (resp. $V_k(1;t,p,j,f)$)  denote the optimal revenue at time $1-t$, with a current price $p$, a remaining capacity $k$, a remaining number of fares $j$ and a density of $f$ for $\xi_{dm}$ (conditional on the current information) if the firm decides to keep the same price (resp. to change it). Then, as \eqref{limited_fares_stopping_time_1}, we have:
\begin{align}
&V_k(0;t,p,j,f)=\int_{0}^{t}f_{\tau_{1-t,p}}(s)\left[p+V^*_{k-1}(t-s,p,j,T(f;q(t,s,p)))\right]ds, \label{eq:V0_inc_M} \\
& V_k(1;t,p,j,f)=\max_{p'>0}\int_{0}^{t}f_{\tau_{1-t,p'}}(s)\left[p'+ V^*_{k-1}(t-s,p',j-1,T(f;q(t,s,p)))\right]ds, \nonumber \\
&V^*_k(t,p,j,f)=\max_{d\in\{0,1\}}V_k(d;t,p,j,f), \nonumber
\end{align}
with the initial conditions $V^*_0(t,p,j,f)=0$. We prove by induction on $k$ that for all $(k,j)\in\{0,...,C_{dm}\}\times \N$,
\begin{equation}
V^*_k(t,p,j,f)=c_{k,j}(q(t,p),f)\left[S_m(1-t,1)\right]^{\frac{1}{\eps}},	
	\label{eq:Vst_inc_M}
\end{equation}
where $c_{k,0}(q,f)=q^{-\frac{1}{\eps}}\int \mathbb{E}[D(qz)\wedge k]f(z)dz$ and
$$\begin{aligned}
	c_{k,j}(q,f)=&\max\Big\{q \int_0^1\int ze^{-qzu}f(z)dz\Big[q^{-1/\eps}+c_{k-1,j\wedge (k-1)}(q(1-u),T(f;qu)) \\
	& \hspace{1cm} (1-u)^{\frac{1}{\eps}}\Big]du, \max_{q'>0} q' \int_0^1\int ze^{-q'zu}f(z)dz\Big[q'{}^{-1/\eps} \\
	& \hspace{1cm} +c_{k-1,j-1}(q'(1-u), T(f;q'u))(1-u)^{\frac{1}{\eps}}\Big]du \Big\}.\\
	\end{aligned}$$
The result holds for $k=0$ since $c_{0,j}=V^*_0(t,p,j,f)=0$. Suppose that it holds for $k-1\geq 0$ and all $j\leq k-1$ (recall that $V^*_k(t,p,j,f)=V^*_k(t,p,m\wedge k,f)$). If $j=0$, the price cannot be changed anymore, so $V^*_k(t,p,j)$ is simply the revenue with price $p$ from $1-t$ to $1$, and \eqref{eq:stopping_time_M_f} holds.

\medskip
If $j\geq 1$, we have, using \eqref{eq:f_tau_marg} and \eqref{eq:V0_inc_M} and the same change of variables as above, we obtain
	\begin{equation*}
	\begin{aligned}
	V_k(0;t,p,j,f) =&\int_{0}^{t}\left[\int_0^\infty(\xi_a+\xi_b)p^{-\eps}\partial_2 S_m(1-t,1-t+s)ze^{-q(t,s,p)z}f(z)dz\right]\\
	&\qquad \Big[p+c_{k-1,j\wedge (k-1)}(q(t-s,p),T(f;q(t,s,p))) [S_m(1-t+s,1)]^{\frac{1}{\eps}}\Big]ds\\
	=&\int_0^1\left[\int_0^\infty q(t,p)ze^{-q(t,p)uz}f(z)dz\right]\Big[p+c_{k-1,j\wedge (k-1)}(q(t,p)(1-u), \\
	&\qquad T(f;q(t,p)u))[S_m(1-t,1)]^{\frac{1}{\eps}}(1-u)^{\frac{1}{\eps}}\Big]du\\
	=&[S_m(1-t,1)]^{\frac{1}{\eps}}q(t,p) \int_0^1\left[\int_0^\infty ze^{-q(t,p)uz}f(z)dz\right]\\
	&\;\times \Big[q(t,p)^{-1/\eps}+c_{k-1,j\wedge (k-1)}(q(t,p)(1-u),T(f;q(t,p)u))(1-u)^{\frac{1}{\eps}}\Big]du.\\
	\end{aligned}
\end{equation*}
By the same reasoning and the change of variable $q=q(t,p)$,
\begin{equation*}
	\begin{aligned}	
	V_k(1;t,p,j,f)	&=[S_m(1-t,1)]^{\frac{1}{\eps}}\max_{q>0}q \int_0^1\left[\int_0^\infty ze^{-qzu}f(z)dz\right]\\
	&\quad \Big[q^{-1/\eps}+c_{k-1,j-1}(q(1-u),T(f;qu))(1-u)^{\frac{1}{\eps}}\Big]du.\\
	\end{aligned}
\end{equation*}
Then, $V^*_k(t,p,j,f)=\max_{d\in\{0,1\}}V_k(d;t,p,j,f)=c_{k,j}(q(p),f)[S_m(1-t,1)]^{\frac{1}{\eps}}$. This concludes the induction step, proving that \eqref{eq:Vst_inc_M} holds for all $k\in\{0,...,C_{dm}\}$.

\subsubsection{Stopping-time pricing with $M$ increasing fares.} The proof follows by making the same changes as those made in the complete information setup.

\subsubsection{Intermediate-$K$ stopping-time pricing.} The proof follows by making the same changes as those made in the complete information setup.

\end{document}